\documentclass[sigconf,nonacm]{acmart}

\def\REVIEW{0}  

\usepackage{amsmath}
\usepackage{amsfonts}
\usepackage{svg}

\usepackage{pifont}
\usepackage{bm}
\usepackage{algorithm}
\usepackage[noend]{algpseudocode}
\usepackage{etoolbox}
\usepackage{enumerate}
\usepackage{subcaption}
\usepackage{relsize}
\usepackage{setspace}
\usepackage{hyperref}
\usepackage{printlen}
\usepackage{newfloat}
\usepackage{bm}
\usepackage{enumitem}
\usepackage{ragged2e}
\usepackage{cuted}
\usepackage{framed}

\usepackage{makecell}
\usepackage{multirow}

\usepackage{placeins}
\usepackage{chngcntr}
\usepackage{balance}

\if \REVIEW 1
    \usepackage[enable,applyremove]{xnotes}
\else
    \usepackage[apply,nonotes]{xnotes}
\fi

\newtheorem{theorem}{Theorem}

\definecolor{darkblue}{RGB}{0,0,200}

\AddXNotesUser{pk}{PK}{brown}
\AddXNotesUser{ya}{YA}{magenta}
\AddXNotesUser{pp}{PP}{red}
\AddXNotesUser{at}{AT}{blue}

\usepackage[continuouslinenumbers,cppcommentstyle,mainfont=small]{smartalgorithmic}

\usepackage{cleveref}

\newcommand{\remove}[1]{}

\NewDeclarationType[prefix=p]{Property}{\text}
\NewDeclarationType[prefix=o]{Operation}[1]{\textsf{\textrm{#1}}}
\NewDeclarationType[prefix=f]{Function}[1]{\textsf{\textrm{#1}}}
\NewDeclarationType[prefix=m]{MessageType}{\textsc}
\NewDeclarationType[prefix=s]{StringLiteral}[1]{\text{``\MakeLowercase#1''}}

\newcommand{\variableBaseStyle}[1]{\mathit{#1}}
\NewDeclarationType[prefix=v]{Variable}[1]{\variableBaseStyle{\MakeLowercase#1}}

\NewDeclarationType[prefix=o]{Object}{\textsc}

\newcommand{\Tuple}[1]{\langle#1\rangle}

\NewDeclarationType{Constant}{\mathit}
\DeclareGlobalConstant[True]{true}
\DeclareGlobalConstant[False]{false}
\DeclareGlobalConstant[Null]{\ensuremath{\perp}}

\newcommand{\tnot}{\text{not}~}
\newcommand{\tand}{\mathbin{\textbf{and}}}
\newcommand{\tor}{\mathbin{\textbf{or}}}

\newcommand{\tis}{\mathbin{\textbf{is}}}

\algnewcommand{\Type}[2]{\textbf{type}~#1~\oassign~#2}
\newcommand{\Epoch}{E}
\newcommand{\Execution}{\mathcal{E}}

\newcommand{\oassign}{\mathbin{:=}}

\newcommand{\ignore}[1]{}

        \newcommand{\AssetTransfer}{Asset Transfer}

        \DeclareGlobalOperation{Transfer}
    \DeclareGlobalOperation{IssueCheque}
    \DeclareGlobalOperation{DepositCheque}
    \DeclareGlobalOperation{GetAccountTransactions}
    
        \DeclareGlobalProperty[TransferSafety]{Transfer Safety}
    \DeclareGlobalProperty[TransferConsistency]{Transfer Consistency}
    \DeclareGlobalProperty[TransferValidity]{Transfer Validity}
    \DeclareGlobalProperty[TransferLiveness]{Transfer Liveness}
    \DeclareGlobalProperty[AccountTransactions]{Account Transactions Completeness}
        \DeclareGlobalProperty[TransferConcurrency]{Transfer Concurrency}
    
        \DeclareGlobalFunction[Start]{start}
    \DeclareGlobalFunction[End]{end}
    \DeclareGlobalFunction[CommitTime]{commitTime}

        \newcommand{\Name}{CryptoConcurrency}
    
        \DeclareGlobalFunction{Recovery}
    
        \DeclareGlobalMessageType{ConfirmInRecovery}
    
        \DeclareGlobalVariable{CommittedDebits}
    \DeclareGlobalVariable*[NEWConfirmCert]{\ensuremath{\sigma_{\mathit{confirm}}}}

        \newcommand{\CODLong}{Closable Overspending Detector} 
    \newcommand{\CODShort}{COD}

        \newcommand{\RODLong}{Recoverable Overspending Detector}

        \DeclareGlobalObject{COD}

        \DeclareGlobalOperation{Submit}
    \DeclareGlobalFunction[VerifyCODCert]{VerifyCODCert}
    \DeclareGlobalOperation[VerifyCODCert]{VerifyCODCert}
    \DeclareGlobalOperation{Close}
    \DeclareGlobalOperation{VerifyCloseStateCert}

        \DeclareGlobalProperty[CODProposeValidity]{COD-Submit Validity}
    \DeclareGlobalProperty[CODProposeSafety]{COD-Submit Safety}
    \DeclareGlobalProperty[CODProposeSuccess]{COD-Submit Success}
    \DeclareGlobalProperty[CODCloseValidity]{COD-Close Validity}
    \DeclareGlobalProperty[CODCloseSafety]{COD-Close Safety}
    \DeclareGlobalProperty[CODLiveness]{COD-Liveness}
    
            \DeclareGlobalFunction{Accept}
    \DeclareGlobalMessageType{AlreadyPrepared}
    
        \DeclareGlobalMessageType{Deps}      \DeclareGlobalMessageType{AcceptRequest}
    \DeclareGlobalMessageType{AcceptAck}
    \DeclareGlobalMessageType{ConfirmState}
    \DeclareGlobalMessageType{ConfirmStateResp}
    
        \DeclareGlobalVariable*[CODState]{\variableBaseStyle{CODState}}
    \DeclareGlobalVariable[NextCODState]{NextCODState}
    \DeclareGlobalVariable[NextCODStateCert]{NextCODStateCert}
    \DeclareGlobalVariable{SelectedDebits}
    \DeclareGlobalVariable{Deps}
    \DeclareGlobalVariable{SubmittedDebits}
    \DeclareGlobalVariable*[AcceptCert]{\ensuremath{\sigma_{\mathit{accept}}}}
    \DeclareGlobalVariable*[CryptoCDCert]{\ensuremath{\sigma_{\mathit{COD}}}}
    \DeclareGlobalVariable*[RecoveryCert]{\ensuremath{\sigma_{\mathit{recovery}}}}
    
    \DeclareGlobalVariable{InitCredits}
    \DeclareGlobalVariable{OutCredits}
    \DeclareGlobalVariable{NewCredits}
    \DeclareGlobalVariable{AcceptedDebits}

        \newcommand{\GlobalStorage}{Global Storage}

        \DeclareGlobalObject{GlobalStorage}

        \DeclareGlobalStringLiteral{Txs}

        \newcommand{\AccountStorage}{Account Storage}
        \DeclareGlobalObject{AccountStorage}

        \DeclareGlobalStringLiteral{Debits}
    \DeclareGlobalStringLiteral{State}

        \newcommand{\KeyValStorage}{Append-Only Storage}
    
        \DeclareGlobalObject{AOStorage}

    \newcommand{\AppendOnlyStorage}{Append-Only Storage}
    
        \DeclareGlobalOperation{AppendKey}
    \DeclareGlobalOperation{ReadKey}
    \DeclareGlobalFunction[KVSVerifyStoredCert]{VerifyStoredCert}

    \DeclareGlobalFunction{WriteValuesToKey}
  
        \DeclareGlobalProperty[KVSConsistency]{AOS-Consistency}
    \DeclareGlobalProperty[KVSInputValidity]{AOS-Input Validity}
    \DeclareGlobalProperty[KVSOutputValidity]{AOS-Output Validity}
    \DeclareGlobalProperty[KVSLiveness]{AOS-Liveness}

        \DeclareGlobalMessageType{AppendKey}
    \DeclareGlobalMessageType{AppendKeyResp}
    
        \DeclareGlobalVariable{Key}
    \DeclareGlobalVariable{Val}
    \DeclareGlobalVariable{Log}
    \DeclareGlobalVariable*[InitV]{\ensuremath{v_{init}}}
    \DeclareGlobalVariable*[InitVs]{\ensuremath{vs_{init}}}

    \newcommand{\Consensus}{Consensus}

        \DeclareGlobalObject{Consensus}
    
        \DeclareGlobalOperation[ConsPropose]{Propose}
    
        \DeclareGlobalProperty[ConsValidity]{C-Validity}
    \DeclareGlobalProperty[ConsConsistency]{C-Consistency}
    \DeclareGlobalProperty[ConsLiveness]{C-Liveness}

        \DeclareGlobalVariable[Cons]{Consensus}

        \DeclareGlobalFunction[CreateThresholdSignature]{CreateTS}
    \DeclareGlobalFunction[VerifyThresholdSignature]{VerifyTS}
    
            \DeclareGlobalFunction{MerkleTree}
    \DeclareGlobalFunction{GetItemProof}
    \DeclareGlobalFunction{VerifyItemProof}
    \DeclareGlobalVariable[MTRoot]{root}
        \DeclareGlobalVariable{ItemProof}
    \DeclareGlobalVariable{MerkleTree}
    \DeclareGlobalVariable*[MTCert]{\ensuremath{\sigma_{MT}}}
    \DeclareGlobalVariable*[VCert]{\ensuremath{\sigma_{v}}}
    \DeclareGlobalVariable*[TSMrkTree]{\vMTCert}

    \DeclareGlobalFunction[Balance]{balance}
    
    \DeclareGlobalFunction[Credits]{credits}
    \DeclareGlobalFunction[Debits]{debits}
    \DeclareGlobalFunction[EpochStart]{epochStart}
    \DeclareGlobalFunction{Sign}
    \DeclareGlobalFunction{VerifyCommitCertificate}
    \DeclareGlobalFunction[Transactions]{Txs}
    \DeclareGlobalFunction{Verify}
    \DeclareGlobalFunction{VerifyCert}
    \DeclareGlobalFunction{VerifyInitTxs}
    \DeclareGlobalFunction{VerifyTxCertForGlobalStorage}
        \DeclareGlobalFunction{Prepare}
            
    \DeclareGlobalFunction[ExtractSafeTransactions]{SplitDebits}
    \DeclareGlobalFunction{ExtractDebits}
    \DeclareGlobalFunction{UpdateCommittedState}
    \DeclareGlobalFunction{VerifyVcProof}
    \DeclareGlobalFunction{VerifyDebitTx}
    
    \DeclareGlobalFunction{SignTx}
    \DeclareGlobalFunction{VerifySignedTx}
    \DeclareGlobalFunction{VerifyCommittedTx}
    \DeclareGlobalFunction{VerifyTx}
    \DeclareGlobalFunction{VerifyEpochState}
    \DeclareGlobalFunction{VerifyAcceptTxs}
    \DeclareGlobalFunction{IsValidPrepareResp}
    \DeclareGlobalFunction{VerifyCreditTx}
    \DeclareGlobalFunction{VerifySignature}
    \DeclareGlobalFunction{VerifyTxSignature}
    
    \DeclareGlobalFunction{VerifyStartRecoveryAckMessages}
    \DeclareGlobalFunction{GetUniqueId}
    \DeclareGlobalFunction{AdversaryBalance}
    \DeclareGlobalFunction{GenesisBalance}
    \DeclareGlobalFunction{AccState}
    \DeclareGlobalFunction{FetchCommittedTxs}
    \DeclareGlobalFunction{VerifyCloseRespMessages}
    \DeclareGlobalFunction{VerifyPreparedDebits}
            \DeclareGlobalFunction{TotalValue}
    \DeclareGlobalFunction{Hash}
    \DeclareGlobalFunction{CommittedBalance}

    \DeclareGlobalFunction{VerifyConfirmedTx}
    \DeclareGlobalFunction{VerifyRecoveryCert}
    \DeclareGlobalFunction[VerifyValue]{VerifyInput}
    \DeclareGlobalFunction{ReadLatestCODState}

    \DeclareGlobalVariable{State}
    \DeclareGlobalVariable{States}
    \DeclareGlobalVariable{Amount}
    \DeclareGlobalVariable*[StateCert]{\ensuremath{\sigma_{\textit{state}}}}

    \newcommand{\WithCert}{^{\sigma}}
    \DeclareGlobalVariable{CommittedTxs}
    
    \DeclareGlobalVariable{Txs}
        
    \DeclareGlobalVariable{Sender}
    \DeclareGlobalVariable{ReadDebits}
    \DeclareGlobalVariable{Recipient}
    \DeclareGlobalVariable{Acc}
    \DeclareGlobalVariable{Msg}
    \DeclareGlobalVariable{A}
    \DeclareGlobalVariable{Tx}
    \DeclareGlobalVariable{SignedStates}
    \DeclareGlobalVariable{SequenceNumber}
    \DeclareGlobalVariable{Acks}
    \DeclareGlobalVariable*[AcksTransfer]{\vAcks_{\mathrm{transfer}}}
    \DeclareGlobalVariable*[AcksChange]{\vAcks_{\mathrm{change}}}
    \DeclareGlobalVariable{StoredTxs}
    \DeclareGlobalVariable{LearnedTxs}
    \DeclareGlobalVariable{StartedTxs}
    \DeclareGlobalVariable{StartedDebits}
    \DeclareGlobalVariable{PendingDebits}
    \DeclareGlobalVariable{SignedCredits}
        \DeclareGlobalVariable{Credits}
    \DeclareGlobalVariable{Debit}
    \DeclareGlobalVariable{Debits}
    \DeclareGlobalVariable{HelpingDebits}
    \DeclareGlobalVariable{Dbts}
    \DeclareGlobalVariable{PrepareResult}
            \DeclareGlobalVariable{Used}
    \DeclareGlobalVariable{Unused}
    \DeclareGlobalVariable{Total}
    \DeclareGlobalVariable{Id}
    \DeclareGlobalVariable{TxId}
    \DeclareGlobalVariable*[vSigTransfer]{\ensuremath{\sigma_{\mathrm{transfer}}}}
    \DeclareGlobalVariable*[vSigChange]{\ensuremath{\sigma_{\mathrm{change}}}}
    \DeclareGlobalVariable{ChangeAmount}
        \DeclareGlobalVariable*[Epoch]{epoch}
    \DeclareGlobalVariable*[Ep]{e}
    \DeclareGlobalVariable[Signature]{sig}
    \DeclareGlobalVariable[Signatures]{sigs}
    \DeclareGlobalVariable[SignatureState]{sigState}
    \DeclareGlobalVariable[SignatureTxs]{sigTxs}
    \DeclareGlobalVariable{LatState}
    \DeclareGlobalVariable{LatStates}
    \DeclareGlobalVariable{Status}
    \DeclareGlobalVariable{Opened}
    \DeclareGlobalVariable{Closed}
    \DeclareGlobalVariable{IsClosed}
    \DeclareGlobalVariable{Var}
    \DeclareGlobalVariable{Crds}
    \DeclareGlobalVariable{AllDebits}
    \DeclareGlobalVariable{ReceivedTxs}
    \DeclareGlobalVariable{SubmitDebits}
    \DeclareGlobalVariable{AcceptState}
    \DeclareGlobalVariable{PreparedTxs}
    \DeclareGlobalVariable{PreparedDebits}
    \DeclareGlobalVariable*[PreparedCert]{\ensuremath{\sigma_{prepare}}}
    \DeclareGlobalVariable*[AcceptedTxsBalCert]{\textit{acceptedTxsCert}}
    \DeclareGlobalVariable*[StartEpochMessage]{startEpochMsg}
    \DeclareGlobalVariable{TransientErrorEpoch}
    \DeclareGlobalVariable{CorrectAccs}
    \DeclareGlobalVariable{ByzantineAccs}
    \DeclareGlobalVariable{KnownCommitedTxs}
    \DeclareGlobalVariable{ReceivedDebits}
    \DeclareGlobalVariable{CancelledDebits}
    \DeclareGlobalVariable{RestrictedDebits}
    \DeclareGlobalVariable*[CODResult]{\variableBaseStyle{CODResult}}
    \DeclareGlobalVariable*[TxInitForA]{\ensuremath{\vTx_{
    \textit{init, a}}}}
    \DeclareGlobalVariable*[TxInitForAcc]{\ensuremath{\vTx_{
    \textit{init, acc}}}}
        \DeclareGlobalVariable{AllCredits}
    \DeclareGlobalVariable{ReceivedCredits}

    \DeclareGlobalVariable{Obj}
    \DeclareGlobalVariable{Vs}
    \DeclareGlobalVariable*[VsKVSCert]{\ensuremath{vs_\textit{AOS}^{\sigma}}}
    \DeclareGlobalVariable{Vc}
        \DeclareGlobalVariable{ExtraTxs}
    \DeclareGlobalVariable{SafeExtraTxs}
    \DeclareGlobalVariable{AllTxs}
    \DeclareGlobalVariable*[Cert]{\ensuremath{\sigma}}
    \DeclareGlobalVariable*[PrepareCert]{\ensuremath{\sigma_{\mathit{prepare}}}}
    \DeclareGlobalVariable*[CommitCert]{\ensuremath{\sigma_{\mathit{commit}}}}
    \DeclareGlobalVariable*[CODCert]{\ensuremath{\sigma_{\mathit{COD}}}}
    \DeclareGlobalVariable*[DebitCert]{\ensuremath{\sigma_{\mathit{debit}}}}
    \DeclareGlobalVariable*[CreditCert]{\ensuremath{\sigma_{\mathit{credit}}}}
        \DeclareGlobalVariable*[TxCert]{\ensuremath{\sigma_{\mathit{\vTx}}}}
    \DeclareGlobalVariable*[TxSigned]{\ensuremath{tx_{\sigma}}}
    \DeclareGlobalVariable*[TxsSigned]{\ensuremath{txs_{\sigma}}}
    \DeclareGlobalVariable*[AllTxsSigned]{\ensuremath{allTxs_{\sigma}}}
    
    \DeclareGlobalVariable*[ObjCert]{\ensuremath{\sigma_{\textit{obj}}}}
    \DeclareGlobalVariable{CommittedState}
    \DeclareGlobalVariable{NecessaryTxs}
        \DeclareGlobalVariable*[EpochCert]{\ensuremath{\sigma_{\mathit{epoch}}}}
    \DeclareGlobalVariable*[ECert]{\ensuremath{\sigma_{\mathit{e}}}}
    \DeclareGlobalVariable[ProgramCounter]{Pc}
    \DeclareGlobalVariable{Messages}
    \DeclareGlobalVariable{First}
    \DeclareGlobalVariable{Second}
    \DeclareGlobalVariable{NewState}
    \DeclareGlobalVariable{OutgoingTxs}
    \DeclareGlobalVariable{IncomingTxs}
    \DeclareGlobalVariable{EpochStartTxs}
    \DeclareGlobalVariable{EpochStartTxsCert}
    \DeclareGlobalVariable{StartEpoch}
    \DeclareGlobalVariable{GenesisTxs}
    \DeclareGlobalVariable{InitTxs}
    \DeclareGlobalVariable{InitDebits}
        \DeclareGlobalVariable*[InitTxsCert]{\ensuremath{\sigma_{\mathit{init}}}}
    \DeclareGlobalVariable[CloseStateCert]{closedStateCert}
    \DeclareGlobalVariable{MsgTxs}
    \DeclareGlobalVariable{AllMsgTxs}
    \DeclareGlobalVariable{Expr}
    \DeclareGlobalVariable*[SigTransfer]{\vSignature_{\textit{transfer}}}
    \DeclareGlobalVariable*[SigChange]{\vSignature_{\textit{change}}}
    \DeclareGlobalVariable*[vSigTx]{\ensuremath{\sigma_{\vTx}}}
    \DeclareGlobalVariable{Account}
    \DeclareGlobalVariable{NewEpoch}
    \DeclareGlobalVariable{InitialState}
    \DeclareGlobalVariable*[StatusCert]{\ensuremath{\sigma_{\textit{status}}}}
    \DeclareGlobalVariable*[ClosedCert]{\ensuremath{\sigma_{\textit{closed}}}}
    \DeclareGlobalVariable[CODStateCert]{CryptoCDStateCert}
    \DeclareGlobalVariable{DebitsHash}
    \DeclareGlobalVariable{ClosedState}
    \DeclareGlobalVariable{Value}
    \DeclareGlobalVariable*[KVSCert]{\ensuremath{\sigma_{\textit{AOS}}}}
    \DeclareGlobalVariable*[KVSVs]{\ensuremath{vs_{\textit{AOS}}}}
    \DeclareGlobalVariable*[STime]{t_{\textrm{stable}}}

    \DeclareGlobalMessageType{InitCOD}
    \DeclareGlobalMessageType[CommitCODInitialState]{CommitInitState}
    \DeclareGlobalMessageType[CommitCODInitialStateResp]{CommitInitStateResp}
    \DeclareGlobalMessageType{CommitCODInitialStateTx}
    \DeclareGlobalMessageType{Transfer}
    \DeclareGlobalMessageType{TransferPrepare}
    \DeclareGlobalMessageType{Payment}
    \DeclareGlobalMessageType{TransferAck}
    \DeclareGlobalMessageType{GlobalStorage}
    \DeclareGlobalMessageType{Receipt}
    \DeclareGlobalMessageType{Promise}
    \DeclareGlobalMessageType{Accepted}
    \DeclareGlobalMessageType{Prepare}
    \DeclareGlobalMessageType{PrepareResp}
    \DeclareGlobalMessageType{Propose}
    \DeclareGlobalMessageType{ProposeResp}
    \DeclareGlobalMessageType{PrepareEpoch}
    \DeclareGlobalMessageType{PrepareEpochResp}
    \DeclareGlobalMessageType{Restart}
    \DeclareGlobalMessageType{RequestState}
    \DeclareGlobalMessageType{UpdateEpoch}
    \DeclareGlobalMessageType{StartEpoch}
    \DeclareGlobalMessageType{StartEpochResp}
    \DeclareGlobalMessageType{ReadStorage}
    \DeclareGlobalMessageType{ReadStorageResp}
    \DeclareGlobalMessageType{WriteStorage}
    \DeclareGlobalMessageType{WriteStorageResp}
    \DeclareGlobalMessageType{WriteDebitTxs}
    \DeclareGlobalMessageType{WriteDebitTxsResp}
    \DeclareGlobalMessageType{WriteStorageStartEpochMsg}
    \DeclareGlobalMessageType{WriteStorageStartEpochMsgResp}
    \DeclareGlobalMessageType{ReadDebits}
    \DeclareGlobalMessageType{ReadDebitsResp}
    \DeclareGlobalMessageType{WriteDebits}
    \DeclareGlobalMessageType{WriteDebitsResp}
    \DeclareGlobalMessageType{Close}
    \DeclareGlobalMessageType{CloseResp}
    \DeclareGlobalMessageType{Closed}
    \DeclareGlobalMessageType{ReadPendingDebits}
    \DeclareGlobalMessageType{ReadPendingDebitsResp}
    \DeclareGlobalMessageType{ReadLatestCODState}
    \DeclareGlobalMessageType{WriteCODState}
    \DeclareGlobalMessageType{ReadLatestCODStateResp}
    \DeclareGlobalMessageType{WriteCODStateResp}
    \DeclareGlobalMessageType{WriteTransaction}
    \DeclareGlobalMessageType{Epoch}      \DeclareGlobalMessageType{ReadKey}
    \DeclareGlobalMessageType{ReadKeyResp}
    \DeclareGlobalMessageType{WriteKey}
    \DeclareGlobalMessageType{WriteKeyResp}

    \newcommand{\aA}{\mathcal{A}}
    
    \newcommand{\tT}{\mathcal{T}}
        
    \newcommand{\typeStyle}[1]{\text{#1}}
    
    \newcommand{\tSet}[1]{\typeStyle{Set}\langle#1\rangle}
    
    \newcommand{\tCert}{\Sigma}
    \newcommand{\tPair}[2]{\typeStyle{Pair}\langle#1, #2\rangle}
        \newcommand{\tTuple}[1]{\typeStyle{Tuple}\langle#1\rangle}

    \newcommand{\tBoolean}{\typeStyle{Boolean}}
    
    \newcommand{\tEpochNum}{\typeStyle{EpochNum}}
    \newcommand{\tSignedTx}{\tT_{\Sigma}}

    \newcommand{\tOK}{\typeStyle{OK}}
    \newcommand{\tFAIL}{\typeStyle{FAIL}}
        
    \newcommand{\tCloseState}{\typeStyle{CloseState}}

\newenvironment{properties}
{    \begin{description}%
        \newcommand{\propertyitem}[1]{\item[##1:]}
}{    \undef\propertyitem
    \end{description}%
}

\DeclareFloatingEnvironment{myalgorithmfloat}

\newenvironment{myalgorithm}[1][htbp]
{                \begin{algorithm*}[#1]
}{    \end{algorithm*}
            }

\newenvironment{myalgorithmic}
{    \begin{smartalgorithmic}[1]
}{    \end{smartalgorithmic}
}

\floatstyle{ruled}
\newfloat{nonfloatalgorithmfloat}{H}{loa}
\floatname{nonfloatalgorithmfloat}{Algorithm}
\crefname{nonfloatalgorithmfloat}{algorithm}{algorithms}
\Crefname{nonfloatalgorithmfloat}{Algorithm}{Algorithms}

\makeatletter
\let\c@nonfloatalgorithmfloat\c@algorithm
\makeatother

\newenvironment{nonfloatalgorithm}{
            \noindent
        \begin{minipage}{\textwidth}
    \begin{nonfloatalgorithmfloat}[H]
}{
    \end{nonfloatalgorithmfloat}
    \end{minipage}
    }

\algnewcommand{\Returns}[1]{\textbf{returns} {#1}}
\algnewcommand{\Where}[1]{\Statex \textbf{where} {#1}}

\algnewcommand{\Throw}{\textbf{throw}}

\algblockdefx[ProcessTypes]{ProcessTypes}{EndProcessTypes}{\textbf{Types:}}{}
\algtext*{EndProcessTypes} 
\algblockdefx[ProcessParameters]{ProcessParameters}{EndProcessParameters}{\textbf{Parameters:}}{}
\algtext*{EndProcessParameters} 
\algblockdefx[ProcessObjects]{ProcessObjects}{EndProcessObjects}{\textbf{Storage Objects:}}{}
\algtext*{EndProcessObjects}  
\algblockdefx[ProcessState]{ProcessState}{EndProcessState}{\textbf{State:}}{}
\algtext*{EndProcessState}  
\algblockdefx[ReplicaState]{ReplicaState}{EndReplicaState}{\textbf{Replica State:}}{}
\algtext*{EndReplicaState}  

\algblockdefx[Function]{Function}{EndFunction}[2]{\textbf{function} #1(#2)}{}
\algtext*{EndFunction}  
\algblockdefx[PublicFunction]{PublicFunction}{EndPublicFunction}[2]{\textbf{public function} #1(#2)}{}
\algtext*{EndPublicFunction}  

\algblockdefx[Procedure]{Procedure}{EndProcedure}[2]{\textbf{procedure} #1(#2)}{}
\algtext*{EndProcedure}  
\algblockdefx[Operation]{Operation}{EndOperation}[2]{\textbf{operation} #1(#2)}{}
\algtext*{EndOperation}  
\algblockdefx[Upon]{Upon}{EndHandler}[1]{\textbf{upon} #1}{}
\algtext*{EndHandler}  
\algblockdefx[UponReceive]{UponReceive}{EndHandler}[2]{\textbf{upon receive} $\langle$#1$\rangle$ \textbf{from} #2}{}
\algtext*{EndHandler}  

\algblockdefx[UponReceiveWhen]{UponReceiveWhen}{EndHandler}[3]{\textbf{upon receive} $\langle$#1$\rangle$ \textbf{from} #2 \textbf{when} #3}{}
\algtext*{EndHandler}

\algblockdefx[UponReceiveAnyMessage]{UponReceiveAnyMessage}{EndHandler}[2]{\textbf{upon receive} any message \textbf{from} #1 \textbf{where} #2}{}
\algtext*{EndHandler}  
\algblockdefx[UponDeliver]{UponDeliver}{EndHandler}[1]{\textbf{upon deliver} $\langle$#1$\rangle$}{}
\algtext*{EndHandler}  
\algblockdefx[UponReceive]{UponWRBDeliver}{EndHandler}[2]{\textbf{upon WRB-deliver} $\langle$#1$\rangle$ \textbf{from} #2}{}
\algtext*{EndHandler}  
\algblock[TryCatchFinally]{try}{EndTry}
\algcblock[TryCatchFinally]{TryCatchFinally}{finally}{EndTry}
\algcblockdefx[TryCatchFinally]{TryCatchFinally}{catch}{EndTry}
	[1]{\textbf{catch} #1}
	{}
\algtext*{EndTry}

\algnewcommand{\WaitFor}{\textbf{wait for} }
\algnewcommand{\Send}[2]{\textbf{send} $\langle$#1$\rangle$ \textbf{to} #2}
\algnewcommand{\ForwardMessage}[2]{\textbf{forward message} #1 \textbf{to} #2}
\algnewcommand{\Broadcast}[1]{\textbf{broadcast} $\langle$#1$\rangle$}
\algnewcommand{\Assert}[1]{\textbf{assert}~#1}

\algnewcommand{\IfThenElse}[3]{  \State \algorithmicif\ #1\ \algorithmicthen\ #2\ \algorithmicelse\ #3}
 \algnewcommand{\IfThen}[2]{\State \algorithmicif\ #1\ \algorithmicthen\ #2\ }

\newcommand{\Message}[1]{\langle#1\rangle}

\newcommand{\myparagraph}[1]{\vspace{-1pt}\subsubsection*{#1.}}
\newcommand{\mysubparagraph}[1]{\vspace{-1pt}\paragraph*{#1.}}

\begin{document}

\settopmatter{printfolios=true}

\title{\Name: (Almost) Consensusless Asset Transfer with Shared Accounts}

\if \REVIEW 0
    \author{Andrei Tonkikh}
    \authornote{Andrei Tonkikh and Pavel Ponomarev share first authorship.}
    \email{tonkikh@telecom-paris.fr}
    \affiliation{%
      \institution{T\'el\'ecom Paris, Institut Polytechnique de Paris}
      \country{France}
    }

    \author{Pavel Ponomarev}
    \authornotemark[1]
        \email{pavponn@gmail.com}
    \affiliation{%
      \institution{Georgia Institute of Technology}
      \country{USA}
    }

    \author{Petr Kuznetsov}
        \email{petr.kuznetsov@telecom-paris.fr}
    \affiliation{%
      \institution{T\'el\'ecom Paris, Institut Polytechnique de Paris}
      \country{France}
    }
    
    \author{Yvonne-Anne Pignolet}
    \email{yvonneanne@dfinity.org}
    \affiliation{%
      \institution{DFINITY}
      \country{Switzerland}
    }
\fi
    
\begin{abstract}
    A typical blockchain protocol uses consensus to make sure that mutually mistrusting users agree on the order in which their operations on shared data are executed.
    \atreplace{It is known, however,}{However, it is known} that \emph{asset transfer} systems, by far the most popular application of blockchains, can be implemented without consensus.
    Assuming that no account can be accessed \atreplace{\emph{concurrently}}{concurrently}\atreplace{, i.e., that}{ and} every account belongs to a single owner,
    one can efficiently implement an asset transfer system in a purely asynchronous, consensus-free manner.
    It has also been shown that\atadd{ implementing} asset transfer with shared accounts is impossible\atremove{ to implement} without consensus.
    
    In this paper, we propose {\Name}, an asset transfer protocol that allows concurrent accesses to be processed \emph{in parallel}, without involving consensus, \emph{whenever possible}.
    More precisely, if concurrent transfer operations on a given account do not lead to overspending, i.e. can all be applied without the account balance going below zero, they proceed in parallel.
    Otherwise, the account's owners may have to access an external consensus object.
    \atrev{Notably, we avoid relying on a central, universally-trusted, consensus mechanism and allow each account to use its own consensus implementation, which only the owners of this account trust. This provides greater decentralization and flexibility.}
\end{abstract}

\maketitle

\section{Introduction} \label{sec:intro}

\atrev{The ability to transfer assets from one user's account to another user's account despite the potential presence of malicious parties comes naturally when the users are able to solve Byzantine fault-tolerant consensus~\cite{lamport1982byzantine} in order to reach an agreement on the evolution of the system state.}
They can simply agree on the order in which their transactions are executed.
Indeed, for a long time consensus-based blockchain protocols~\cite{bitcoin,ethereum} have remained the \emph{de facto} standard to implement \emph{asset transfer} (also known as \emph{cryptocurrency}).

\atrev{However, Byzantine fault-tolerant consensus is a notoriously hard synchronization problem.
Not only is it impossible to solve deterministically in asynchronous systems~\cite{flp}, there are also harsh lower bounds on its costs even with stronger synchrony assumptions: at least $\Omega(f^2)$ messages~\cite{dolev1985bounds} and $\Omega(f)$ rounds of communication~\cite{aguilera1999simple,dolev1983authenticated,dutta2002inherent}, even in the synchronous model (i.e., when there is a known upper bound $\Delta$ on the time it takes for a message sent by a correct process to reach its destination).
Despite the efforts of many brilliant researchers and engineers, existing consensus-based blockchain implementations still struggle to achieve latency, throughput, and transaction fees acceptable for widely applicable payment systems.}

The good news is that consensus is not necessary to implement an asset transfer system~\cite{cons-crypto,Gup16}. 
This observation led to a series of purely asynchronous, \emph{consensus-free} cryptocurrencies~\cite{astro-dsn,fastpay,sliwinski2022consensus,pastro21disc}.
Practical evaluations have confirmed that such solutions have significant advantages over consensus-based protocols in terms of scalability, performance, and robustness~\cite{astro-dsn,fastpay}.

However, all existing consensus-free implementations share certain limitations.
\atreplace{Notably}{In particular}, they assume that each account is controlled by a single user that never issues multiple transactions in parallel.
This assumption precludes sharing an account by multiple users, e.g., by family members, or safely accessing it from multiple devices.
If an honest user \atreplace{does accidentally issue}{accidentally issues} several concurrent transactions, the existing implementations may block the account forever, without any possibility to recover it.
Consensus-free systems based on the UTXO model~\cite{wiki:utxo}, such as~\cite{sliwinski2022consensus}, share the same restriction.

In this paper, we propose \emph{\Name}, a hybrid protocol that combines the benefits of both approaches. 
It allows accounts to be shared by multiple users and avoids using consensus in most cases.
Indeed, as demonstrated in~\cite{cons-crypto}, in certain cases, consensus is unavoidable. Therefore, the challenge is to minimize its use.

Informally, in our implementation, if transactions concurrently issued on the same account can \emph{all} be applied without exhausting the account's balance, they are processed in parallel, in a purely asynchronous way (i.e., without invoking consensus).
This property appears natural as such transactions can be ordered arbitrarily, and the order will not affect the resulting account's state. %
In contrast, when the account balance does not allow for accepting all of the concurrent transactions, the account owners may use consensus to agree which transactions should be accepted and which ones should be considered failed due to the lack of funds.

Our protocol dynamically detects the cases when consensus should be used.
This distinguishes our approach from earlier work on combining weak and strong synchronization in one implementation, where conflicts were defined in a \emph{static} way, i.e., any \emph{potentially} conflicting concurrent operations incur the use of consensus, both in general-purpose systems~\cite{lamport2010generalized,li2012redblue,byzgenpaxos,bazzi2022clairvoyant} and in systems specialized for asset transfer~\cite{sliwinski2022consensus}.   

Interestingly, every account can be associated with distinct consensus instances that only need to be accessed by the account's owners. 
In practice, consensus instances for different accounts can be implemented in different ways and on different hardware, depending on the trust assumptions of their owners.

We believe that the results of this paper can be further generalized to applications beyond asset transfer and that this work can be a step towards devising \emph{optimally-concurrent} protocols that dynamically determine the cases when falling back to stronger synchronization primitives is unavoidable.
Intuitively, it seems to be possible for an object with a sequential specification~\cite{linearizability} to operate in a purely asynchronous manner without resorting to consensus in any executions where reordering of the concurrent operations does not affect their outcomes.
This enables the development of lightweight, adaptive implementations that can avoid the costs of heavy synchronization primitives in most cases without compromising functionality or sacrificing liveness even in highly concurrent scenarios.

\myparagraph{Roadmap}
The rest of this paper is organized as follows.
We overview related work in \Cref{sec:related}.
In \Cref{sec:model}, we introduce our system model. 
In \Cref{sec:architecture}, we overview the basic principles of our algorithm.  
We state the problem\atadd{ and the main theorem of the paper} formally in \Cref{sec:problem-statement}, describe\atadd{ the key} building blocks in \Cref{sec:cod-main,sec:keyval-storage} and \atreplace{outline our implementation in \Cref{sec:implementation}}{provide the complete protocol in \Cref{sec:implementation}}.
We conclude the paper in \Cref{sec:discussion}.
Details on the algorithm and its proof of correctness are delegated to the appendix.

\section{Related Work} \label{sec:related}

\begin{table*}[htb]
    \renewcommand\theadfont{\bfseries}

    \definecolor{mediumgreen}{RGB}{0,128,0}
    \definecolor{darkorange}{RGB}{240, 130, 0}
    
        \newcommand{\GoodMathEntry}[1]{\textcolor{mediumgreen}{\ensuremath{\bm{#1}}}}
    \newcommand{\AlmostGoodMathEntry}[1]{\textcolor{darkorange}{\ensuremath{\bm{#1}}}}
    \newcommand{\BadMathEntry}[1]{\textcolor{red}{\ensuremath{#1}}}

    \newcommand{\AccountConsensus}{\text{Account Consensus}}
    \newcommand{\GlobalConsensus}{\text{Global Consensus}}
    
    \centering
    \small
    
    \begin{tabular}{|c|c|c|c|c|}
        \hline
                    \multirowcell{2}{\thead{protocol}}
            & \multirowcell{2}{\thead{resilience}}
            & \multicolumn{3}{c|}{\textbf{worst-case end-to-end latency} (in round-trip times)}
        \\ \cline{3-5}
                                &             & \thead{no concurrency\\on the account}
            & \thead{$k$ concurrent requests,\\\atremove{same account, }no overspending}
            & \thead{$k$ concurrent requests,\\\atremove{same account, }with overspending}
        \\ \hline\hline
                    \thead{Consensus-based}
            & \GoodMathEntry{f < n/3}
            & \BadMathEntry{\GlobalConsensus}
            & \BadMathEntry{\GlobalConsensus}
            & \BadMathEntry{\GlobalConsensus}
        \\ \hline
                    \thead{$k$-shared AT~\cite{cons-crypto}}
            & \GoodMathEntry{f < n/3}
            & \AlmostGoodMathEntry{\AccountConsensus}
            & \AlmostGoodMathEntry{\AccountConsensus}
            & \GoodMathEntry{\AccountConsensus}
        \\ \hline
                    \thead{Astro II~\cite{astro-dsn} / FastPay~\cite{fastpay}}
            & \GoodMathEntry{f < n/3}
            & \GoodMathEntry{2} RTTs with \GoodMathEntry{O(n)} msgs
            & \BadMathEntry{\text{Not supported}}
            & \BadMathEntry{\text{Not supported}}
        \\ \hline
                                                                            \thead{Consensus on Demand~\cite{sliwinski2022consensus}}
            & \BadMathEntry{f < n/5}
            & \GoodMathEntry{2} RTTs with \BadMathEntry{O(n^2)} msgs ${}^{\text{\ref{itm:cons-od:latency}}}$
            & \BadMathEntry{\GlobalConsensus} ${}^{\text{\ref{itm:cons-od:consensus-type}}}$
            & \BadMathEntry{\GlobalConsensus} ${}^{\text{\ref{itm:cons-od:consensus-type}}}$
        \\ \hline
                    \thead{CryptoConcurrency}
            & \GoodMathEntry{f < n/3}
            & \AlmostGoodMathEntry{5} RTTs with \GoodMathEntry{O(n)} msgs ${}^{\text{\ref{itm:crypto-concurrency:latency}}}$
            & \GoodMathEntry{k+4} RTTs
            & \GoodMathEntry{\AccountConsensus}
        \\ \hline
    \end{tabular}

    \justify
    \smallskip
    \begin{atreview}
    \begin{enumerate}[label={\arabic*},leftmargin=5pt]
        \item\label{itm:cons-od:latency} The original paper does not consider how the client learns a relevant sequence number. Hence, we added one round-trip for the client to fetch it. Note that the local client's sequence number can be outdated unless the client is also required to act as a replica and to stay online observing all other transactions.

        \item\label{itm:cons-od:consensus-type} The original paper considers only Global Consensus, trusted by all parties. However, we believe that it is possible to make a version of~\cite{sliwinski2022consensus} that relies only on Account Consensus without affecting latency in case of absence of concurrency, using techniques similar to those used in {\Name}.

        \item\label{itm:crypto-concurrency:latency} 2 out of 5 RTTs are used to fetch an up-to-date initial state. We discuss potential ways to avoid it as well as other directions for optimizations in \Cref{subsec:total-latency-and-optimizations}.
            \end{enumerate}
    \end{atreview}

    \vspace{0.1cm}
    \caption{Asset transfer protocol comparison.}
    \vspace{-0.5cm}

    \label{table:comparison-table}
\end{table*}

Conventionally, asset transfer systems (or cryptocurrencies) were considered to be primary applications of blockchains~\cite{bitcoin,ethereum,hyperledger-fabric}, consensus-based protocols implementing replicated state machines.
In~\cite{cons-crypto,Gup16}, it has been observed that asset transfer \emph{per se} does not in general require consensus. 
This observation gave rise to simpler, more efficient and more robust implementations than consensus-based solutions~\cite{fastpay,astro-dsn,sliwinski2022consensus,pastro21disc}.
These implementations, however, assume that no account can be concurrently debited, i.e., no conflicting transactions must ever be issued by honest account owners.

In this paper, we propose an asset-transfer \atreplace{implementations}{implementation} in the setting where users can share an account and, thus, potentially issue conflicting transactions.
Our implementation does resort to consensus in some executions, which is, formally speaking,  inevitable~\cite{cons-crypto}.
Indeed, as demonstrated in~\cite{cons-crypto}, a fully consensus-free solution would be impossible, as there is a reduction from consensus to asset transfer with shared accounts.
\yarev{The algorithm for asset transfer with account sharing presented in~\cite{cons-crypto}  uses consensus for all transfers issued by accounts owned by multiple clients, even if the account owners never try to access the account concurrently.}

\atremove[We say it a few paragraphs later.]{In contrast, {\Name} dynamically detects overspending attempts and falls back to consensus only when such attempts are detected (which \atreplace{are}{is} assumed to be rare).}

Lamport's Generalized Paxos~\cite{lamport2010generalized} describes a state-machine replication algorithm for executing concurrently applied non-conflicting commands in a \emph{fast way}, i.e., within two message delays.  
The algorithm involves reaching agreement on a partially ordered  \emph{command-structure} set with a well-defined least upper bound.
The approach, however, cannot be applied directly in our case, as it assumes that every set of pairwise compatible operations can be executed concurrently~\cite[p. 11]{lamport2010generalized}.
This is not the case with asset transfer systems: imagine three transactions operating on the same account, such that applying all three of them drain the account to a negative balance, but every two of them do not. 
In order to account for such transactions, we therefore have to further generalize Generalized Paxos, in addition to taking care of Byzantine faults.  
(The original protocol was designed for the crash-fault model, though an interesting Byzantine version has been recently proposed~\cite{byzgenpaxos}).

\atadd{Byblos~\cite{bazzi2022clairvoyant}, a ``clairvoyant'' state machine replication protocol, further improves upon Byzantine Generalized Paxos by making it leaderless and compatible with a more general class of consensus protocols for the fallback at the cost of sub-optimal resilience ($n \ge 4f + 1$). However, it also considers a static definition of conflicts.}

In~\cite{li2012redblue}, RedBlue consistency was introduced. It manifests a different approach to \atreplace{combine}{combining} weak (asynchronous) and strong (consensus-based) synchronization in one implementation.
In defining the sequential specification of the object to be implemented,\atadd{ the} operations are partitioned \emph{a priori} into blue (parallelizable) and red (requiring consensus).
In the case of asset transfer, transfer operations would be declared red, which would incur using consensus among all clients all the time.

\pprev{Generalized Lattice Agreement~\cite{gla} has emerged as a useful abstraction for achieving agreement on comparable outputs among clients without resorting to consensus. One can use it to build a fully asynchronous state-machine replication protocol assuming that \atreplace{input requests form a \textit{lattice order} and thus can be merged together}{\emph{all} operations are commutative (can be executed in arbitrary order, without affecting the result)}.}
\atadd{In this paper, we further extend these ideas to operations that are not always commutative.}

\atrev{Recent work by Sliwinski, Vonlanthen, and Wattenhofer}~\cite{sliwinski2022consensus} aims to achieve the same goal of combining the consensus-free and consensus-based approaches as we do in this paper.
It describes an asset-transfer implementation that uses consensus whenever there are two concurrent transactions on the same account, regardless of the account's balance.
The algorithm assumes $5f+1$ replicas, where up to $f$ can be Byzantine, and a central consensus mechanism trusted by all participants.
In contrast, {\Name} implements dynamic (balance-based) \atreplace{conflict}{overspending} detection with the optimal number of $3f+1$ replicas and without universally trusted consensus, but has higher latency in conflict-free executions. 
\pprev{We achieve this by introducing a new abstraction that combines and then extends key ideas from Generalized Lattice Agreement~\cite{gla} and \atremove{Generalized }Paxos~\cite{lamport2010generalized}\atadd{ as will be further elaborated in \Cref{sec:cod-main}}.}

\atrev{We summarize the performance \atreplace{comparison}{of {\Name} compared to similar protocols} in \Cref{table:comparison-table}.
In the absence of concurrent transactions on the same account, the end-to-end latency of {\Name} is $5$ round-trips,
compared to $2$ in~\cite{astro-dsn} and~\cite{sliwinski2022consensus}.
If $k$ concurrent transactions on the same account can all be satisfied without overspending, the worst-case latency will be $k+4$ round-trips, whereas \atreplace{in \cite{sliwinski2022consensus} it would be $k$ times the latency of consensus}{\cite{sliwinski2022consensus} would fall back to consensus}, and \cite{astro-dsn} would lose liveness.}

\atadd{The higher latency of CryptoConcurrency 
is mainly due to the fact that we do not assume that clients have an up-to-date state when they start executing a transaction (in principle, a ``client'' can be simply a smart card storing a private key connected to a mobile point-of-sale device). Furthermore, since the main goal was to demonstrate the possibility rather than to achieve the best performance, we preferred simplicity over efficiency and opted for a highly consistent storage system (see \Cref{sec:keyval-storage} for details). Hence, we spend 2 round-trips to obtain the relevant state at the beginning of each operation.
We provide a more detailed latency breakdown and discuss potential ways to decrease it in \Cref{subsec:total-latency-and-optimizations}.}

\section{System Model} \label{sec:model}

\subsection{Processes, clients and replicas}

Let $\Pi$ be a (possibly infinite) set of potentially participating 
\emph{processes}. 
We assume that there is a fixed subset of $n$ processes \atreplace{$\rR \subseteq \Pi$}{$\Pi$}, called \emph{replicas}, that verify operations and maintain the \emph{system state}. 
Every process can also act as a \emph{client} that invokes \emph{operations} on the shared state. We make no assumptions on the number of clients in the system or on their availability, i.e., a client can go offline between its requests to the system.
In the definitions and proofs, we impose the standard assumption of the existence of a \emph{global clock}, not accessible to the processes.

We assume an adaptive adversary that can corrupt any process at any moment in time.
Once corrupted, the process falls under complete control of the adversary.
For simplicity of presentation, we say that a process is \emph{correct} if it is never corrupted during the whole execution and is \emph{Byzantine} otherwise.

The adversary can perfectly coordinate all Byzantine processes, and it is aware of the entire system state at any point of time except for the private cryptographic keys of the correct processes. We rely on the standard assumption that the computational power of the adversary is bounded so that it cannot break cryptographic primitives, such as digital signatures.

We assume that any number of clients and $f<n/3$ replicas can be Byzantine and that each pair of correct processes can communicate over a \emph{reliable authenticated channel}.

\subsection{Accounts} \label{subsec:accounts}
We assume a set of \emph{accounts} $\mathcal{A}$ across which assets are exchanged in the system. As every account can be owned by multiple processes, we equip accounts with a map $\mu: \mathcal{A} \rightarrow 2^\Pi$ that associates each account with a finite set of clients that can perform debit operations on it.
We say that client $q \in \mu(a)$ is an owner of an account $a$ ($q$ owns account $a$).
To simplify the model, we suppose that no client owns more than one account. 
To use two or more accounts, one is required to have multiple client instances.

Account $a$ is called \textit{correct} if it is owned by correct clients, i.e., $\forall q \in \mu(a):$ $q$ is correct.
We assume that the owners of an account trust each other and, if any owner of an account is corrupted, the other owners are also considered corrupted (i.e., Byzantine) and thus lose any guarantees provided by the system.

\subsection{Consensus objects}

The owners of each correct account share an unbounded supply of \emph{consensus objects} $\oConsensus[\vAcc][1,2,\dots]$. 
Each consensus object exports a single operation $\oConsPropose(v)$, 
which satisfies the following three properties:
    (\pConsLiveness:) each invocation of $\oConsPropose(v)$ by a correct client eventually returns a value;
    (\pConsConsistency:) no two correct clients return different values; and
    (\pConsValidity:) if a correct client returns value $v$, then some client invoked $\oConsPropose(v)$.

The particular implementation of consensus can be chosen by the owners of the account and does not have to be trusted by other participants.
As an extreme example, if an account is shared by two people, they could resolve a conflict via a phone call.
Another option would be to use any consensus-based blockchain of choice.

\subsection{Protocols and executions}

A \emph{protocol} equips every process $p$ (client or replica) with an automaton $A_p$ that, given an \emph{input} (a received message or, if $p$ is a client, an invocation of an operation), changes its state according to its transition function and produces an output (a message to send, a consensus invocation\atadd{,} or, if $p$ is a client, a response to an operation). %
An \emph{execution} of a protocol is a sequence of \emph{events}, where each event is a received message, a sent message, an invocation\atadd{,} or a response.
We assume that every invocation or response carries a unique identifier of the corresponding \emph{operation instance} (we simply say operation).

\subsection{Cryptographic Primitives} \label{sec:crypto}

\atrev{For simplicity, we assume the cryptographic primitives to be ``oracles'' implementing their ideal functionalities.}

\myparagraph{Digital signatures} \pprev{In our algorithms we extensively use digital signatures to ensure that every participant can verify the authenticity of the received messages.} 
We model digital signatures using two functions:
\begin{itemize}
    \item $\fSign(m)$ -- returns a signature for message $m$;
    \item $\fVerify(m, \vSignature, p)$ returns $\True$ iff 
        $\vSignature$ was obtained with $\fSign(m)$ invoked by process $p$.
\end{itemize}

\myparagraph{Threshold signatures}

Additionally, we assume that $n-f$ valid signatures on the same message $m$ can be efficiently aggregated via a \emph{threshold signature scheme}~\cite{threshold-sig,boldyreva2003threshold} or a
\emph{multi-signature scheme}~\cite{ohta1999multi,boldyreva2003threshold}.
Namely, the following operations are available to all processes:
\begin{itemize}
        
    \item $\fCreateThresholdSignature(m, S)$ -- returns a threshold signature given a message $m$ and a set $S$ of valid digital signatures on message $m$ issued by $n-f$ distinct \atreplace{processes}{replicas};

    \item $\fVerifyThresholdSignature(m, s)$ --  returns $\True$ iff signature $s$ was obtained by invoking $\fCreateThresholdSignature(m, S)$ for some set $S$.
\end{itemize}

\myparagraph{Merkle trees}

A Merkle tree (or a hash tree) is a binary tree in which every leaf node is labeled with a value (or a hash of a value)\atadd{,} and every internal node is the hash of its two child nodes.
\atrev{One can use it to efficiently create a short commitment (namely, the root of the tree) to a set of values $M = \{m_1, \ldots, m_k\}$.
Then, for any of the original values $m_i$, it is easy to prove that $m_i$ belongs to $M$ to \atreplace{somebody}{anybody} who knows the root.
\atremove{The proof consists of $\log_2 k$ sibling nodes of all nodes on the path from $m_i$ to the root in the tree.}
We model this primitive with the \atreplace{below}{following} functions available to all processes:}
\begin{itemize}
    \item \atreplace{$\fMerkleTree(ms)$}{$\fMerkleTree(M)$} -- returns a Merkle tree $\vMerkleTree$ for the set of values \atreplace{$ms$}{$M$}. One can access the root of the tree using the notation $\vMerkleTree.\vMTRoot$;
    
    \item $\fGetItemProof(\vMerkleTree, m)$ -- returns a proof for item $m$ iff \atreplace{$\vMerkleTree = \fMerkleTree(ms)$}{$\vMerkleTree = \fMerkleTree(M)$} for some \atreplace{$ms$}{$M$} s.t. \atreplace{$m \in ms$}{$m \in M$};

    \item $\fVerifyItemProof(\vMTRoot, \vItemProof, m)$ -- returns $\True$ iff $\vItemProof = \fGetItemProof(\vMerkleTree, m)$ for some $\vMerkleTree$ with $\vMerkleTree.\vMTRoot = \vMTRoot$.
\end{itemize}

In our protocols, Merkle \yareplace{tress}{trees} could be replaced by more communication-efficient cryptographic primitives such as \emph{set accumulators}~\cite{set-accumulators} or \emph{vector commitments}~\cite{vector-commitments-1,vector-commitments-2}.
However, they typically require more expensive computation and a trusted setup.

\section{{\Name} Architecture}\label{sec:architecture}

\begin{figure}[htbp]
    \newcommand{\coin}{}
    \small
    \newcommand{\xs}[1]{{\smaller#1}}

    \centering
    \includesvg[scale=1]{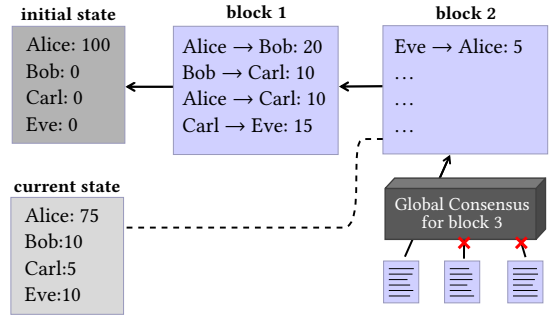}
    \caption{Total order asset transfer architecture}
    \label{fig:asset-transfer-traditional}
\end{figure}

In a traditional, consensus-based, asset transfer system~\cite{bitcoin,ethereum}, the participating processes agree on a (totally ordered) sequence of transactions, usually split into discrete blocks and applied to some initial state (often called \emph{the genesis block}), as illustrated in \Cref{fig:asset-transfer-traditional}.

\begin{figure}[htbp]
    \newcommand{\coin}{}
    \small
    \newcommand{\xs}[1]{{\smaller#1}}

    \centering
    \includesvg[scale=1]{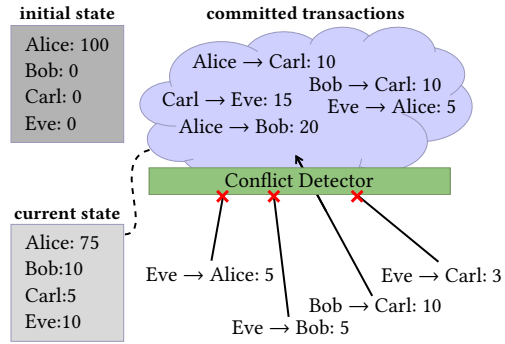}
    \caption{Consensus-free asset transfer architecture}
    \label{fig:asset-transfer-consensus-free}
\end{figure}

A crucial observation is that, as long as the final balance of each account is non-negative, the resulting state does not depend on the order in which the transactions are applied.
This provides the core insight for the so-called \emph{consensus-free} asset transfer systems~\cite{fastpay,astro-dsn,cons-crypto,Gup16,pastro21disc,abc-tr}.
At a high level, 
such systems maintain an \emph{unordered set} of committed transactions.
In order to be added to the set, a new transaction must pass a special \emph{Conflict Detector object}.
The object maintains the invariant of non-negative balances by imposing a notion of pairwise conflicts on the transactions and preventing multiple conflicting transactions from being accepted (see \Cref{fig:asset-transfer-consensus-free}).
Intuitively, two transactions are considered conflicting when they are trying to move the same assets.
The Conflict Detector object operates in a way similar to Byzantine Consistent Broadcast~\cite{textbook,astro-dsn}.
Namely, a quorum of replicas must acknowledge a transaction in order for it to pass the Conflict Detector and each replica acknowledges at most one of the conflicting transactions.

As discussed in the introduction, the main downside of such systems is that they preclude any concurrent use of an account.
Moreover, the existing solutions may actually punish even accidental attempts to issue several conflicting transactions concurrently by not letting \emph{any} of them to pass the Conflict Detector and, hence, effectively blocking the entire account.

\begin{figure}[htbp]
    \newcommand{\coin}{}
    \small
    \newcommand{\xs}[1]{{\smaller#1}}

    \centering
    \includesvg[scale=1]{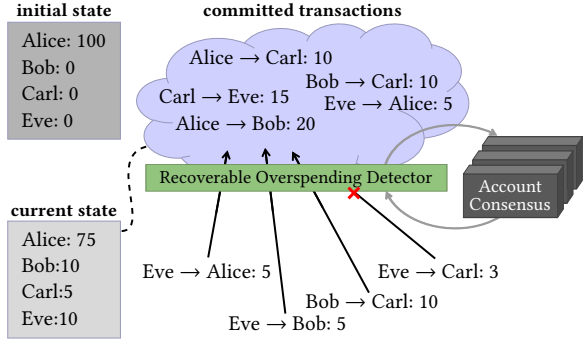}
    \caption{CryptoConcurrency architecture}
    \label{fig:asset-transfer-cryptoconcurrency}
\end{figure}

To mitigate this issue and enable new use-cases such as shared accounts or periodic subscription payments, we propose a hybrid approach: we replace the Conflict Detector with a more advanced \emph{Recoverable Overspending Detector} abstraction and 
use external consensus objects to perform the recovery procedure in case an attempt to overspend is detected.
We build an adaptive asset transfer system that goes through a consensus-free ``fast path'' whenever
possible.
The system supports shared accounts and avoids blocking the funds because of an accidental attempt of overspending on an account (see \Cref{fig:asset-transfer-cryptoconcurrency}).

In order to preserve the efficiency and robustness of consensus-free solutions, the consensus objects are used as rarely as possible.
More precisely, our protocol only accesses consensus objects when the total volume of all ongoing transactions on the account exceeds the balance of the account.
Our implementation of the Recoverable Overspending Detector \yaadd{(see \Cref{sec:cod-main})} inherits the key ideas from the Lattice Agreement protocol of~\cite{gla} and Paxos~\cite{Lam98,lamport2001paxos}.

Finally, we avoid the need of a central, universally trusted consensus mechanism by allowing the owners of each account to use their own consensus protocol of choice.
To this end, after obtaining a consensus output, the owners send ``notarization'' requests to the replicas and the replicas will refuse to notarize diverging outputs for the same consensus instance.

\section{Formal Problem Statement} \label{sec:problem-statement}

Now we formally define the asset transfer abstraction that {\Name} implements.

\myparagraph{Transactions}

In asset transfer systems, clients move funds between accounts by issuing \emph{transactions}.
A \emph{transaction} is a tuple $tx = \Tuple{\vSender, \vRecipient, \vAmount, \vId, \vSignature}$. 
The $\vAmount$ value specifies the funds transferred from account $\vSender$ to account $\vRecipient$.
In order to distinguish transactions with identical accounts and transferred amounts, 
each transaction is equipped with a special unique element called $id$.
In practice, one can use a long ({e.g., }128 bits) randomly generated string or a sequence number concatenated with the id of the client.
Each transaction contains a digital signature $\vSignature$ of one of the owners of the account $\vTx.\vSender$ to confirm the transaction's authenticity.
We denote the set of all possible well-formed transactions as $\tT$.
Ill-formed transactions (including transactions with invalid signatures) are ignored by the participants.
 
For every account $a \in \aA$, there exists a \emph{genesis transaction} $\vTxInitForA = \Tuple{\Null, a, \vAmount_a, 0, \Null}$, which specifies the initial balance $\vAmount_a$ of account $a$.
All genesis transactions are publicly known and are considered to be well-formed by definition.

In addition, from the perspective of an account $a \in \aA$, we distinguish two types of transactions: \emph{debits} and \emph{credits} (on $a$).
A \emph{debit transaction} (or simply a \emph{debit}) is a transaction $\vTx$, for which  $\vTx.\vSender = a$, and a \emph{credit transaction} (or simply a \emph{credit}) is a transaction $\vTx$, such that $\vTx.\vRecipient = a$.
In other words, debits ``spend money'' and credits ``add money to the account''.
Let us also define a helper function $\fTotalValue$ that, given a set of transactions, returns the sum of funds they transfer: $\fTotalValue(\vTxs) = \sum\limits_{\vTx \in \vTxs}\vTx.\vAmount$.
Let $\fCredits(\vTxs,\vAcc)$ and $\fDebits(\vTxs,\vAcc)$ denote the sets of credit and debit transactions on $\vAcc$  in $\vTxs$, respectively (i.e., $\fCredits(\vTxs,\vAcc) = \{ \vTx \in \vTxs \mid \vTx.\vRecipient = \vAcc\}$ and $\fDebits(\vTxs, \vAcc) = \{ \vTx \in \vTxs \mid \vTx.\vSender = \vAcc\}$).
Now the \emph{balance} of $\vAcc$ in $\vTxs$ is defined as: $\fBalance(\vTxs,\vAcc)=\fTotalValue(\fCredits(\vTxs,\vAcc))-\fTotalValue(\fDebits(\vTxs,\vAcc))$.

\myparagraph{Interface} 
Clients interact with the asset transfer system using operations $\oTransfer(\vTx)$ and $\oGetAccountTransactions()$.
The system also provides a function $\fVerifyCommitCertificate(\vTx, \vCommitCert).$
Operation $\oTransfer(\vTx)$, $\vTx \in \tT$, is used by the clients to move  assets as stipulated by the transaction $\vTx$.
The operation may return one of the following responses:

\begin{itemize}
    \item $\tOK(\vCommitCert)$, indicating that the transfer has been completed successfully, $\vCommitCert$ is a certificate proving this. $\vCommitCert$ can be verified by any process using the $\fVerifyCommitCertificate$ function;
    \item $\tFAIL$, indicating that the transfer failed due to insufficient balance.
\end{itemize}
Operation $\oGetAccountTransactions()$ can be used to obtain the current set $\{\Tuple{\vTx, \vCommitCert}\}$ of debit and credit transactions $\vTx$ applied to the client's account with their commit certificates $\vCommitCert$.

\myparagraph{Committed and active transactions} 
A transaction $\vTx$ is called \emph{committed} iff there exists a certificate $\vCommitCert$ such that $\fVerifyCommitCertificate(\vTx, \vCommitCert) = \True$.
For the purposes of this paper, by the existence of a cryptographic certificate\atadd{,} we mean that some process or the adversary is capable\atadd{ of} computing it with non-negligible probability using the available information in polynomial time.
Let $C(t)$ denote the set of all such transactions at time $t$.
Note that for all $t'>t$, we have $C(t) \subseteq C(t')$. 
We define $\fCommitTime(\vTx)$ as the moment of time $t$ when $\vTx$ gets committed, i.e., $t = \fCommitTime(\vTx) \Leftrightarrow \vTx \in C(t)$ and $\forall t' < t: \vTx \notin C(t')$.

We assume that every transaction has a unique identifier and
that the owners of a correct account will never invoke $\oTransfer$ more than once with the same transaction $\vTx$.
Hence, for a correct account $\vAcc$, there is a one-to-one mapping between the debit transactions issued on $\vAcc$ and $\oTransfer$ operations invoked by the owners of $\vAcc$ (including the ones that return $\tFAIL$).

From the perspective of the owners of a correct account $\vAcc$, a debit transaction starts when the corresponding $\oTransfer$ operation is invoked and ends when the operation terminates (with either $\tOK(\vCommitCert)$ or $\tFAIL$).
Hence, for a debit transaction $\vTx$, we define $\fStart_{\vAcc}(\vTx)$ and $\fEnd_{\vAcc}(\vTx)$ as the moments in time when the corresponding operation is invoked and returns, respectively.

A debit transaction $\vTx$ on a correct account is called \emph{active at time $t$} iff $\fStart_{\vAcc}(\vTx) \le t \le \fEnd_{\vAcc}(\vTx)$.
Let $O(t,\vAcc)$ denote the set of all active \emph{debit} transactions on account $\vAcc$ at time $t$.

As for the credit transactions, the owners of $\vAcc$ have no insight into the execution of the corresponding operations (which could be performed by Byzantine clients).
Instead, a credit transaction appears to happen instantly at the moment it is committed.
Hence, for a credit transaction $\vTx$, we define $\fStart_{\vAcc}(\vTx) = \fEnd_{\vAcc}(\vTx) = \fCommitTime(\vTx)$.

\myparagraph{Properties}

Given an execution $\Execution$ and a correct account $\vAcc$, we define $\tT(\Execution,\vAcc)$ as the set of all debit transactions and \emph{committed} credit transactions on $\vAcc$ that appear in $\Execution$.
The map $\rho_{\Execution,\vAcc}$ associates each debit transaction in $\tT(\Execution,\vAcc)$ with its response in $\Execution$ (if any).
We say that a debit transaction $\vTx \in \tT(\Execution, \vAcc)$ is \emph{successful} iff $\rho_{\Execution, \vAcc}(\vTx) = \tOK(\vCert)$ (for some $\vCert$).

We define a \emph{real-time} partial order on transactions in $\tT(\Execution,\vAcc)$  as follows:
we say that $\vTx_1$ \emph{precedes} $\vTx_2$ in execution $\Execution$ from the point of view of account $\vAcc$, and we write $\vTx_1\prec_{\Execution,\vAcc} \vTx_2$ iff $\fEnd_{\vAcc}(\vTx_1) < \fStart_{\vAcc}(\vTx_2)$ in $\Execution$.

Let $H$ be a permutation (i.e., a totally ordered sequence) of transactions in $\tT(\Execution,\vAcc)$.
Given $\vTx$, a debit transaction on $\vAcc$ in $\Execution$, let $S(H,\vTx)$ denote the set of credit and successful debit transactions in the prefix of $H$ up to, but not including, $\vTx$.
We say that permutation $H$ is \emph{legal} if and only if, for every debit transaction $\vTx \in \tT(\Execution, \vAcc)$, 
\pprev{$\rho_{\Execution,\vAcc}(\vTx) = \tOK(\vCert) \Leftrightarrow \vTx.\vAmount \le \fBalance(S(H,\vTx), \vAcc)$}.

We say that $H$ is \emph{consistent with $\prec_{\Execution,\vAcc}$} iff for all $\vTx_1,\vTx_2\in\tT(\Execution,\vAcc)$, $\vTx_1\prec_{\Execution,\vAcc} \vTx_2$ implies that $\vTx_1$ precedes $\vTx_2$ in $H$.

Now we are ready to formally state the properties that every execution $\Execution$ of our \emph{asset transfer implementation} must satisfy.
First, no account (be it correct or Byzantine) can exhibit a negative balance:
\begin{properties}
    \propertyitem{\pTransferSafety} At any time $t$, for all $ \vAcc \in \aA$: $\fBalance(C(t), \vAcc) \geq 0$. \end{properties}

Furthermore, for every correct account $\vAcc$, from the point of view of the owners of the account, the outputs of the $\oTransfer$ operations are as if they were executed sequentially, one at a time, with no concurrency and the certificates returned by $\oTransfer$ are valid. More formally, the following properties hold for each correct account $\vAcc$ and each protocol execution $\Execution$:

\begin{properties}
    \propertyitem{\pTransferConsistency} There exists a legal permutation of transactions in $\tT(\Execution,\vAcc)$ that is consistent with $\prec_{\Execution,\vAcc}$.
       
    \propertyitem{\pTransferValidity}
    If $\oTransfer(\vTx)$ on $\vAcc$ returns $\tOK(\vCommitCert)$, then $\fVerifyCommitCertificate(\vTx, \vCommitCert) = \True$.
\end{properties}

The second operation, $\oGetAccountTransactions$, must return the set of committed transactions related to the account\atreplace{. More formally:}{:}

\begin{properties}
    \propertyitem{\pAccountTransactions} $\oGetAccountTransactions()$ invoked by an owner of a correct account $\vAcc$ at time $t_0$ returns a set $\{\Tuple{\vTx_i, \vCert_i}\}_{i=1}^l$ such that $\forall i: \fVerifyCommitCertificate(\vTx_i, \vCert_i) = \True$ and $\fDebits(C(t_0), \vAcc) \cup \fCredits(C(t_0), \vAcc) \subseteq \{\vTx_i\}_{i=1}^l$.
                \end{properties}

\atrev{An asset transfer system should also satisfy the following liveness property:}
\begin{properties}
    \propertyitem{\pTransferLiveness} Every operation invoked by a correct client eventually returns.
\end{properties}

\atadd{Finally, }{\Name} satisfies one more important property that we \atreplace{take}{consider} as one of the key contributions of this paper. Intuitively, if the owners of a correct account do not try to overspend, eventually the system will stabilize from their \atreplace{past}{previous} overspending attempts (if any)\atadd{,} and the clients will not need to invoke consensus from that point on. \pprev{Let us now define this property formally.}

\atrev{We say that there is an \emph{overspending attempt at time $t$} iff $\fTotalValue(O(t,\vAcc)\setminus C(t)) > \fBalance(C(t), \vAcc)$.}

\begin{properties}
    \propertyitem{\pTransferConcurrency}
    Let $\vAcc$ be a correct account.
    \atrev{If there is no overspending attempt at any time $t > t_0$, for some $t_0$,}
    then there exists a time $t_1$ such that the owners of $\vAcc$ do not invoke consensus objects after $t_1$.%
        \footnote{Note that, \ppreplace{by}{if} the {\pTransferConsistency} property\ppadd{ holds}, all $\oTransfer$ operations invoked after $t_1$ must return $\tOK(\vCommitCert)$ and cannot return $\tFAIL$.}
\end{properties}

\pkadd{In terms of algorithmic complexity, this paper is focused on the latency exhibited by an asset-transfer implementation in the absence of overspending attempts, i.e., when consensus objects are not involved. We measure the latency in \emph{round-trip times} (RTTs). Informally, RTT is the time it takes for a given process to send a \emph{request} message to another process and receive a \emph{response} message.~\footnote {A more precise definition of time complexity of an asynchronous algorithm can be found in~\cite{canetti-rabin-93}.}} 
\atrev{For an algorithm satisfying the {\pTransferConcurrency} property, we say that it exhibits \emph{$k$-overspending-free latency $f(n, k)$} if, after the time $t_1$ (defined in {\pTransferConcurrency}), any transfer operation that runs in the absence of overspending concurrently with at most $k-1$ other transfer operations on the same account in a system with $n$ replicas completes in at most $f(n, k)$ RTTs.}

Given the above, the main theorem of this paper is as follows.
\begin{theorem}\label{thm:main}
        \pkrev{There exists a deterministic asynchronous protocol that implements an asset transfer system (as formally defined in this section), satisfies the {\pTransferConcurrency} property and exhibits $k$-\atreplace{conflict}{overspending}-free latency of $k+4$ RTTs.}
\end{theorem}

\algnewcommand{\IfReceivedClosed}{
    \If {received a valid $\Message{\mClosed, \vSignature}$ reply} \Return $\tFAIL$ \EndIf
}

\section{{\CODLong}} \label{sec:cod-main}

We implement the {\RODLong} layer illustrated in \Cref{fig:asset-transfer-cryptoconcurrency} in \Cref{sec:architecture} as a collection of slightly simpler objects, which we call \emph{\CODLong} (or \emph{\CODShort} for short).
{\CODShort} objects are account-specific.
At each moment in time\atadd{,} there is at most one {\CODShort} object per account that is capable of accepting \atreplace{clients'}{client} transactions.
The mission of a single {\CODShort} object is to ensure operation under normal conditions when there are no overspending attempts.
In this case, {\CODShort} will accept every transaction that a client submits to it using the $\oSubmit$ operation.

However, because of concurrency, even the owners of a correct account might accidentally try to overspend.
In this case, some of the \atreplace{clients'}{client} requests submitted to {\CODShort} may fail.
\atreplace{The clients}{Clients} can then use the $\oClose$ operation that deactivates this {\CODShort} instance and provides a snapshot of its final state.
This allows the \atreplace{owners of the account}{account owners} to gracefully recover from overspending by instantiating a new instance of {\CODShort} from this snapshot.
To this end, each account $\vAcc$ is provided with a list of {\CODShort} objects $\oCOD[\vAcc][1,2,\ldots]$.
Object $\oCOD[\vAcc][e]$ is said to be associated with \emph{epoch number} $e$.
The procedure of migrating from one {\CODShort} object to another, i.e., from one \emph{epoch} to another, is called \emph{recovery} \atreplace{(described in detail in \Cref{subsec:composing-crypto-la-main})}{and will be described in detail in \Cref{subsec:composing-crypto-la-main}}. 
\subsection{{\CODShort} Protocol Overview} \label{subsec:cod-overview}

\atrev{We define the interface and the properties of {\CODShort} formally and provide pseudocode in \Cref{app:cryptoCD}. However, \yareplace{we believe that understanding the}{learning about this} concept alongside a high-level overview of the algorithm, as presented in the rest of this section, may be more accessible.}
\atadd{In order to keep the explanation simple and highlight the main ideas, we omit some minor implementation details.}\yaremove{Please refer to \Cref{app:cryptoCD} for the formal specification of the protocol.}

\myparagraph{\atadd{Initial state}}

\atadd{As described before, each account is associated with a sequence of {\CODShort} objects, each serving for one \emph{epoch} -- a period of time without overspending attempts.
However, if an overspending attempt is detected, a {\CODShort} object is closed, and, after a procedure that we call \emph{recovery}, a new {\CODShort} object is initialized.
Thus, a {\CODShort} object needs to be initialized from some initial state, namely: $\vInitDebits$ -- the set of debits accepted in prior epochs, $\vInitCredits\WithCert$ -- a set of credits with commit certificates sufficient to cover $\vInitDebits$, and $\vRestrictedDebits$ -- the set of debits canceled in prior epochs (the {\CODShort} object must not accept the transactions from $\vRestrictedDebits$).}

\myparagraph{\atadd{$\oSubmit$ operation}}

\atrev{The main operation of {\CODShort} is $\oSubmit(\vDebits, \vCredits\WithCert)$, through which clients inform replicas of new incoming transactions (credits) and request approval for new outgoing transactions (debits).
Credits must be accompanied by valid commit certificates.
Hence, $\vCredits\WithCert$ is a set of pairs $\Tuple{\vTx, \vCommitCert}$ such that $\vTx.\vRecipient = \vAcc$ and $\fVerifyCommitCertificate(\vTx, \vCert) = \True$.}

\atrev{As a convention\atadd{, throughout the rest of the paper}, we use 
variable names with a superscript ``${}\WithCert$'' (e.g., $\vTxs\WithCert$, $\vCredits\WithCert$, $\vDebits\WithCert$, etc.) to denote sets of transactions\atadd{ paired} with\atadd{ some kind of cryptographic} certificates (e.g., $\{ \Tuple{\vTx_1, \vCert_1}, \dots, \Tuple{\vTx_n, \vCert_n} \}$).
We also define \atreplace{a}{an auxiliary} function $\fTransactions(\vTxs\WithCert)$ that, given a set of pairs $\vTxs\WithCert = \{\Tuple{\vTx_1, \vCert_1},\ldots \Tuple{\vTx_n, \vCert_n}\}$, returns only the transactions $\{ \vTx_1,\ldots, \vTx_n\}$, without the certificates.}

\atrev{In the optimistic scenario, when there are no overspending attempts, $\oSubmit$ returns $\tOK(\vDebits\WithCert, \vOutCredits\WithCert)$.
Here, $\vDebits\WithCert$ contains the same set of transactions as in the input to $\oSubmit$, augmented with certificates that we will call \emph{accept certificates}, confirming that the transactions passed through the {\CODLong} object.
The {\CODShort} implementation exposes a boolean function $\oVerifyCODCert(\vTx, \vCert)$ that can be used to verify the accept certificates.
We say that a transaction $\vTx$ is \emph{accepted by $\oCOD[\vAcc][\vEp]$} iff $\vTx \in \oCOD[\vAcc][\vEp].\vInitDebits$ or there exists $\vCert$ such that $\oCOD[\vAcc][\vEp].\oVerifyCODCert(\vTx, \vCert) = \True$.}

\atrev{The second value returned by a successful invocation of $\oSubmit$, $\vOutCredits\WithCert$, is, intuitively, the set of credits used to ``cover'' the debits.
For correct accounts, {\CODShort} maintains the property that, at any time $t$, the total value of all debits accepted by {\CODShort} does not exceed the total value of all credits returned from successful invocations of $\oSubmit$.\footnote{\atadd{For Byzantine clients, we cannot formally use the language of ``returned values''. Instead, for Byzantine accounts, {\CODShort} guarantees that the total value of accepted debits does not exceed the total value of \emph{all} committed credits for the account, ensuring non-overspending.}}}

\myparagraph{\atadd{The ``Prepare'' phase}}

\begin{figure}[htb]
  \centering
    \small

  \newcommand{\textLineOne}{Alice, Bob, and Carl share an account with the balance of 2 coins.}
  \newcommand{\textLineTwo}{Each of the three transactions ($\vTx_A$, $\vTx_B$, and $\vTx_C$) spends 1 coin.}
  \includesvg[scale=0.95]{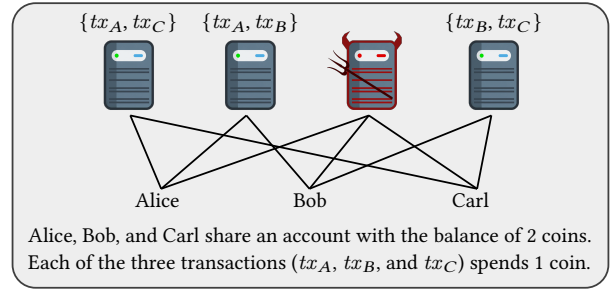}
  \undef\textLineOne
  \undef\textLineTwo
  
  \vspace{-.2cm}
  \caption{\atrev{An example where the naive 1 RTT algorithm fails.}}

  \label{fig:strawman-algo-overspending-example}
\end{figure}

\pprev{To detect overspending attempts, a correct replica $r$ maintains a set\atadd{ $\vCredits\WithCert_{r}$} of all credits\atremove{ $\vCredits\WithCert_{r}$} it has seen so far\atadd{ (with the corresponding commit certificates)} and \atreplace{an ever-growing}{a} set\atadd{ $\vDebits\WithCert_{r}$} of debit transactions\atremove{ $\vDebits\WithCert_{r}$} it acknowledged\atadd{ (with client signatures)}, preserving the invariant that $\fTotalValue(\fTransactions(\vCredits\WithCert_{r})) \ge \fTotalValue(\fTransactions(\vDebits\WithCert_{r}))$.
To process a client request, it adds all the committed credits\atadd{ the} client attached to the message to \atreplace{a}{the} local set
$\vCredits\WithCert_{r}$ (given they come with valid commit certificates) and then adds the received debits to $\vDebits\WithCert_{r}$ \atreplace{if their total value does not exceed the total value of $\vCredits\WithCert_{r}$}{if possible without violating the invariant}. \atadd{The }replica then responds to \atreplace{a}{the} client with both\atremove{ these} sets and a signature\atremove{ $\vSignature$} \atreplace{of}{on} \atreplace{a}{the} debits set.}
\pprev{Then, on \atreplace{a}{the} client side, it is\atremove{ very} tempting to wait for a quorum of responses, such that each will contain client debit transactions and assume that this means that the set of transactions client sent to replicas does not lead to overspending. Indeed, it would mean that at least $\dfrac{n}{3}$ correct replicas added all input debits to their local sets. 
However, this approach is not sufficient to prevent potential overspending.}
\atadd{Consider the example illustrated in \Cref{fig:strawman-algo-overspending-example}.}
\atadd{In this example, each of the 3 correct replicas (1,2, and 4) acknowledged 2 transactions each and thus did not detect overspending.
The third replica is Byzantine and it acknowledged all 3 transactions.
In the end, all three transactions would manage to pass the overspending detector even though the account they share contains funds only for 2 of them.}

\atrev{To deal with such situations, we follow a similar approach to the one proposed to solve Generalized Lattice Agreement~\cite{gla}: we retry requests to replicas until we receive identical sets of debits in the responses, updating the inputs with the new transactions we learn with every attempt. In the above example, Alice, after receiving $\{\vTx_A, \vTx_C\}$ from replica $1$ and $\{\vTx_A, \vTx_B\}$ from replica $2$, would retry its request as these two responses are different even though both contain $\vTx_A$.} %
\atadd{However, unlike in Generalized Lattice Agreement, where the assumption is that any combination of inputs is ``mergeable'' (i.e., commutative), in our case, due to the non-overspending invariant maintained by the replicas, the object may reach a state where convergence would not be reached regardless of how many retries the client performs. We will explain how to deal with such situations later in this section. For now, let us consider the ``good'' scenario when there are no overspending attempts.}

\atadd{For all requests that successfully passed this phase, we can guarantee the \emph{comparability} property: suppose that client $A$ obtains a quorum of signatures for a set of transaction $S_A$ and client $B$ -- for a set $S_B$. Then we can claim that either $S_A \subseteq S_B$ or vice versa. Indeed, it is sufficient to consider the quorum intersection property~\cite{byz-quorums}: there must be a correct replica that signed both sets and the one it signed later must be a superset of the one it signed earlier.}
\atadd{We can also guarantee that for all transactions in both sets, there are enough committed credits to cover all of them.
Indeed, consider the largest of the two sets. It was acknowledged by a quorum of replicas that would only acknowledge it if they saw enough credits.}

\atadd{We call this process of repeatedly trying to get a quorum of replicas to converge on the same set of debits the \emph{Prepare phase}.}
\atadd{After a client successfully passes the Prepare phase in the course of executing a $\oSubmit$ operation, the client obtains a set of debits $\vPreparedDebits$ with signatures from a quorum of replicas confirming that they acknowledged this set. We use a \emph{threshold signature scheme} (as defined in \Cref{sec:crypto}) in order to compress these signatures into one small signature $\vPreparedCert$.}

\myparagraph{\atadd{The ``Accept'' phase}}
\atadd{The purpose of the second phase of the $\oSubmit$ operation, called the \emph{Accept phase}, is to ensure \emph{recoverability}, i.e., to make it possible to transfer state from one {\CODShort} object to the next without reverting transactions that could have been accepted. The phase consists of just one round-trip: the client simply sends $\vPreparedDebits$ along with the threshold signature $\vPreparedCert$ to the replicas, the replicas check the validity of $\vPreparedCert$ and, \emph{unless they have previously received a request to close this {\CODShort} instance}, acknowledge the client's request with a signature on the \emph{Merkle tree root} (as defined in \Cref{sec:crypto}) of the set $\vPreparedDebits$.
The client can then extract individual Merkle tree proofs for each of the transactions, thus obtaining a short accept certificate for each individual transaction.}

\myparagraph{\atadd{Detecting overspending}}

\begin{atreview}
The client may not be able to terminate its request on the ``good'' path for one of two reasons:
\begin{enumerate}
    \item In the Prepare phase, the client does not have enough committed credits to cover the union of sets of debits returned by the replicas. This means that there is an overspending attempt;

    \item In either phase, a replica refuses to process the client's request because it has already acknowledged a request to close this {\CODShort} instance. This means that some other owner of the account observed an overspending attempt and started migrating the state from this instance to the next.
\end{enumerate}

In either of these two cases, the client returns $\tFAIL$ from the $\oSubmit$ operation, indicating a potential overspending attempt and that a (consensus-based) recovery is necessary.
\end{atreview}

\myparagraph{\atadd{Closing an instance}}

\atrev{The $\oClose$ operation is designed to deactivate the {\CODShort} object and to collect a snapshot of its state to facilitate the recovery.
During this operation, the client solicits from a quorum of replicas the sets of transactions they accepted and then asks the replicas to sign the accumulated joint state in order to obtain a short proof (a single threshold signature) of validity of the resulting snapshot.
Accessing a quorum guarantees that the client gathers all the debits that have been previously returned from the $\oSubmit$ operations.
As a result, \atreplace{confirmed}{accepted} transactions are never lost.}

\atrev{Upon careful examination, one can notice that the interaction between the Accept phase of the $\oSubmit$ operation and the first message of the $\oClose$ operation is similar to that of ``propose'' messages and ``prepare'' messages with a larger ballot number in Paxos~\cite{lamport2001paxos}.
Intuitively, it guarantees that, if there is a $\oSubmit$ concurrent with a $\oClose$, either the client executing $\oSubmit$ will ``see'' the $\oClose$ operation and return $\tFAIL$ or the client executing the $\oClose$ operation will ``see'' that the debits were accepted by some replicas.}

\myparagraph{\atadd{Achieving liveness}}

\atadd{Special care is necessary to ensure liveness of all clients in such a protocol, especially in the Prepare phase. First, we only wait for convergence on debits in the replicas' responses and not on credits, as otherwise, it would be possible to prevent progress on a correct account by sending lots of small credits to it. Second, it is not hard to see that a certain unlucky client may never reach convergence due to a constant inflow of new debits. In order to avoid such situations, a replica that already accepted a set of debits that includes all the debits submitted by a client will notify that client, and it will be able to move on directly to the Accept phase.}

\subsection{{\CODShort} performance} \label{subsec:cod-performance}

\atrev{We are mostly interested in the latency of the $\oSubmit$ operation as the $\oClose$ operation is only used in case of an overspending attempt, which is assumed to happen rarely.
The protocol for the $\oSubmit$ operation consists of two phases: Prepare and Accept, where the latter always consists of just one rount-trip while the former may involve multiple retries, until the client obtains a quorum of identical replies.
However, the main strength of this implementation is that it is adaptive.
Indeed, in absence of other concurrent requests, the client will be able to finish the Prepare phase in just one round-trip. Otherwise, it may take up to $k$ round-trips in presence of $k-1$ other concurrent requests.}

\atrev{Put differently, we pay one extra round-trip compared to purely asynchronous solutions such as~\cite{astro-dsn,fastpay} in order to ensure recoverability (the Accept phase), but the overspending detection part comes ``for free''.
The main difference is that {\Name} retries where others would give up, until it reaches a state where further retries would be pointless.}

\section{\AppendOnlyStorage}\label{sec:keyval-storage}

\pprev{In this section, we present an abstraction called \emph{\AppendOnlyStorage}, which allows us to implement two distinct algorithm building blocks. The first one, \emph{\GlobalStorage}, represents a layer in the architecture of {\Name} that stores all committed transactions (as illustrated in Figure~\ref{fig:asset-transfer-cryptoconcurrency} in Section~\ref{sec:architecture}). \pkrev{The second one, {\AccountStorage}, is associated with each account\atadd{,} and its primary purpose is to facilitate communication between its owners and ensure liveness of its operations}.}

\pkrev{{\AppendOnlyStorage} can be seen as a distributed implementation of\atadd{ an} indexed collection of sets: with each \emph{key} $k$, the abstraction associates an unordered set of \emph{values} $\vVs$.}
\atreplace{In addition}{Furthermore}, {\AppendOnlyStorage} is capable of (i)~verifying that a given value is allowed to be added to a specified key and (ii)~providing a proof of the fact that the values appended or read by a client for a specific key $k$ are stored persistently (i.e., any later read operation with key $k$ will return a set \atreplace{including}{that includes} these values).

To add a value $v$ to a set of values stored for a key $k$, a client calls $\oAppendKey(k, v, \vVCert)$, where $\vVCert$ is a validity certificate for value $v$ and key $k$. 
To read\atadd{ the} values associated with a key $k$, a \atreplace{process}{client} invokes the $\oReadKey(k)$ operation.

\atrev{We provide a formal definition of {\AppendOnlyStorage} and discuss \atreplace{its implementation}{the protocol implementing it} in Appendix~\ref{app:storage}.}
\atadd{For simplicity, we use a highly consistent storage system that requires only one round-trip for the $\oAppendKey$ operation, but two round-trips for the $\oReadKey$ operation: one to fetch the values and one for the ``write-back'' phase~\cite{abd} to ensure that any subsequent read will see at least as many elements in the set.}

\subsection{Global and Account Storage}

\pprev{We use {\AppendOnlyStorage} as a generalized implementation for both \emph{\GlobalStorage} and \emph{\AccountStorage}. Let us explain how we use each of these objects in the protocol.}

\mysubparagraph{{\GlobalStorage}}

In {\Name}, all clients have access to one common  {\AppendOnlyStorage} instance called \emph{\GlobalStorage}.
After a transaction is accepted by a {\CODShort} object or the recovery procedure (discussed in more detail below), it is written to \emph{Global Storage} to make it publicly available to all clients.
\atreplace{From the implementation's perspective}{In {\Name}}, the moment a transaction is written to \atreplace{this storage}{{\GlobalStorage}}, it becomes committed and the persistence certificate\atremove{ from {\GlobalStorage}} \atreplace{doubles as the}{plays the role of a} commit certificate for the transaction.
One can view the {\GlobalStorage} abstraction as the ``final'' public transaction ledger.
However, unlike\atadd{ in} consensus-based cryptocurrencies, this ledger is not a sequence, but just a set, and the transactions in it can be applied in any order.
In the pseudocode, we denote this instance by $\oGlobalStorage$.

\mysubparagraph{{\AccountStorage}}

Every account $\vAcc$ is equipped with an instance of {\AppendOnlyStorage} called \textit{\AccountStorage}.
Inside this instance, clients that share account $\vAcc$ store information about started debit transactions on $\vAcc$ and initial states for {\CODShort} objects.
This allows us to ensure progress of all operations on the account: 
i.e., any operation invoked by an owner of $\vAcc$ terminates (assuming $\vAcc$ is correct).
More specifically, we use the technique known as \emph{helping}, common to concurrent algorithms~\cite{afek1993atomic,guerraoui2018algorithms,Her91}.
Communicating via storage allows clients to temporarily go offline (lose connection)  and still preserve all system guarantees, which would be hard to achieve with a broadcast-like algorithm.
In the code, we denote an instance of an {\AccountStorage} for an account $\vAcc$ as $\oAccountStorage[\vAcc]$. 

Similarly to the consensus objects, the implementation of {\AccountStorage} can be account-specific and only the owners of the account need to trust it.

\section{{\Name}: Algorithm}\label{sec:implementation}

\begin{figure*}[thb]
  \centering
    \small

  \includesvg[scale=0.95]{images/tx-lifecycle-algo.svg}
  
  \vspace{-.2cm}
  \caption{Transaction lifecycle in {\Name}}

  \label{fig:transaction-lifecycle-algo-main}
\end{figure*}

\atadd{In this section, we demonstrate how to combine {\CODLong}, {\AccountStorage}, {\GlobalStorage}, and {\Consensus} objects in order to obtain a protocol implementing an asset transfer system with the {\pTransferConcurrency} property as defined in \Cref{sec:problem-statement}, thus providing a constructive proof for \Cref{thm:main}.}

\subsection{Composing {\CODShort} and {\Consensus} instances} \label{subsec:composing-crypto-la-main}

In {\Name}, {\CODShort} serves as a fundamental building block: it allows the owners of an account $\vAcc$ to issue concurrent transactions.
As long as there are enough committed credit transactions to cover all\atremove{ the} debit transactions submitted to {\CODShort}, all owners of a correct account will be able to confirm their transactions and get matching certificates, i.e., the {\CODShort} object will accept them.

However, \atreplace{it is possible that correct clients try to overspend}{even correct clients may accidentally try to overspend}.
In this case, a correct client might return $\tFAIL$ from a $\oSubmit$ operation.
Intuitively, when this happens, we consider the internal state of the {\CODShort} instance being ``broken'' and we need to ``recover'' from this.

Let us overview the recovery procedure \atreplace{executed}{performed} after an overspending attempt was detected in $\oCOD[\vAcc][e]$.
First, the client invokes the $\oClose$ operation on $\oCOD[\vAcc][e]$, which will return a snapshot with the following data:
\begin{enumerate}
    \item a set of \emph{selected debits} that includes (but is not limited to) all debit transactions accepted by $\oCOD[\vAcc][e]$;
    \item a set of committed credit transactions sufficient to cover the selected debits;
    \item a set of \emph{cancelled debits}, for which there are not enough known credits.
\end{enumerate}

If multiple clients invoke $\oClose$, the snapshots they receive might vary due to the asynchronous nature of {\CODShort}. 
However, \emph{every} valid\atremove{ such} snapshot will include \emph{all} debit transactions that are or ever will be accepted by $\oCOD[\vAcc][e]$. %
To this end, the $\oClose$ operation ``invalidates'' $\oCOD[\vAcc][e]$, so that no transaction can be accepted after \atreplace{the}{a} snapshot was made (in case of concurrency, either the transaction will make it to the snapshot or it will not be accepted).

In order to initialize $\oCOD[\vAcc][e+1]$, the clients must agree on its initial state.
Hence, they will propose the snapshots they received from the $\oClose$ operation invoked on $\oCOD[\vAcc][e]$ to $\oConsensus[\vAcc][e+1]$.
There will be exactly one snapshot selected by the consensus instance, from which the initial state for $\oCOD[\vAcc][e+1]$ will be derived.
\atadd{We would like to emphasize that {\Name} preserves both safety and liveness regardless of which of the (potentially multiple) valid snapshots is selected by the consensus object as long as the owners of the account agree on it (which is guaranteed by the {\pConsConsistency} property of consensus).}

We do not bind clients to any specific consensus protocol:
every instance of consensus can be implemented differently.
To preserve safety, the consensus output must be \emph{notarized} (i.e., signed) by a quorum of replicas that run {\Name}\atadd{,} and no correct replica will sign two different consensus outputs.
Since the owners of a correct account will never try to get multiple different outputs notarized for the same consensus instance, they will always be able to get the notarization.
The consensus output signed by a quorum can be used to initialize $\oCOD[\vAcc][e+1]$.
This concludes the recovery procedure.

\subsection{Transfer algorithm}

\mysubparagraph{Overview}

At a very high level, when the $\oTransfer(\vTx)$ operation is invoked, $\vTx$ is submitted to the latest {\CODShort} instance.
If there are sufficient credit transactions on the account to cover $\vTx$, it is accepted by the {\CODShort} instance and registered in the {\GlobalStorage}. 
Otherwise, there is a risk of overspending, and the account's state is recovered as described in \Cref{subsec:composing-crypto-la-main}.
A transaction is considered \emph{failed} if it ends up in the set of cancelled debits in the snapshot selected by the consensus object.
Similarly, a transaction can be accepted by the recovery procedure if it ends up in the set of selected debits.
\Cref{fig:transaction-lifecycle-algo-main} depicts the lifecycle of a transaction in our protocol.

\mysubparagraph{Preparation}
The algorithm proceeds in consecutive epochs for each account.
For a correct account $\vAcc$, epoch $e$ corresponds to the period of time when $\oCOD[\vAcc][e]$ is active (i.e.,\atadd{ is} initialized\atremove{,} but is not yet closed).
Let us consider a correct client $p$, one of the owners of an account $\vAcc$.
When $p$ invokes $\oTransfer(\vTx)$, it first fetches the current epoch number $e$ and the initial state of $\oCOD[\vAcc][e]$ from the $\oAccountStorage[\vAcc]$.
The client also accesses the {\GlobalStorage} to fetch all \atreplace{new}{newly} committed credit \atreplace{transaction}{transactions} for account $\vAcc$. 
It will later submit these credit transactions to the {\CODShort}\atadd{ object} along with $\vTx$. 
Lastly, $p$ writes $\vTx$ to $\oAccountStorage[\vAcc]$.
This way, other owners of $\vAcc$ will be able to help\atremove{ to} commit $\vTx$ when they read it from the {\AccountStorage}.
This step is crucial to avoid starvation of slow clients.

\mysubparagraph{Main loop}
After the preparation, the client enters a loop.
In each iteration (corresponding to one \emph{epoch} $e$), the client reads\atadd{ the} pending debits (all debit transactions on account $\vAcc$ that are not yet committed) from the {\AccountStorage}.
After this, $p$ invokes the $\oSubmit$ operation on $\oCOD[\vAcc][e]$ with the pending debits it just read from the {\AccountStorage} and the committed credits it read during the preparation phase from the {\GlobalStorage}.
If the invocation returns $\tOK(\dots)$, then the only thing left to do is to commit $\vTx$ by writing it to the {\GlobalStorage} (as a result, the recipients of the transaction can learn about it by reading the {\GlobalStorage}).
Finally, $p$ returns $\tOK$ from the $\oTransfer$ operation with the certificate of a successful write to the {\GlobalStorage} acting as the commit certificate $\vCommitCert$ for transaction $\vTx$.

If, however, $\oCOD[\vAcc][e].\oSubmit$ returns $\tFAIL$, it indicates that either the volume of all known credits was insufficient to cover all existing debits at some point of time, or some other client invoked the $\oClose$ operation on this {\CODShort} object.
Either way\atadd{,} the client then goes through the recovery procedure as described in \Cref{subsec:composing-crypto-la-main} in order to obtain the initial state for $\oCOD[\vAcc][e+1]$.

Finally, the client checks whether\atadd{,} during the $\fRecovery$ procedure\atadd{,} $\vTx$ ended up in the set of selected or cancelled debits (see \Cref{subsec:composing-crypto-la-main}).
In the former case, the client commits the transaction by writing it to the {\GlobalStorage} and returns $\tOK$.
In the latter case, the client simply returns $\tFAIL$ from the $\oTransfer$ operation.
However, it might also happen that $\vTx$ is neither accepted nor canceled (e.g., if\atadd{,} due to concurrency, the initial state for $\oCOD[\vAcc][e+1]$ was selected before any process other than $p$ learned about $\vTx$).
In this case, $p$ increments its local epoch number $e$ and proceeds to the next iteration of the loop, proposing $\vTx$ to the next $\oCOD$ instance.

This loop eventually terminates due to the helping mechanism: once $\vTx$ is stored in the {\AccountStorage} in the preparation phase, every other owner of $\vAcc$ executing $\oTransfer$ submits $\vTx$ to {\CODShort} along with their own transactions.

\subsection{{\Name} pseudocode}

\atadd{The }pseudocode for {\Name} is provided in \Cref{alg:crypto-main-client,alg:crypto-main-client-recovery,alg:crypto-main-replica,alg:objects-init}.
We assume that each block of code (a function, an operation, a procedure or a callback) is generally executed sequentially to completion. 
However, a block may contain a $\WaitFor$ operator, which interrupts the execution until the wait condition is satisfied.
Some events, e.g., receiving a message, might trigger callbacks (marked with $\textbf{upon}$ keyword).
They are not executed immediately but are first placed in an event queue, waiting for their turn (we assume a fair scheduler). 
We denote an assignment of an expression $\vExpr$ to a variable $\vVar$  as $\vVar \oassign \vExpr$.

In the code, we assume that each message sent by a client carries a distinct sequence number.
When a replica replies to a client, it also implicitly includes the same sequence number in its response.
This allows clients to match replies with the corresponding request messages and ignore outdated replies.
A message from process $p$ is considered \emph{valid} by process $q$ if there exists a possible execution of the protocol in which $p$ is correct and it sends this message to $q$.
In most cases, this boils down to the correct number and order of message attachments as well as all the attached signatures and certificates being valid. 
In our protocols, we implicitly assume that invalid messages are simply ignored by correct processes. 
Hence, for the adversary, sending an invalid message is equivalent to not sending anything at all.

Recall that the $m$-th consensus instance associated with account $\vAcc$ is denoted by $\oConsensus[\vAcc][m]$.

\paragraph{Clients and replicas in {\Name}.}
The client's protocol is described in \Cref{alg:crypto-main-client,alg:crypto-main-client-recovery} and the replica's protocol is described in Algorithm~\ref{alg:crypto-main-replica}.
The algorithm uses instances of {\CODShort}, and instances of {\KeyValStorage}: {\GlobalStorage} (one per system, shared by all clients), and {\AccountStorage} (one per account, shared by the owners of the account). 
We describe objects initialization in \Cref{alg:objects-init}.
The implementations of {\CODShort} and {\KeyValStorage} are delegated to \Cref{app:cryptoCD,app:storage}, respectively.
The algorithm proceeds in consecutive epochs for each account. 
Let us consider a correct client $p$, one of the owners of a correct account $\vAcc$.

When $\oTransfer(\vTx)$ is invoked, client $p$ first fetches the current epoch number $\vEp$, the initial state $\vCODState$ of $\oCOD[\vAcc][\vEp]$ and a certificate $\vCert$ for $\vCODState$ from $\oAccountStorage[\vAcc]$ (line~\ref{line:main:fetch-cryptocd-state}). 
Then it reads committed transactions from {\GlobalStorage} at line~\ref{line:main:read-global-storage}.
After this, the client forms a set of credits (line~\ref{line:main:form-new-credits}) that are later submitted to the $\oCOD$ and writes $\vTx$ to {\AccountStorage} (line~\ref{line:main:put-tx-to-account-storage}): this way, other owners of $\vAcc$ will be able to help to commit $\vTx$.

\begin{algorithm*}[t]

    \caption{{\Name} (for client $p$, account $\vAcc$)}
    \label{alg:crypto-main-client}
    
    \begin{myalgorithmic}
        \State $\Type{\tCert}{\{0,1\}^*}$ \Comment{set of all possible cryptographic certificates}
        \State $\Type{\tSignedTx}{\tPair{\tT}{\tCert}}$ \Comment{set of all possible pairs of form $\Tuple{\vTx,\vCert}$, where $\vTx \in \tT$ and $\vCert \in \tCert$}
        
        \algspace
        \Operation{\oTransfer}{$\vTx$} \Returns{{$\tOK(\tCert)$} or {\tFAIL}}
                        \LineComment{Read the latest up-to-date state of the account}
            \State $\Tuple{\vEp, \vCODState, \vStateCert} \oassign \fReadLatestCODState()$ \label{line:main:fetch-cryptocd-state}
            
            \LineComment{Read all committed transactions (with their commit certificates) related to this account}
            \State $\vCommittedTxs\WithCert \oassign \oGetAccountTransactions()$ \label{line:main:read-global-storage}

            \State $\vNewCredits\WithCert \oassign \{\Tuple{\vTx', \vCert_{\vTx'}} \mid \Tuple{\vTx', \vCert_{\vTx'}} \in \vCommittedTxs\WithCert, \vTx'.\vRecipient = \vAcc \} \setminus \vCODState.\vInitCredits\WithCert$ \label{line:main:form-new-credits}
            
            \LineComment{Make sure that all other owners of this account will eventually see this transaction to prevent starvation}
            \State $\oAccountStorage[\vAcc].\oAppendKey(\sDebits, \vTx, \Null)$ \label{line:main:put-tx-to-account-storage}
                        \While {$\True$} \label{line:transfer-loop-begin}

                    \LineComment{Read debits from the {\AccountStorage}, no need for certificates}
                    \State $\vDebits  \oassign \fTransactions(\oAccountStorage[\vAcc].\oReadKey(\sDebits))$ \label{line:main:read-debits}
                    \LineComment{The client must help to commit other pending transactions in order to avoid starvation}
                                        \State \atrev{$\vPendingDebits \oassign \vDebits \setminus \vCODState.\vCancelledDebits$} \label{line:main:compute-pending-debits}
                                        
                    \LineComment{Help all replicas to catch up with the current epoch and initialize $\oCOD[\vAcc][\vEp]$}
                    \State \Send{$\mInitCOD, \vEp, \vCODState, \vStateCert$}{all replicas} \label{line:main:init-cryptocd}
                                        
                    \LineComment{Try to commit the pending transactions (including $\vTx$) in this epoch}
                    \State $\vCODResult \oassign \oCOD[\vAcc][\vEp].\oSubmit(\vPendingDebits, \vNewCredits\WithCert)$ \label{line:main:crypto-la-propose}
                    \If {$\vCODResult \tis \tOK(\vAcceptedDebits\WithCert, \vOutCredits\WithCert)$}
                        \LineComment{Extract the certificate for $\vTx$ from the $\oSubmit$ response}
                        \State let $\vCryptoCDCert$ be a certificate such that $\Tuple{\vTx, \vCryptoCDCert} \in \vAcceptedDebits\WithCert$
                                                \State $\vCommitCert \oassign \oGlobalStorage[\vAcc].\oAppendKey(\sTxs, \vTx, \Tuple{\vAcc, \vEp, \vCryptoCDCert})$ \label{line:main:write-transaction-after-crypto-la}
                        \State \Return{$\tOK(\vCommitCert)$} \label{line:main:return-ok-after-crypto-la}
                    \EndIf
                    
                    \LineComment{$\oCOD[\vAcc][\vEp].\oSubmit$ has failed}
                    \State $\Tuple{\vCODState, \vStateCert, \vSelectedDebits\WithCert} \oassign \fRecovery(\vEp\atremove{, \vTx}, \vPendingDebits)$ \label{line:main:call-recovery}

                    \If {$\vTx \in \vCODState.\vSelectedDebits$} \label{line:main:check-recovery-results}
                        \State let $\vRecoveryCert$ be a certificate such that $\Tuple{\vTx, \vRecoveryCert} \in \vSelectedDebits\WithCert$
                        \State $\vCommitCert  \oassign \oGlobalStorage[\vAcc].\oAppendKey(\sTxs, \vTx, \vRecoveryCert)$ \label{line:main:write-transaction-after-recovery}

                        \State \Return{$\tOK(\vCommitCert)$} \label{line:main:return-ok-after-recovery}
                    \EndIf
                    
                    \If {$\vTx \in \vCODState.\vCancelledDebits$}  \label{line:main:check-if-cancelled}
                                                                        
                        \State \Return $\tFAIL$  \Comment{The transaction is cancelled due to insufficient balance} \label{line:main:return-fail-after-recovery}
                    \EndIf

                    \State $\vEp$ \texttt{+=} 1
                        \label{line:main:tx-not-found-after-recovery}
                        \label{line:main:increment-epoch} 
                    \LineComment{$\vCODState$ and $\vStateCert$ from the recovery are used for the next iteration of the loop}
                    \label{line:transfer-loop-end}
            \EndWhile
        \EndOperation
          
        \algspace
        \Operation{$\oGetAccountTransactions$}{} \Returns{$\tSet{\tSignedTx}$}
            \State $\vCommittedTxs\WithCert \oassign  \oGlobalStorage[\vAcc].\oReadKey(\sTxs)$
                        \State \Return $\{\Tuple{\vTx, \vCommitCert} \in \vCommittedTxs\WithCert \mid \vTx.\vSender = \vAcc \tor \vTx.\vRecipient = \vAcc \}$
        \EndOperation
        
        \algspace
        \PublicFunction{$\fVerifyCommitCertificate$}{$\vTx$, $\vCommitCert$} \Returns{$\tBoolean$} \label{line:verify-commit-certificate}
            \State \Return $\oGlobalStorage.\fKVSVerifyStoredCert(\sTxs, \vTx, \vCommitCert)$
        \EndPublicFunction

        \algspace
        \Function{$\fReadLatestCODState$}{} \Returns{$\Tuple{\tEpochNum, \tCloseState, \tCert}$}
                                                                        \State \Return $\Tuple{\vEp, \vCODState, \vCert}$ such that $\Tuple{\Tuple{\vEp, \vCODState, \vCert}, \bot} \in \oAccountStorage[\vAcc].\oReadKey(\sState)$ and $\vEp$ is maximum
        \EndFunction
        
    \end{myalgorithmic}

\end{algorithm*}

\begin{algorithm*}[t]
    
    \caption{{\Name}: $\fRecovery$ (code for client $p$, account $\vAcc$)}
    \label{alg:crypto-main-client-recovery}

    \begin{myalgorithmic}
        \Function{\fRecovery}{$\vEp$, $\vPendingDebits$} \Returns{$\tTuple{\tCloseState, \tCert, \tSet{\tSignedTx}}$}
            \LineComment{Close the current instance of {\CODShort} and get the closing state and a certificate confirming that this state is valid}
            \State $\Tuple{\vCODState, \vCloseStateCert} \oassign \oCOD[\vAcc][\vEp].\oClose(\vPendingDebits)$ \label{line:recovery:invoke-close}
            
                        \State $\Tuple{\vAllCredits\WithCert, \vSelectedDebits, \vCancelledDebits} \oassign \vCODState$
                        
            \LineComment{The consensus mechanism resolves the overspending attempts that led to the recovery.}
                        \State $\Tuple{\vNextCODState, \vNextCODStateCert} \oassign \oConsensus[\vAcc][\vEp + 1].\oConsPropose(\Tuple{\vCODState, \vCloseStateCert})$ \label{line:recovery:consensus-call}
           
            \LineComment{Since consensus is only trusted by the owners of the account, the client needs to commit to the consensus output.}
            \LineComment{This prevents malicious clients from creating multiple different initial states for the same COD.}             \State \Send{$\mCommitCODInitialState, \vEp+1, \vNextCODState, \vNextCODStateCert$}{all replicas} \label{line:recovery:sign-begin}
            
            \State \WaitFor valid $\Message{\mCommitCODInitialStateResp, \vSignatureState_i, \vSignatureTxs_i}$ replies from a quorum $Q$ \label{line:recovery:wait-for-commit-state-resp}

            \State $\vStateCert \oassign \fCreateThresholdSignature(\Message{\mCommitCODInitialStateResp, \vNextCODState}, \{\vSignatureState_i\}_{i \in Q})$ \label{line:recovery:sign-end}

                        \LineComment{Let other owners of the account know about the new epoch and its initial state.}
            \State $\oAccountStorage.\oAppendKey(\sState, \Tuple{e + 1, \vNextCODState, \vStateCert}, \bot)$
            
            \LineComment{Create confirm certificates for the selected debits.}
            \State $\vMerkleTree  \oassign \fMerkleTree(\vNextCODState.\vSelectedDebits)$
            \State $\vMTCert \oassign \fCreateThresholdSignature(\Message{\mConfirmInRecovery, e, \vMerkleTree.\vMTRoot}, \{\vSignatureTxs_i\}_{i \in Q})$
            \State $\vSelectedDebits\WithCert  = \{ \Tuple{\vTx, \Tuple{\vMerkleTree.\vMTRoot, \fGetItemProof(\vMerkleTree, \vTx), \vMTCert, e}} \mid \vTx \in \vNextCODState.\vSelectedDebits \}$

            \State \Return $\Tuple{\vNextCODState, \vStateCert, \vSelectedDebits\WithCert}$
        \EndFunction
        
        \algspace
        \PublicFunction{$\fVerifyRecoveryCert$}{$\vTx$, $\vRecoveryCert$} \Returns{$\tBoolean$} \label{line:verify-recovery-certificate}
            \State $\Tuple{\vMTRoot, \vItemProof, \vTSMrkTree, e} \oassign \vRecoveryCert$
            \State \Return {$\fVerifyItemProof(\vMTRoot, \vItemProof, \vTx) \tand \fVerifyThresholdSignature(\Tuple{\mConfirmInRecovery, e, \vMTRoot}, \vTSMrkTree)$}
        \EndPublicFunction

                                                    \end{myalgorithmic}

\end{algorithm*}

\begin{algorithm*}[t]
        \caption{{\Name} (code for replica $r$)}
    \label{alg:crypto-main-replica}
            
    \begin{myalgorithmic}

        \ProcessState
                                                            
            \begin{ppreview}
            \BreakableLine[$\vSignedStates$~--~] mapping from an account and an epoch number to the hash of a signed state,
                \RaggedBreak 
                initially: $\forall a \in \aA, s \ge 0: \vSignedStates[a][s] = \Null$
            \EndBreakableLine
            \end{ppreview}

        \EndProcessState
        
        \algspace
        \UponReceive{$\mInitCOD, \vEp, \vCODState, \vCert$}{owner $q$ of account $\vAcc$}
                                    \If {$\oCOD[\vAcc][\vEp]$ is already initialized} \Return \EndIf
                                    \If {\tnot $\fVerifyThresholdSignature(\Message{\mCommitCODInitialStateResp, \vCODState}, \vStateCert)$} \Return \EndIf
                        \State $\Tuple{\vAllCredits\WithCert, \vSelectedDebits, \vCancelledDebits} \oassign \vCODState$
                        \BreakableLine \textbf{initialize} $\oCOD[\vAcc][\vEp]$ \textbf{with}
                \label{line:main:initialize-next-cryptocd}
                \RaggedBreak $\vInitDebits \oassign \vSelectedDebits$,
                \RaggedBreak $\vInitCredits\WithCert \oassign \vAllCredits\WithCert$,
                \RaggedBreak $\vRestrictedDebits \oassign \vCancelledDebits$
            \EndBreakableLine
        \EndHandler

        \algspace
            \UponReceive{$\mCommitCODInitialState, \vEp, \vCODState, \vCODStateCert$}{owner $q$ of account $\vAcc$}
            \label{line:receive-commit-crypo-la-initial-state-begin}
            
            \If {$\tnot \oCOD[\vAcc][e - 1].\oVerifyCloseStateCert(\vCODState, \vCODStateCert)$} \Return \EndIf \label{line:verify-close-state-cert}
            \LineComment{Check if previously signed a different initial state for $\oCOD[\vAcc][e]$}
            \If{$\vSignedStates[\vAcc][\vEp] \neq \Null \tand \vSignedStates[\vAcc][\vEp] \neq \fHash(\vCODState)$} \Return \EndIf
            \State $\vSignedStates[\vAcc][\vEp] \oassign \fHash(\vCODState)$
            \State $\vSignatureState \oassign \fSign(\Message{\mCommitCODInitialStateResp, \vCODState})$
            
                                                \State $\vSignatureTxs \oassign \fSign(\Tuple{\mConfirmInRecovery,e - 1, \fMerkleTree(\vCODState.\vSelectedDebits).\vMTRoot})$
            
            \State \Send{$\mCommitCODInitialStateResp, \vSignatureState, \vSignatureTxs$}{$q$}
            \label{line:receive-commit-crypo-la-initial-state-end}
        \EndHandler
       
    \end{myalgorithmic}

\end{algorithm*}

\paragraph{Accessing {\CODShort}}
After this, the client enters a while loop. 
In each iteration of the loop (corresponding to one \emph{epoch} $e$), the client reads pending debits on $\vAcc$ from {\AccountStorage} (\cref{line:main:read-debits,line:main:compute-pending-debits}).
Then, it sends the epoch number $e$, the initial state $\vCODState$\atadd{,} and the matching certificate $\vCert$ to the replicas (line~\ref{line:main:init-cryptocd}). 
This way replicas that fall behind can initiate an up-to-date instance of {\CODShort} object for account $\vAcc$ before receiving messages associated with it.
After this, client $p$ invokes the $\oSubmit$ operation on the corresponding {\CODShort} object.
If the invocation returns $\tOK(\dots)$, then the only thing left to do is to write $\Tuple{\vTx, \vCert}$ to the {\GlobalStorage} at line~\ref{line:main:write-transaction-after-crypto-la} (as a result, the recipients of the transaction can learn about it by performing $\oGetAccountTransactions()$ on their side).
Finally, $p$ returns $\tOK$ with the commit certificate from the $\oTransfer$ operation.

Otherwise, if $p$ returns $\tFAIL$ from \yareplace{$\oCOD[\vAcc][e].\oSubmit(\vDebits, \vCredits\WithCert)$}{\oSubmit}, then either all known credits were not enough to cover all existing debits at some point of time, or some other client invoked $\oClose$ operation on this {\CODShort} object.
Either way the client then goes through the $\fRecovery$ procedure to get the state for the next ${\CODShort}[\vAcc][e+1]$ (line~\ref{line:main:call-recovery}).

\paragraph{Recovery and consensus.}

In the $\fRecovery$ procedure, the client first makes sure that the current instance $\oCOD[\vAcc][e]$ is closed by invoking $\oClose$ operation on it  (line~\ref{line:recovery:invoke-close}).
Then it uses consensus object $\oConsensus[\vAcc][\vEp+1]$ (line~\ref{line:recovery:consensus-call}) in order to initialize the next instance of {\CODShort}.
We allow the owners of each account to use local, distinct consensus objects that   can be handled outside of the {\Name} system.
However, due to this, the outcome of consensus should be signed by a quorum of replicas (lines~\ref{line:recovery:sign-begin}-\ref{line:recovery:sign-end} and \ref{line:receive-commit-crypo-la-initial-state-begin}-\ref{line:receive-commit-crypo-la-initial-state-end}), so that Byzantine clients are not able to break the system and use a diverging result.

The resulting initial state for the {\CODShort} object in epoch $e + 1$ contains $\vSelectedDebits$, a set of transactions that were decided to be used as initial debits in the new transactions. 
It is guaranteed that all previously committed debits and all debits accepted by the $\oCOD[\vAcc][e]$ are in this set, though it may also contain some extra debits.
For convenience, the $\fRecovery$ procedure also returns $\vSelectedDebits\WithCert$ (signed version of  $\vSelectedDebits$) that are ready to be written to Global Storage.
For every $\Tuple{\vTx, \vCert} \in \vSelectedDebits\WithCert: \fVerifyRecoveryCert(\vTx, \vCert) = \True$,
where $\fVerifyRecoveryCert$ is a publicly known function.

\balance

Finally, the client checks if $\vTx$ is in the set $\vCODState.\vSelectedDebits$ (line~\ref{line:main:check-recovery-results})
or $\vCODState.\vCancelledDebits$ (line~\ref{line:main:check-if-cancelled}).
In the former case, the client\atadd{ commits the transaction by appending it to the {\GlobalStorage} (\cref{line:main:write-transaction-after-recovery}) and} returns $\tOK$.
\atrev{In the latter case, it simply returns $\tFAIL$.}

However, it might happen that $\vTx$ is in neither of these sets.
In this case, $p$ proceeds \atreplace{with}{to} the next iteration of the while loop corresponding to the next epoch number $e+1$ (line~\ref{line:main:increment-epoch}).

\clearpage
\nobalance

\begin{strip}
\begin{nonfloatalgorithm}
    
    \caption{{\CODShort}, Global Storage and Account Storage initialization}
    \label{alg:objects-init}
    
    \begin{myalgorithmic}
        \DeclareLocalFunction{VerifyDebit}
        \DeclareLocalFunction{VerifyAccountState}

        \ProcessObjects
            
            \BreakableLine[] $\oGlobalStorage$ is an $\oAOStorage$ object with one key:
                \RaggedBreak\algtab key $\sTxs$ with initial set of value $\{\vTxInitForA\}_{a\in \aA}$ and validity function $\fVerifyTxCertForGlobalStorage$
            \EndBreakableLine

            \BreakableLine[]
                $\forall a \in \aA: \oAccountStorage[a]$ is an $\oAOStorage$ object with the following keys:
                    \RaggedBreak\algtab key $\sDebits$ with initial set of values $\emptyset$ and validity function $\fVerifyDebit_{\vAcc}$
                    \RaggedBreak\algtab key $\sState$ with initial set of values $\{\Tuple{1,\Tuple{\{\vTxInitForA\}, \emptyset, \emptyset}, \Null}\}$ and validity function $\fVerifyAccountState_{\vAcc}$
            \EndBreakableLine
            
            \State $\forall a \in \aA: \oCOD[a][0] = \oCOD(a, 1, \emptyset,\{\vTxInitForA\},\emptyset)$
        \EndProcessObjects
        
        \algspace
        \Function{$\fVerifyTxCertForGlobalStorage$}{$\vTx$, $\vCert$} \Returns{$\tBoolean$}
                                                \If{$\Tuple{\vAcc_{\vCert}, \vEp_{\vCert}, \vCODCert} = \vCert$}
                \State \Return $\oCOD[\vAcc_{\vCert}][\vEp_{\vCert}].\fVerifyCODCert(\vTx, \vCODCert)$
            \EndIf
            \State \Return $\fVerifyRecoveryCert(\vTx, \vCert)$
        \EndFunction
        
        \algspace
        \Function{$\fVerifyDebit_{\vAcc}$}{$\vTx$, $\_$} \Returns{\tBoolean}
            \State \Return{$\vTx.\vSender = \vAcc \tand \vTx\text{ is well-formed (including a valid signature)}$}
        \EndFunction

        \algspace
        \Function{$\fVerifyAccountState_{\vAcc}$}{$\vState$, $\_$} \Returns{\tBoolean}
            \State $\Tuple{\vEp, \vCODState, \vStateCert} \oassign \vState$
            \State \Return{$\fVerifyThresholdSignature(\Message{\mCommitCODInitialStateResp, \vNextCODState}, \vStateCert)$}
        \EndFunction

    \end{myalgorithmic}

\end{nonfloatalgorithm}
\end{strip}

\subsection{Latency breakdown and optimizations} \label{subsec:total-latency-and-optimizations}

\atadd{\Cref{line:main:fetch-cryptocd-state,line:main:read-global-storage,line:main:put-tx-to-account-storage}, as well as \cref{line:main:read-debits} on the first iteration of the loop, can all be executed in parallel. This is crucial to achieve the latency claimed in \Cref{thm:main}.} 
\begin{atreview}
With this optimization applied and with the instantiation of {\CODShort} described in \Cref{sec:cod-main}, the algorithm achieves latency of $5$ round-trips in case of absence of concurrency.

We can further break down the latency costs into 3 categories:
\begin{itemize}
    \item 2 round-trips are necessary for a consistent broadcast protocol with optimal resilience ($n=3f+1$) and a linear number of messages~\cite{textbook}. This is the latency of the purely asynchronous asset transfer protocols with optimal resilience~\cite{astro-dsn,fastpay}, which serve as a baseline for us;

    \item 2 round-trips to read the up-to-date state (\cref{line:main:fetch-cryptocd-state,line:main:read-global-storage,line:main:read-debits} executed in parallel). This facilitates light-weight clients and is especially important in the context of shared accounts, where a client may not always have up-to-date information about its own account;

    \item 1 round-trip to facilitate the recovery.
\end{itemize}
As discussed in \Cref{subsec:cod-performance}, somewhat surprisingly, the {\pTransferConcurrency} property of {\Name} does not have an inherent latency cost, as the {\CODShort} protocol automatically adapts to the current level of contention.
\end{atreview}

\begin{atreview}
\myparagraph{Further optimizations}
We believe that one can achieve a smaller latency with any combination of the following techniques (each coming at its own cost):
\begin{description}
    \item[Use larger quorums:] It is a common pattern in distributed computing to trade resilience for latency~\cite{fast-paxos,fab-paxos,broadcast-latency-orig,ktz21,consensus-latency-cat,broadcast-latency-cat,sliwinski2022consensus}.
    Recently, it was exploited in a context very similar to {\Name} in~\cite{sliwinski2022consensus}.

    \item[Use all-to-all communication between replicas:] Naturally, all-to-all communication often offers smaller latency than client-server communication pattern (as in {\Name}) at the cost of a quadratic number of messages being exchanged. This technique was also adopted by~\cite{sliwinski2022consensus}.

    \item[Execute storage write-back in parallel with $\oCOD.\oSubmit$:] By opening up the storage black box, one may try to perform the write-back phase of $\oReadKey$ (described in \Cref{sec:keyval-storage}) in parallel with the $\oSubmit$ operation of {\CODShort}. This may reduce the latency by 1 RTT at the cost of slightly complicating the protocol.

            \item[Make additional assumptions about the clients:] This may help to avoid the need to interact with storage prior to accessing {\CODShort}.
    However, relying on the client's local state, even if the client actively listens to all events on the network, would likely require a slight relaxation of the {\pTransferConsistency} property (as defined in \Cref{sec:problem-statement}).

    \item[Make a more monolithic design:] In the current design, clients are required to relay information about credits from {\GlobalStorage} to {\CODShort}. It may be possible to avoid this if these objects are maintained by the same set of replicas. If combined with the two preceding suggested optimizations, it may be sufficient to avoid the need to read the storage before accessing {\CODShort}, even without relaxing the consistency requirement.

        \end{description}
\end{atreview}

\section{Proof Outline}

In this section, we sketch the main \yareplace{correctness }{}arguments for \yareplace{our implementation}{Theorem~\ref{thm:main}}. \pprev{In particular, we show that {\Name} protocol satisfies all the conditions imposed by the theorem.} The detailed proof is deferred to \Cref{app:proof}.

\mysubparagraph{\pTransferSafety} This property requires that, for any account, the balance is always non-negative.
This property is ensured by the {\CODShort} objects used on account $\vAcc$, which guarantee that, at any time, the set of accepted debit transactions does not surpass the total value of committed credit transactions.
We also show that this invariant is preserved during the transition between consecutive {\CODShort} objects.

\mysubparagraph{\pTransferConsistency}
To show that our implementation of {\Name} satisfies the {\pTransferConsistency} property, we define an order on the transactions on a correct account $\vAcc$ and then show that this order is both legal and consistent with the real-time order $\prec_{\Execution,\vAcc}$. More precisely, we order transactions by the epochs they are accepted in, and, then, inside each epoch $e$, we divide them \atreplace{in}{into} three consecutive groups: (i) transactions accepted by $\oCOD[\vAcc][e]$, (ii) selected by $\oConsensus[\vAcc][e + 1]$, and (iii) failed debit transactions canceled by $\oConsensus[\vAcc][e + 1]$. 
Inside each group we order transactions by $\fEnd_{\vAcc}(\vTx)$, giving the priority to the credit transactions in case of ties.

\mysubparagraph{\pTransferValidity}
The proof of {\pTransferValidity} property is relatively simple. 
We show that {\Name} satisfies the following two facts: (i) any successful transaction is written to the {\oGlobalStorage}\atadd{,} which produces a commit certificate $\vCommitCert$,  and (ii) $\fVerifyCommitCertificate$ is implemented via verification function of the certificate from the {\GlobalStorage}.

\mysubparagraph{\pTransferLiveness} {\Name} ensures that every operation invoked by a correct client returns. The proof proceeds as follows. We first show that all operations invoked on {\AccountStorage}, {\GlobalStorage}\atadd{,} and {\oCOD} eventually terminate. Then, we show that the number of epochs one $\oTransfer$ operation can span \atreplace{on}{over} is finite. Combining these facts, we demonstrate that any operation invoked by a correct client returns.

\mysubparagraph{\pAccountTransactions} We ensure that {\Name} satisfies the {\pAccountTransactions} property with the help of {\GlobalStorage}. Every time a client invokes $\oGetAccountTransactions$, it essentially reads all committed transactions from $\oGlobalStorage$ and then filters out the ones that are not relevant.

\balance

\mysubparagraph{\pTransferConcurrency} In the {\pTransferConcurrency} property, we show that, if from some moment on, there are no overspending attempts observed on a correct account $\vAcc$ (i.e., the total amount spent by all active debit transactions does not exceed the balance at any time $t$), then there is some moment of time after which no {\oConsensus} object is invoked on this account. 
We prove this by showing that, in such a case, the number of epochs an account goes through is finite\atadd{,} and thus, from some point on, the account does not go through the recovery process and owners do not use $\oConsensus$ objects.

\begin{ppreview}
\mysubparagraph{Latency}
To prove that $k$-overspending-free latency of {\Name} is $k + 4$ RTTs, we show that the upper bound for the latency of $\oCOD.\oSubmit$ is $k + 1$. Here, $k$ round-trips come from the Prepare phase, and $1$ more comes from the Accept phase. We also prove that latencies of $\oAppendKey$ and $\oReadKey$ operations of {\AppendOnlyStorage} are constant and equal to $2$ and $1$ RTTs respectively. 
Finally, we then conclude that with most of the read requests combined as described in \Cref{subsec:total-latency-and-optimizations}, we achieve $k$-overspending-free latency of $k + 4$ RTTs.
\end{ppreview}

\section{Concluding Remarks}\label{sec:discussion}

There are multiple interesting directions for future work.

First, our algorithm leaves space for further optimizations\atadd{ in addition to the ones we discussed in \Cref{subsec:total-latency-and-optimizations}}.
To preclude the system state and protocol messages from growing without bound, one can introduce a checkpointing mechanism.
For example, checkpointing can be implemented using occasional invocations of consensus similar to how our protocol resolves overspending attempts. 
\atrev{Additionally, a practical implementation should probably avoid repeatedly exchanging (potentially, large) sets of transactions between the replicas and the clients and instead should send only the updates (``deltas'') since their last communication.}

Second, to quantify the actual performance gains of {\Name}, a practical implementation and a comparison to other similar protocols such as~\cite{byzgenpaxos} would be of interest.

\yareplace{Thirdly}{Third}, recent results~\cite{dbrb,pastro21disc,bla,sliwinski2022consensus} demonstrate how asynchronous Byzantine fault-tolerant systems can be \emph{reconfigured} without relying on consensus.
Notably, in~\cite{pastro21disc}, it is shown that a permissionless Proof-of-Stake asset transfer system can be implemented using a similar technique.
A fascinating challenge is to combine reconfiguration with the ideas of {\Name} for building a \emph{reconfigurable}, or even \emph{permissionless}, asset transfer system with shared accounts and without a central consensus mechanism.

\pkrev{\atreplace{We}{Finally, we} believe that our work opens the way to \emph{optimally-concurrent} solutions to other, more general problems, such as \yaadd{fungible} \emph{token smart contracts}~\cite{smart-contracts-21}, an abstraction intended to grasp the synchronization requirements of \yaadd{a specific type of} Ethereum smart contracts. 
The objects allow the set of account owners to vary over time, which might require generalizing our notion of conflicts.}

\pkrev{In the general case, it is appealing to address the question of optimally-concurrent state machine replication.
Intuitively, one would like to avoid consensus-based synchronization whenever any reordering of concurrent operations has the same effect.
Formalizing this intuition and implementing this kind of optimality in the context of generic state machine replication is left for future work.}

\begin{ppreview}
\section*{Acknowledgments}
We would like to thank the anonymous reviewers for their insightful comments on the structure of the paper and overall presentation as well as for drawing our attention to the question of the algorithmic complexity of the protocol and for providing references to the important related work we were missing in our original draft.
\end{ppreview}
\if \REVIEW 0
    \atadd{This work was in part supported by TrustShare Innovation Chair (financed by Mazars).}
\fi

\clearpage
\bibliographystyle{abbrv}
\bibliography{main}

\clearpage
\appendix

\nobalance
\section{{\CODLong}}
\label{app:cryptoCD}

\pprev{In this section, we state the properties {\CODLong} should satisfy and how we implement this abstraction.}
\subsection{Formal definition of {\CODLong}}

An instance of {\CODShort} is identified by an account $\vAcc$ and an epoch number $\vEp$.
Additionally, all owners of the account and correct replicas must run the {\CODShort} instance with the same initial state that consists of:
\begin{itemize}
    \item $\vInitDebits: \tSet{\tT}$ -- the initial set of debit transactions for account $\vAcc$, $\forall \vTx \in \vInitDebits: \vTx.\vSender = \vAcc$;

    \item $\vInitCredits\WithCert: \tSet{\tPair{\tT}{\Sigma}}$ -- initial set of credit transactions for account $\vAcc$, $\forall \Tuple{\vTx, \vCert} \in \vInitCredits\WithCert: \vTx.\vRecipient = \vAcc \tand \fVerifyCommitCertificate(\vTx, \vCert) = \True$;
    
    \item $\vRestrictedDebits: \tSet{\tT}$ -- the set of debit transactions for account $\vAcc$ that this {\CODShort} object is prohibited from accepting, $\vRestrictedDebits \cap \vInitDebits = \emptyset$;
\end{itemize}
Moreover, the initial balance of the account must be non-negative (i.e., $\fTotalValue(\fTransactions(\vInitCredits\WithCert)) \ge \fTotalValue(\vInitDebits)$).

The {\CODLong} abstraction exports two operations: $\oSubmit(\vDebits, \vCredits\WithCert)$ and $\oClose(\vPendingDebits)$. 
It also provides two verification functions: $\oVerifyCODCert(\vTx, \vCert)$ and  $\oVerifyCloseStateCert(\vCODState, \vCert)$.

We say that a transaction $\vTx$ is \emph{accepted by an instance $I$ of {\CODShort}} iff $\vTx \in I.\vInitDebits$ or there exists $\vCert$ such that $I.\oVerifyCODCert(\vTx, \vCert) = \True$.
In order to get new debit transactions accepted, correct clients submit them to the {\CODShort} object by invoking $\oSubmit(\vDebits, \vCredits\WithCert)$.
Here, $\vCredits\WithCert$ is the set of committed credit transactions the client is aware of with the corresponding commit certificates.

In the most common case when no overspending \atreplace{is}{attempts are} detected, $\oSubmit(\vDebits, \vCredits\WithCert)$ returns $\tOK(\vDebits\WithCert, \vOutCredits\WithCert)$, where $\vDebits\WithCert$ contains the same transactions as in $\vDebits$ augmented with the cryptographic certificates confirming that these transactions are accepted by a {\CODShort} and $\vOutCredits\WithCert$ contains enough committed credit transaction to cover all of the transactions in $\vDebits$.
More formally, the following two properties are satisfied:

\begin{properties}
    \propertyitem{\pCODProposeValidity} If a correct client obtains $\tOK(\vDebits\WithCert, \vOutCredits\WithCert)$ from $\oSubmit(\vDebits, \vCredits\WithCert)$, then $\fTransactions(\vDebits\WithCert) = \vDebits$ and $\forall \Tuple{\vTx, \vCert} \in \vDebits\WithCert: \oVerifyCODCert(\vTx, \vCert) = \True$. Moreover, $\forall \Tuple{\vTx, \vCert} \in \vOutCredits\WithCert: \fVerifyCommitCertificate(\vTx, \vCert) = \True$;

    \propertyitem{\pCODProposeSafety}~
        \begin{itemize}

            \item At any time $t$, the total value of\atadd{ the} initial debits ($\vInitDebits$) and\atadd{ the} debits accepted by a {\CODShort} object by time $t$ does not exceed the total value of committed credits on\atadd{ the} account $\vAcc$ by time $t$.
            Moreover, if $\vAcc$ is correct, it does not exceed the total amount of all credits returned by the $\oSubmit$ operation by time $t$;

            \item No $\vTx \in \vRestrictedDebits$ is ever accepted by a {\CODShort} object.
            
        \end{itemize}
\end{properties}

$\oCOD.\oSubmit$ may also return $\tFAIL$ if the replicas observe an overspending attempt.
Note, however, that due to communication delays, a replica may observe only a subset of all the $\oSubmit$ operations that are being executed.
Hence, we allow the {\CODShort} object return $\tFAIL$ if \emph{any} subset of $\oSubmit$ operations overspends.
Another case when we allow the {\CODShort} object to return $\tFAIL$ is when some client already invoked the $\oCOD.\oClose$ operation.
More formally:

\begin{properties}
    \propertyitem{\pCODProposeSuccess} 
        If (i) $\vAcc$ is a correct account, (ii) no client invokes the $\oClose$ operation, and (iii) for every subset $S$ of invoked operations $\oSubmit(\vDebits_i, \vCredits\WithCert_i)$,
                $\fTotalValue(\vDebits) \le \fTotalValue(\vCredits)$ (where $\vDebits = \bigcup_{i \in S} \vDebits_i \cup \vInitDebits$ and $\vCredits = \fTransactions(\bigcup_{i \in S} \vCredits\WithCert_i \cup \vInitCredits\WithCert)$, then no $\oSubmit$ operation returns $\tFAIL$\@.
\end{properties}

In {\Name}, once a correct process receives $\tFAIL$ from an invocation of $\oSubmit$,
it proceeds to closing the {\CODShort} instance by invoking $\oClose(\vPendingDebits)$,
where $\vPendingDebits$ is an arbitrary set of debit transactions.
The operation returns $\Tuple{\vCODState, \vStateCert}$, where $\vCODState$ is a snapshot of the accumulated internal state of the {\CODShort} object and $\sigma$ is a certificate confirming the validity of the snapshot.
$\vStateCert$ can be later verified by any third party using the $\oVerifyCloseStateCert$ function.

The snapshot $\vCODState$ contains the following fields:
\begin{enumerate}
    \item $\vCredits\WithCert$ -- the set of credits submitted to this {\CODShort} along with their commit certificates;
    
    \item $\vSelectedDebits$ -- a set of debits submitted to this {\CODShort} such that $\fTotalValue(\vSelectedDebits) \le \fTotalValue(\fTransactions(\vCredits\WithCert))$; 
    
    \item $\vCancelledDebits$ -- a set of debits submitted to this {\CODShort} that cannot be added to $\vSelectedDebits$ without exceeding the amount of funds provided by $\vCredits\WithCert$.
\end{enumerate}
The set $\vSelectedDebits$ must contain all the transactions that have been or ever will be accepted by this {\oCOD} instance.
To this end, as stipulated by the name of the operation, it ``closes'' the instance of {\CODShort} and, as already formalized in the {\pCODProposeSuccess} property, new invocations of the $\oSubmit$ operation may return $\tFAIL$ even there is no overspending.
Formally, operation $\oClose$ must satisfy the following properties:

\begin{properties}
    
    \propertyitem{\pCODCloseValidity} If a correct client obtains $\Tuple{\vCODState, \vStateCert}$ from $\oClose(\vPendingDebits)$, then $\oVerifyCloseStateCert(\vCODState, \vStateCert) = \True$;

    \propertyitem{\pCODCloseSafety} If a correct client obtains $\Tuple{\vCODState, \vStateCert}$  from $\oClose(\vPendingDebits)$, then:
        \begin{itemize}
                        \item $\vInitDebits \subseteq \vCODState.\vSelectedDebits$ and $\vRestrictedDebits \subseteq \vCODState.\vCancelledDebits$;
            
            \item for every transaction $\vTx$ accepted by this {\CODShort} instance: $\vTx \in \vCODState.\vSelectedDebits$;

            \item $\vPendingDebits \subseteq  \vCODState.\vSelectedDebits \cup \vCODState.\vCancelledDebits$;
                        
            \item for any $\oSubmit$ operation that returns  $\tOK(\vDebits\WithCert, \vOutCredits\WithCert)$: $\vOutCredits\WithCert \subset \vCODState.\vCredits\WithCert$ and $\forall \Tuple{\vTx, \vCert} \in \vCODState.\vCredits\WithCert$: $\fVerifyCommitCertificate(\vTx, \vCert) = \True$. Moreover, $\vInitCredits\WithCert \subseteq \vCODState.\vCredits\WithCert$;
            
            \item $\fTotalValue(\fTransactions(\vCODState.\vCredits\WithCert)) \ge \fTotalValue(\vCODState.\vSelectedDebits)$. Moreover, $\nexists \vTx \in \vCancelledDebits$, such that $\fTotalValue(\fTransactions(\vCODState.\vCredits\WithCert)) \ge \fTotalValue(\vCODState.\vSelectedDebits \cup \{\vTx\})$;
            
            \item $\vCODState.\vSelectedDebits \cap \vCODState.\vCancelledDebits = \emptyset$;

        \end{itemize}
\end{properties}

Finally, both operations must eventually terminate:

\begin{properties}
    \propertyitem{\pCODLiveness} Every call to $\oSubmit$ and $\oClose$ operations by a correct client eventually returns.
\end{properties}

In {\Name}, each account $\vAcc$ is provided with a list of {\CODShort} objects $\oCOD[\vAcc][1,2, \ldots]$ and for all $e \ge 1$, the initial state ($\vInitDebits$, $\vInitCredits\WithCert$, and $\vRestrictedDebits$) of $\oCOD[\vAcc][e+1]$ is initialized using a snapshot returned by $\oCOD[\vAcc][e]$ ($\vSelectedDebits$, $\vCredits\WithCert$, and $\vCancelledDebits$).

\subsection{Implementation of {\CODLong}}

The variables defining  a {\CODLong} object and a replica's state are listed in Algorithm~\ref{alg:crypto-la-state}.
The pseudocode of the verifying functions is provided \Cref{alg:crypto-la-verify}, while the protocols of a client and a replica are presented in \Cref{alg:crypto-la-client,alg:crypto-la-replica}, respectively.
Here we implicitly assume that each protocol message and each message being signed carries information on the account $\vAcc$ and the epoch the {\CODShort} object is parameterized with.
This ensures that different instances of {\CODShort} do not interfere with each other and, in particular, that signatures created in one instance cannot be used in another one.

The $\oSubmit$ operation consists of two phases: Prepare and Accept.

\myparagraph{Prepare phase}

The implementation of the Prepare phase inherits the key ideas from the Generalized Lattice Agreement protocol of~\cite{gla}.
In this phase, each debit transaction is appended with a set of \emph{dependencies}, i.e., credit transactions that are submitted to {\CODShort} along with the debit transaction, signed by the client (\cref{line:cod:add-deps}).
Intuitively, this is necessary in order to prevent Byzantine replicas from falsely detecting overspending simply by ignoring credit transactions.
Then, the client sends a {\mPrepare} message containing all the debit transactions (with signed sets of dependencies attached to them) and credit transactions (with commit certificates) it is aware of to all the replicas (\cref{line:cod:send-prepare}) and waits for their replies (\cref{line:cod:prepare:wait-replies}).
It also attaches the set of debit transactions that it started the Prepare phase with ($\vSubmitDebits$).
If any replica replies with any debit transactions the client is unaware of, the client repeats the request with the updated sets of debits and credits (\cref{line:cod:prepare:refine-sets}).

There are multiple ways in which this loop may terminate.
In a successful scenario, the client either manages to collect a quorum of signed replies with an identical set of debits (\cref{line:ccd:prepare:equal-sets}) or receives a notification that some other client already managed to prepare a set of transactions that includes all the transactions from $\vSubmitDebits$ (\cref{line:cod:prepare:if-fast-return}).
In these two cases, the client returns will move on to the Accept phase with a quorum of signatures as a certificate that it performed the Prepare phase correctly (\cref{line:prepare:fast-return,line:prepare:return-ok-quorum}).

If the total value of all debits that the client is aware of (including the ones from ongoing $\oSubmit$ operations) exceeds the total value of committed credits known to the client (\cref{line:prepare:balance-check}), this indicates a potential overspending attempt. 
As a result, the client will return $\tFAIL$ both from the Prepare phase (\cref{line:ccd:prepare:balance-fail}) and from the $\oSubmit$ operation (\cref{line:cod:submit:fail-if-prepare-failed}). 
Finally, if the client is notified (with a valid signature from another owner of the account) that this {\CODShort} instance is being closed, it also returns {\tFAIL} (\cref{line:prepare:if-closed}).

When a correct replica receives a {\mPrepare} message from a client, it first checks that the {\CODShort} instance is not yet closed (\cref{line:cod:replica:upon-receive-prepare:check-closed}) and that the transactions in the $\vSubmitDebits$ set are not yet prepared (\cref{line:cod:replica:upon-receive-prepare:check-already-prepared}).
If one of these conditions does not hold, the replica notifies the client with a proper certificate and stops processing the message.
Otherwise, the replica proceeds to check that the client's message is well-formed (\crefrange{line:cod:replica:upon-receive-prepare:check-validity:start}{line:cod:replica:upon-receive-prepare:check-validity:end}) and, if it is, the replica adds the received transactions to its local state (\cref{line:ccd:prepare:add-debits-replica,line:ccd:prepare:check-replica}).
Finally, the replica sends to the client all debit transactions (with signed dependencies) and all credit transactions (with commit certificates) it is aware of.
It also attaches a digital signature on the set of debits if the account's balance is non-negative after all the transactions the replica is aware of are applied (\crefrange{line:ccd:prepare:check-replica}{line:cod:replica:upon-receive-prepare:end}).
Intuitively, the signature indicates that the replica acknowledged these debits.
As described above, if a client collects a quorum of such signatures for an identical set of debits, it can move on to the Accept phase.

\myparagraph{Accept phase}

In the Accept phase (\crefrange{line:cod:accept:start}{line:cod:accept:end}), the client gathers signatures for the Merkle tree root of the set of debits it obtained during the Prepare phase.
Once it has collected signatures from a quorum of replicas, the client constructs certificates for the debit transactions with which it invoked the $\oSubmit$ operation and returns them along with all credits used to cover these transactions.
The client may return $\tFAIL$ in the Accept phase if another client has invoked the $\oClose$ operation.

The purpose of the Accept phase is, intuitively, to ensure that each transaction accepted by this {\CODShort} instance is stored in the $\vPreparedDebits$ set on at least a quorum of replicas.
This is necessary to guarantee that, in the $\oClose$ operation, the clients will be able to reliably identify which transactions could have been accepted (by looking at the $\vPreparedDebits$ sets reported by the replicas).

\myparagraph{The $\oClose$ operation}
The $\oClose$ operation is designed to deactivate the {\CODShort} object and to collect a snapshot of its state. 
During this operation, the client collects the states of a quorum of replicas and then asks the replicas to sign the accumulated joint state.
Accessing a quorum  guarantees that the client gathers all the debits that have been previously returned from the $\oSubmit$ operations.
As a result, confirmed transactions are never lost.

The interaction between the {\mAcceptRequest} and {\mClose} messages in the {\CODShort} implementation is similar to that of ``propose'' messages and ``prepare'' messages with a larger ballot number in Paxos~\cite{lamport2001paxos}.
Intuitively, it guarantees that, if there is a $\oSubmit$ concurrent with a $\oClose$, either the client executing $\oSubmit$ will ``see'' the $\oClose$ operation (i.e., will receive a {\mClosed} message) and will return $\tFAIL$ or the client executing the $\oClose$ operation will ``see'' the debits in a replica's $\vPreparedDebits$ set.

\begin{myalgorithm}
    \caption{{\CODLong} (Parameters and replica state)}
    \label{alg:crypto-la-state}

    \begin{myalgorithmic}

                \ProcessParameters
            \State $\vAcc$ -- account for which this {\CODShort} is used 
            \State $\vEp$ -- epoch number of this {\CODShort}
                        \State $\vInitDebits$ -- set of initial debits used by this {\CODShort}
            \State $\vInitCredits\WithCert$ -- set of initial credits used by this  {\CODShort}
            \State $\vRestrictedDebits$ -- set of debits that should not be accepted by this  {\CODShort}
            
        \EndProcessParameters
        
        \ReplicaState
                        \State $\vDebits\WithCert$ -- the set of debits \ppreplace{accepted}{acknowledged} by this replica, initially $\{ \Tuple{\vTx, \Null} \mid \vTx \in \vInitDebits \}$
            \State $\vCredits\WithCert$ --  the set of known credits, initially $\vInitCredits\WithCert$
            \State $\vPreparedDebits$ -- prepared debits received in $\mAcceptRequest$ messages, initially $\emptyset$
            \State $\vPreparedCert$ -- certificate for $\vPreparedDebits$, initially $\Null$
            \State $\vIsClosed$ -- current status of this {\CODShort}, initially $\False$
            \State $\vClosedCert$ -- proof of the fact that {\CODShort} was closed, initially $\Null$
        \EndReplicaState
      
    \end{myalgorithmic}
\end{myalgorithm}

\begin{myalgorithm}
    \caption{{\CODShort} verifying and helper functions}
    \label{alg:crypto-la-verify}

    \begin{myalgorithmic}
        \PublicFunction{\oVerifyCODCert}{$\vTx$, $\vCert$}
            \State $\Tuple{\vMTRoot, \vItemProof, \vTSMrkTree, \vAcc_{\vCert}, \vEp_{\vCert}} \oassign \vCert$
            \If{$\vAcc_{\vCert} \neq \vAcc \tor \vEp_{\vCert} \neq \vEp$} \Return{$\False$}\EndIf
            \State \Return {$\fVerifyItemProof(\vMTRoot, \vItemProof, \vTx) \tand \fVerifyThresholdSignature(\Message{\mAcceptAck, \vMTRoot}, \vTSMrkTree)$}
        \EndPublicFunction
        
        \algspace
        \PublicFunction{\oVerifyCloseStateCert}{$\vClosedState$, $\vCloseStateCert$} \Returns{\tBoolean}
            \State let $\Tuple{\vCredits\WithCert, \vSelectedDebits, \vCancelledDebits}$ be $\vClosedState$
            \If {$\vCredits\WithCert$ are not sufficient to cover all debits in $\vSelectedDebits$} \Return{$\False$} \EndIf
                                    \State \Return {$\fVerifyThresholdSignature(\Tuple{\mConfirmStateResp, \vSelectedDebits, \vCancelledDebits}, \vCloseStateCert)$}
        \EndPublicFunction

        \algspace
 
        \Function{\fExtractSafeTransactions}{$\vPendingDebits$, $\vMessages$} \Returns{$\tPair{\tSet{\tT}}{\tSet{\tT}}$} \label{line:cod:split-debits:start}
            \State \textbf{assert} $\vMessages$ is a set of tuples $\{\Message{\mCloseResp, \vCredits\WithCert_{i}, \vPreparedDebits_i, \vPreparedCert_i, \vSignature_i}\}_{i \in Q}$
            \State let $\vPreparedDebits \oassign \bigcup_{i \in Q} \vPreparedDebits_i$
            \State let $\vAllCredits \oassign \fTransactions(\vInitCredits\WithCert) \cup (\bigcup_{i \in Q} \fTransactions(\vCredits\WithCert_{i}))$
            \State \textbf{assert} $\fTotalValue(\vAllCredits) \ge \fTotalValue(\vInitDebits \cup \vPreparedDebits)$
            \LineComment{At least all initial debits and all prepared debits must be selected}
            \State $\vSelectedDebits \oassign \vInitDebits \cup \vPreparedDebits$
            \For {$\vTx \in \vPendingDebits \setminus \vRestrictedDebits$}
                \If {$\fTotalValue(\vSelectedDebits \cup \{\vTx\}) \le \fTotalValue(\vAllCredits)$}
                    \State $\vSelectedDebits \oassign \vSelectedDebits \cup \{\vTx\}$
                \EndIf
            \EndFor
            \State \Return $\Tuple{\vSelectedDebits, \vRestrictedDebits \cup (\vPendingDebits \setminus \vSelectedDebits)}$ \label{line:cod:split-debits:end}
        \EndFunction

    \end{myalgorithmic}
\end{myalgorithm}

\begin{myalgorithm}
    \caption{\CODShort ~(code for client $p$)}
    \label{alg:crypto-la-client}

        \begin{myalgorithmic}
  
        \Operation{\oSubmit}{$\vDebits, \vCredits\WithCert$} 
        \Returns{$\tOK(\tPair{\tSet{\tSignedTx}}{\tSet{\tSignedTx}})$ or $\tFAIL$}
            \State $\vPrepareResult = \fPrepare(\vDebits, \vCredits\WithCert)$
            \If {$\vPrepareResult$ $\tis$  {\tFAIL}} \Return {\tFAIL} \label{line:cod:submit:fail-if-prepare-failed} \EndIf
            \State let $\tOK(\Tuple{\vPreparedDebits, \vAllCredits\WithCert, \vPreparedCert})$ be $\vPrepareResult$
                                    \State \Return $\fAccept(\vDebits, \vPreparedDebits, \vAllCredits\WithCert, \vPreparedCert)$
        \EndOperation
        
        \algspace
        \Function{\fPrepare}{$\vDebits$, $\vCredits\WithCert$} 
                \Returns{$\tOK(\tTuple{\tSet{\tT}, \tSet{\tSignedTx}, \tCert})$ or $\tFAIL$}
            \State $\vSubmitDebits \oassign \vDebits$
            \State $\vDeps \oassign \{ \vTx.\vId \mid \vTx \in \fTransactions(\vCredits\WithCert) \}$
            \State $\vDebits\WithCert \oassign \{\Tuple{\vTx, \Tuple{\vDeps, \fSign(\Message{\mDeps, \vTx, \vDeps})}} \mid \vTx \in \vDebits \}$ \label{line:cod:add-deps}
                        \While {$\True$} \label{line:prepare:while-loop}
                \State \Send{\mPrepare, $\vDebits\WithCert$, $\vCredits\WithCert$, $\vSubmitDebits$}{all replicas} \label{line:cod:send-prepare}
                \BreakableLine[\WaitFor] quorum $Q$ of valid $\mPrepareResp$ replies $\Message{\mPrepareResp, \vDebits\WithCert_{i}, \vCredits\WithCert_{i}, \vSignature_i}$ \label{line:cod:prepare:wait-replies}
                    \RaggedBreak or $1$ valid $\mAlreadyPrepared$ reply or $1$ valid $\mClosed$ reply
                \EndBreakableLine
                \IfReceivedClosed \label{line:prepare:if-closed}
                                \If {received a valid $\Message{\mAlreadyPrepared, \vPreparedDebits, \vPreparedCert}$ reply} \label{line:cod:prepare:if-fast-return}
                                        \State \Return $\Tuple{\vPreparedDebits, \vPreparedCert}$  \Comment{Someone has already prepared a set with our transactions.}
                    \label{line:prepare:fast-return}
                \EndIf
                \State $\vAllDebits\WithCert \oassign \vDebits\WithCert \cup \left(\bigcup_{i \in Q} \vDebits\WithCert_{i}\right)$; $\vAllCredits\WithCert \oassign \vCredits\WithCert \cup \left(\bigcup_{i \in Q} \vCredits\WithCert_{i}\right)$
                                                \If {$\fBalance(\fTransactions(\vAllDebits\WithCert) \cup \fTransactions(\vAllCredits\WithCert), \vAcc) < 0$} \label{line:prepare:balance-check}
                    \State \Return $\tFAIL$   \label{line:ccd:prepare:balance-fail}  \Comment{Impossible to add all transactions without violating the invariant.}
                                \ElsIf {$\forall i \in Q: \fTransactions(\vDebits\WithCert_{i}) = \fTransactions(\vAllDebits\WithCert)~\tand \vSignature_i \neq \Null$} \label{line:ccd:prepare:equal-sets}
                                        \State $\vPreparedCert \oassign \fCreateThresholdSignature
                    (\Tuple{\mPrepareResp, \fTransactions(\vAllDebits\WithCert)}, \{ \vSignature_{i}\}_{i\in Q})$ \label{line:ccd:prepare:threshold-sign}
                    \State \Return $\tOK(\Tuple{\fTransactions(\vAllDebits\WithCert),\vAllCredits\WithCert, \vPreparedCert})$ \label{line:prepare:return-ok-quorum}
                \EndIf
                \State $\vDebits\WithCert \oassign \vAllDebits\WithCert$; $\vCredits\WithCert \oassign \vAllCredits\WithCert$ \label{line:cod:prepare:refine-sets}
            \EndWhile
        \EndFunction

        \algspace
        \Function{\fAccept}{$\vSubmittedDebits$, $\vPreparedDebits$, $\vAllCredits\WithCert$, $\vPreparedCert$} \label{line:cod:accept:start}
            \Returns{$\tOK(\tPair{\tSet{\tSignedTx}}{\tSet{\tSignedTx}})$ or $\tFAIL$}
    
            \State \Send{\mAcceptRequest, $\vPreparedDebits$, $\vAllCredits\WithCert$, $\vPreparedCert$}{all replicas}
            
            \State \WaitFor valid $\Message{\mAcceptAck, \vSignature_i}$ replies from a quorum $Q$ or 1 valid $\mClosed$ reply

            \IfReceivedClosed \label{line:accept-update-ballot}
            \State $\vMerkleTree = \fMerkleTree(\vPreparedDebits)$
            \State $\vTSMrkTree = \fCreateThresholdSignature(\Tuple{\mAcceptAck, \vMerkleTree.\vMTRoot}, \{\vSignature_i\}_{i \in Q})$
            
            \State $\vSubmittedDebits\WithCert  = \{ \Tuple{\vTx, \Tuple{\vMerkleTree.\vMTRoot, \fGetItemProof(\vMerkleTree, \vTx), \vTSMrkTree, \vAcc, \vEp}} \mid \vTx \in \vSubmittedDebits \}$ \label{line:ccd:confirm:form-result}
            
            \State \Return $\tOK(\vSubmittedDebits\WithCert, \vAllCredits\WithCert)$ \label{line:cod:accept:end}
        \EndFunction

        \algspace
        \Operation{\oClose}{$\vPendingDebits$} \Returns{$\tPair{\tCloseState}{\Sigma}$} \label{line:cod:close:start}
                                    \State \Send{\mClose, $\fSign(\Tuple{\mClose, \vEp})$}{all replicas}
            \State \WaitFor a quorum $Q$ of valid $\mCloseResp$ replies
            \State $\vMessages~\oassign $ replies $\Message{\mCloseResp, \vCredits\WithCert_{i}, \vPreparedDebits_i, \vPreparedCert_i, \vSignature_i}$ from quorum $Q$
            \State $\vCredits\WithCert \oassign \bigcup \vCredits\WithCert_{i}$
                                                                                    \State $\Tuple{\vSelectedDebits, \vCancelledDebits} \oassign \fExtractSafeTransactions(\vPendingDebits, \vMessages)$ \label{line:ccd:close:split}
                        \State \Send{\mConfirmState, $\vPendingDebits$,$\vMessages$}{all replicas}
            \State \WaitFor{a quorum $Q$ of valid $\Message{\mConfirmStateResp, \vSignature_i}$ replies}
            \State $\vCloseStateCert \oassign \fCreateThresholdSignature(\Message{\mConfirmStateResp\atremove{, \vEp}, \vSelectedDebits, \vCancelledDebits}, \{\vSignature_i\}_{i \in Q})$
            \State \Return{$\Tuple{\Tuple{\vCredits\WithCert, \vSelectedDebits, \vCancelledDebits}, \vCloseStateCert}$} \label{line:cod:close:end}
        \EndOperation
    \end{myalgorithmic}

\end{myalgorithm}

\algnewcommand{\CheckStatus}{
    \If {$\vIsClosed$} \Send{\mClosed, $\vClosedCert$}{$q$} and \Return \EndIf
}

\begin{myalgorithm}
    \caption{\CODShort ~(code for replica $r$)}
    \label{alg:crypto-la-replica}

    \begin{myalgorithmic}

        \UponReceive{\mPrepare, $\vReceivedDebits\WithCert$, $\vReceivedCredits\WithCert$, $\vSubmitDebits$}{client $q$} \label{line:cod:replica:upon-receive-prepare:start}
            \CheckStatus \label{line:cod:replica:upon-receive-prepare:check-closed}
            \If {$\vSubmitDebits \subseteq \vPreparedDebits$} \label{line:cod:replica:upon-receive-prepare:check-already-prepared}
                \State \Send{$\mAlreadyPrepared$, $\vPreparedDebits$, $\vPreparedCert$}{$q$}
                \State \Return
            \EndIf
           
            \For {all $\vTx \in \fTransactions(\vReceivedDebits\WithCert)$} \label{line:cod:replica:upon-receive-prepare:check-validity:start}
                \If {$\vTx.\vSender \neq \vAcc$} \Return \EndIf
                \If {$\vTx \in \vRestrictedDebits$} \Return \EndIf \label{line:ccd:prepare:no-restricted}
            \EndFor

            \For {all $\Tuple{\vTx, \vCreditCert} \in \vReceivedCredits\WithCert$}
                \If {$\vTx.\vRecipient \neq \vAcc \tor \tnot \fVerifyCommitCertificate(\vTx, \vCreditCert)$} \Return \EndIf
            \EndFor

            \If {$(\bigcup_{\Tuple{\vTx, \Tuple{\vDeps, \vCert}} \in \vReceivedDebits\WithCert} \vDeps) \not\subseteq \fTransactions(\vReceivedCredits\WithCert)$}
                \Return \label{line:cod:replica:upon-receive-prepare:check-validity:end}
            \EndIf 
            
            \State $\vCredits\WithCert \oassign \vCredits\WithCert \cup \vReceivedCredits\WithCert$
            \State $\vDebits\WithCert \oassign \vDebits\WithCert \cup \vReceivedDebits\WithCert$ \label{line:ccd:prepare:add-debits-replica}
            \If {$\fBalance(\fTransactions(\vDebits\WithCert \cup \vCredits\WithCert), \vAcc) \geq 0$} \label{line:ccd:prepare:check-replica}
                                            \State $\vSignature \oassign \fSign(\Tuple{\mPrepareResp, \fTransactions(\vDebits\WithCert)})$
            \Else ~$\vSignature \oassign \Null$
            \EndIf

            \State \Send{\mPrepareResp, $\vDebits\WithCert$, $\vCredits\WithCert$, $\vSignature$}{$q$} \label{line:cod:replica:upon-receive-prepare:end}
        \EndHandler

        \algspace
        \UponReceive{\mAcceptRequest, $\vReceivedDebits$, $\vReceivedCredits\WithCert$, $\vCert$}{client $q$}
            \CheckStatus
                        \If{$ \tnot \fVerifyThresholdSignature(\Tuple{\mPrepareResp, \vReceivedDebits}, \vCert)$} \Return \EndIf
            \For{all $\Tuple{\vTx, \vCert} \in \vReceivedCredits\WithCert$}
                \If{$ \tnot \fVerifyCommitCertificate(\vTx, \vCert)$} \Return \EndIf \label{line:ccd:accept:ignore-fake-credits}
            \EndFor
            \State $\vCredits\WithCert \oassign \vCredits\WithCert \cup \vReceivedCredits\WithCert$ \label{line:ccd:accept:add-credits}
            \If {$\vPreparedDebits \subset \vDebits$}
                \State $\vPreparedDebits \oassign \vDebits$; $\vPreparedCert \oassign \vCert$
            \EndIf

                        \State $\vSignature \oassign \fSign(\Tuple{\mAcceptAck, \fMerkleTree(\vDebits).\vMTRoot})$
            \State \Send{\mAcceptAck, $\vSignature$}{$q$}
        \EndHandler

        \algspace
        \UponReceive{\mClose, $\vSignature$}{client $q$}
            \If{$\tnot \fVerifySignature(\Tuple{\mClose, \vEp}, \vSignature, q)$} \Return \EndIf
            \State $\vIsClosed \oassign \True$; $\vClosedCert \oassign \vSignature$
            \State $\vSignature \oassign \fSign(\Tuple{\mCloseResp\atremove{, \vEp}, \vPreparedDebits})$
            \State \Send{\mCloseResp, $\vCredits\WithCert$, $\vPreparedDebits, \vPreparedCert$, $\vSignature$}{$q$}
        \EndHandler

        \algspace
        \UponReceive{\mConfirmState, $\vPendingDebits$, $\vMessages$}{client $q$}
            \If {$\vMessages$ contains invalid messages} \Return{} \EndIf
                        \State $\Tuple{\vSelectedDebits, \vCancelledDebits} \oassign \fExtractSafeTransactions(\vPendingDebits, \vMessages)$
            \State $\vSignature \oassign \fSign(\Message{\mConfirmStateResp\atremove{, \vEp}, \vSelectedDebits, \vCancelledDebits})$
    
            \State \Send{$\mConfirmStateResp$, $\vSignature$}{client $q$}
            \label{line:ccd:send:initstate}
        \EndHandler

    \end{myalgorithmic}
\end{myalgorithm}

\section{{\AppendOnlyStorage}}\label{app:storage}

First, we give a formal definition for {\AppendOnlyStorage} in~\ref{app:storage-formal} and then proceed with its detailed implementation in~\ref{app:storage-impl}. 

\subsection{Formal definition of {\AppendOnlyStorage}}\label{app:storage-formal}

The abstraction is parameterized with a set of tuples $\{\Tuple{k_1, \vInitVs_{k_1}, \fVerifyValue_{k_1}}, \ldots, \Tuple{k_n, \vInitVs_{k_n}, \fVerifyValue_{k_n}}\}$. Each tuple consists of a key $k_i$, initial set of values $\vInitVs_{k_i}$ for this key, and a verifying function for this key $\fVerifyValue_{k_i}$, which takes a value $v$ and a certificate $\vVCert$ for this value and returns $\True$ if $\Tuple{v, \vVCert}$ is valid input for key $k_i$, and $\False$ otherwise.\footnote{Sometimes, we define the validity based only on the values themselves and use $\Null$ for the certificates.}
Also, any initial value for a given key is valid: $\forall v \in \vInitVs_{k_i}: \fVerifyValue_{k_i}(k_i, v, \Null) = \True$.
The abstraction exports two operations: $\oAppendKey(k, v, \vVCert)$ and  $\oReadKey(k)$. It also provides one boolean function: $\fKVSVerifyStoredCert(k, v, \vKVSCert)$.%

The operation $\oAppendKey(k, v, \vVCert)$ accepts a key $k$ and a value $v$ together with its certificate $\vVCert$ and adds $v$ to the set of stored values for $k$ in {\AppendOnlyStorage}, but only if $\fVerifyValue_k(v, \vVCert) = \True$.
As a result, the $\oAppendKey(k, v, \vVCert)$ operation outputs a  certificate $\vKVSCert$, which is an evidence of the fact that value $v$ is stored in the set corresponding to the key $k$ in {\AppendOnlyStorage}.
The $\oReadKey(k)$ operation returns $\vVsKVSCert$, where  $\vVsKVSCert$ is a set of pairs $\Tuple{v, \vKVSCert}$, such that: $v$ is a valid value for a key $k$ (i.e., there exists a certificate $\vVCert$, such that $\fVerifyValue_k(v, \vVCert) = \True$), and $\vKVSCert$ is a certificate that proves that $v$\atremove{ is} belongs to the set of values for the key $k$ in the {\AppendOnlyStorage}.

Let us formally define the properties of {\AppendOnlyStorage}:

\begin{properties}
    \propertyitem{\pKVSConsistency} If there exists $\vKVSCert$ such that $\fKVSVerifyStoredCert(k, v, \vKVSCert)= \True$ at the moment when $\oReadKey(k)$ was invoked by a correct client, then the output of this operation will contain $v$ (paired with a certificate);
        
    \propertyitem{\pKVSInputValidity} If there exists a certificate $\vKVSCert$ such that $\fKVSVerifyStoredCert(k, v, \vKVSCert) = \True$, then there exists $\vVCert$, such that $\fVerifyValue_k(v, \vVCert)$;

     \propertyitem{\pKVSOutputValidity} If a correct client returns $\vKVSCert$ from $\oAppendKey(k, v, \vVCert)$, 
        then $\fKVSVerifyStoredCert(k, v, \vKVSCert) = \True$.
        Moreover, if a correct client returns $\vVsKVSCert$ from $\oReadKey(k)$, then $\forall \Tuple{v, \vKVSCert} \in \vVsKVSCert: \fKVSVerifyStoredCert(k, v, \vKVSCert) = \True$;

    \propertyitem{\pKVSLiveness} All operations eventually terminate.
\end{properties}

\subsection{Implementation of {\AppendOnlyStorage}}\label{app:storage-impl}

The pseudocode for the {\AppendOnlyStorage} can be found in \Cref{alg:keyval-storage}.
In operation $\oAppendKey(k, v, \vVCert)$, the client calls function $\fWriteValuesToKey$ with a key $k$ and a singleton set $\{\Tuple{v, \vVCert}\}$ as parameters.
In $\fWriteValuesToKey$, given a key $k$ and a set of values with certificates $\vVs\WithCert$, the client first sends $k$ and $\vVs\WithCert$ to all replicas and waits for a quorum of valid replies.
The client then returns the submitted values, provided with their aggregated Merkle Tree signatures, and extracted certificates for all $v$ in $\vVs\WithCert$ ($\Tuple{v, *} \in \vVs\WithCert$).

In operation $\oAppendKey(k, v, \vVCert)$, the call to $\fWriteValuesToKey$ returns a set consisting of only one pair $\Tuple{v, \vKVSCert}$ (due to the input being a singleton set). Finally, $\oAppendKey$ returns $\vKVSCert$.
In the $\oReadKey(k)$ operation, the client first requests the values stored by a quorum of replicas for this key.
The aggregated set of values $\vVs\WithCert$ is then passed to the $\fWriteValuesToKey$ function to make sure that any value $v$ in $\vVs\WithCert$ is written to a quorum of processes. This guarantees that any value read from the {\KeyValStorage} will be read from it again later. Finally, $\vVs\WithCert$ is returned by the operation.

In {\Name}, we use two types of the {\KeyValStorage}. The first is {\GlobalStorage} which allows different accounts to interact with each other (i.e., receive incoming transactions). The other one -- {\AccountStorage} is used per account, i.e., only clients that share an account can communicate with it and every account has an {\AccountStorage} associated with it.
As there is one {\GlobalStorage}\yaremove{ is one} per system, it makes sense to implement it on the same set of replicas as all of the other parts of the algorithm. At the same time, {\AccountStorage} serves only one account, and, in fact, can be implemented on a different, \textit{local} set of replicas for each account.

\begin{myalgorithm}
    \caption{\KeyValStorage}
    \label{alg:keyval-storage}
        
    \begin{myalgorithmic}
 
    \LineCommentx{Code for client $p$}
        \Operation{\oReadKey}{$k$} \Returns{$\tSet{\tPair{V}{\Sigma}}$}
            \State \Send{\mReadKey, $k$}{all replicas}
            \State \WaitFor valid $\Message{\mReadKeyResp, \vVs\WithCert_{i}}$ replies from a quorum $Q$ \label{line:gs:read:wait}
            \State $\vVs\WithCert \oassign \bigcup_i \vVs\WithCert_{i}$
            \State \Return {$\fWriteValuesToKey(\vVs\WithCert)$}
        \EndOperation
        
        \algspace
        \Operation{\oAppendKey}{$k$, $v$, $\vVCert$} \Returns{$\Sigma$} \label{line:aos:append-key-start}
            \State $\{\Tuple{v, \vKVSCert}\} \oassign \fWriteValuesToKey(k, \{\Tuple{v, \vVCert}\})$
            \State \Return $\vKVSCert$
        \EndOperation
        
        \algspace
        \Operation{\fWriteValuesToKey}{$k$, $\vVs\WithCert$}  \Returns{$\tSet{\tPair{V}{\Sigma}}$}
            \State \Send{\mAppendKey, $k$, $\vVs\WithCert$}{all replicas}
            \State \WaitFor{valid $\Message{\mAppendKeyResp, \vSignature_i}$ replies from a quorum $Q$} \label{line:gs:write-fun:wait}
            \State  $\vVs  \oassign \{v \mid \Tuple{v, \vVCert} \in \vVs\WithCert\}$
            \State $\vMerkleTree  \oassign \fMerkleTree(\vVs)$
            \State $\vTSMrkTree  \oassign \fCreateThresholdSignature(\Tuple{\mAppendKeyResp, k, \vMerkleTree.\vMTRoot}, \{\vSignature_i\}_{i \in Q})$
            \State \Return $\{ \Tuple{v, \Tuple{\vMerkleTree.\vMTRoot, \fGetItemProof(\vMerkleTree, v), \vTSMrkTree}} \mid v \in \vVs \}$
        \EndOperation \label{line:aos:write-values-end}
        
        \algspace
        \PublicFunction{\fKVSVerifyStoredCert}{$k$, $v$,
        $\vKVSCert$} \Returns{$\tBoolean$}
            \State $\Tuple{\vMTRoot, \vItemProof, \vTSMrkTree} \oassign \vKVSCert$
            \State \Return {$\fVerifyItemProof(\vMTRoot, \vItemProof, v) \tand \fVerifyThresholdSignature(\Message{\mAppendKeyResp, k, \vMTRoot}, \vTSMrkTree)$}
        \EndPublicFunction
        
    \algspace 
    \LineCommentx{Code for replica $r$}
        \ProcessState 
            \State $\vLog$ -- mapping from a key to a set of values for this key, initially $\forall k: \vLog[k] = \emptyset$
        \EndProcessState
    
        \UponReceive{\mReadKey, $k$}{client $p$}
            \State \Send{\mReadKeyResp, $\vLog[k]$}{$p$}
        \EndHandler

        \algspace
        \UponReceive{\mAppendKey, $k$, $\vVs\WithCert$}{client $p$}
             \For {all $\Tuple{v,\vVCert} \in \vVs\WithCert$}
                \If {$\tnot \fVerifyValue_k(v, \vVCert)$} \Return \EndIf
            \EndFor
            \State $\vLog[k] \oassign \vLog[k] \cup \vVs\WithCert$
            \State $\vVs  \oassign \{v \mid \Tuple{v, \vVCert} \in \vVs\WithCert\}$
            \State $\vSignature  \oassign \fSign(\Message{\mAppendKeyResp, k, \fMerkleTree(\vVs)\atremove{)}.\vMTRoot})$
            \State \Send{$\mAppendKeyResp, \vSignature$}{$p$}
        \EndHandler
    \end{myalgorithmic}
\end{myalgorithm}

\section{Proofs of correctness} \label{app:proof}

In~\ref{app:proof-cryptoconcurrency} we prove the correctness of the {\Name} protocol, assuming the correctness of the underlying building blocks. 
Then, in \ref{subsection:proof:cryptocd} and \ref{subsection:proof:keyval}, we show that the implementations of {\CODLong} and {\KeyValStorage} are correct.

\subsection{Proof of Correctness: {\Name}}~\label{app:proof-cryptoconcurrency}

In this subsection, we prove that {\Name} satisfies the six {\AssetTransfer} properties {\pTransferLiveness}, {\pTransferValidity}, {\pTransferSafety}, {\pTransferConsistency}, {\pAccountTransactions} and {\pTransferConcurrency}, as defined in \Cref{sec:problem-statement}.

\myparagraph{\pTransferLiveness}

Let us start with the proof of the {\pTransferLiveness} property.
First of all, it is important to note that all invocations of operations of {\CODShort}, {\KeyValStorage} ({\AccountStorage} and {\GlobalStorage}), and {\Consensus} will eventually terminate due to the liveness properties of these objects (namely, {\pCODLiveness}, {\pKVSLiveness} and {\pConsLiveness}).
Moreover, in the implementation of {\Name} presented in \Cref{alg:crypto-main-client}, there is just one ``{\WaitFor}'' statement (\cref{line:recovery:wait-for-commit-state-resp}) and one potentially infinite loop (\crefrange{line:transfer-loop-begin}{line:transfer-loop-end}).
Hence, we need to prove their eventual termination.

\begin{lemma} \label{lem:recovery:finite-wait}
    \Cref{line:recovery:wait-for-commit-state-resp}
    invoked by a correct client always eventually terminates.
\end{lemma}
\begin{proof}
    Let us say that two $\mCommitCODInitialState$ messages are \emph{conflicting} if they contain the same epoch, but different $\vNextCODState$ fields.
    By inspecting \crefrange{line:receive-commit-crypo-la-initial-state-begin}{line:receive-commit-crypo-la-initial-state-end}, it is easy to verify that, unless the owners of some account issue conflicting $\mCommitCODInitialState$ messages, the correct replicas will reply to each $\mCommitCODInitialState$ message that contains a $\vCODState$ with a valid certificate.
    
    Thanks to the {\pConsConsistency} property of consensus, correct owners of the same account will never send conflicting $\mCommitCODInitialState$ messages.
    Moreover, according to our assumptions, correct clients never share their account with Byzantine clients (see \Cref{subsec:accounts}).

    Hence, a correct client that reached \cref{line:recovery:wait-for-commit-state-resp} will always eventually collect a quorum of replies and will move on to the next line.
\end{proof}

\newcommand{\vCurTx}{\vTx^*}
\begin{lemma} \label{lem:main:finite-iterations}
    A correct client $p$ executing $\oTransfer(\vCurTx)$ enters the loop of \crefrange{line:transfer-loop-begin}{line:transfer-loop-end} \yaremove{only} a finite number of times.
\end{lemma}
\begin{proof}
    \newcommand{\vEpMax}{\vEp_{\mathit{max}}}

    Let $\vAcc$ be the account of client $p$.

    Let us consider the moment $t$ when $p$ returns from the invocation on \cref{line:main:put-tx-to-account-storage}.
    Let us consider all clients that are executing the $\oTransfer$ operation on $\vAcc$ at time $t$.
    Let $\vEpMax$ be the maximum of their epoch numbers.
    
    By the {\pKVSConsistency} property of {\AccountStorage}, whenever any owner of $\vAcc$ enters epoch $\vEpMax+1$, it will have $\vCurTx$ in its variable $\vPendingDebits$ on \cref{line:main:compute-pending-debits}.
    This implies that, whenever any process invokes $\fRecovery(\vEpMax+1, \vPendingDebits)$ on \cref{line:main:call-recovery}, $\vCurTx \in \vPendingDebits$.
    By the {\pCODCloseSafety} property of {\CODShort}, $\vCurTx$ will belong to the $\vCODState$ (either to $\vCODState.\vSelectedDebits$ or $\vCODState.\vCancelledDebits$) state received by any owner of $\vAcc$ invoking $\oCOD[\vAcc][\vEpMax+1].\oClose(\vPendingDebits)$ on \cref{line:recovery:invoke-close}. 
    Finally, by the {\pConsValidity} property of {\Consensus}, $\vCurTx$ will also belong to $\vNextCODState$ on \cref{line:recovery:consensus-call}.
    
    Since with each iteration of the loop $p$ increments its own epoch number, it will either eventually exit the loop or reach \cref{line:main:check-recovery-results} with $\vEp = \vEpMax+1$
    and, as we just established, will find $\vCurTx$ in either $\vCODState.\vSelectedDebits$ or $\vCODState.\vCancelledDebits$.
    In any case, the client will terminate $\oTransfer(\vCurTx)$ with either $\tOK(\vCommitCert)$ on \cref{line:main:return-ok-after-recovery} or $\tFAIL$ on \cref{line:main:return-fail-after-recovery}.
    
    \undef{\vEpMax}
\end{proof}
\undef{\vCurTx}

\begin{theorem} \label{theorem:asset-transfer:liveness}
    {\Name} satisfies the {\pTransferLiveness} property of {\AssetTransfer}.
\end{theorem}
\begin{proof}
    Liveness of $\oTransfer$ operation follows directly from \Cref{lem:recovery:finite-wait,lem:main:finite-iterations} and the liveness properties of the underlying building blocks.
    Liveness of $\oGetAccountTransactions$ operation follows from the liveness property of {\AppendOnlyStorage} ({\pKVSLiveness}).
\end{proof}

\myparagraph{\pTransferValidity}

Now we proceed  by proving the {\pTransferValidity} property. 

\begin{theorem} \label{theorem:asset-transfer:validity}
    {\Name} satisfies the {\pTransferValidity} property of {\AssetTransfer}.
\end{theorem}
\begin{proof}
    This theorem follows from the implementation of the algorithm and the {\pKVSOutputValidity} property of the {\KeyValStorage}. 
    If a correct client returns $\tOK(\vCommitCert)$ from the $\oTransfer(\vTx)$ operation, then it successfully returned  $\vCommitCert$ from  $\oGlobalStorage.\oAppendKey('\vTxs', \vTx, \vCert)$ (at either line~\ref{line:main:write-transaction-after-crypto-la} or line~\ref{line:main:write-transaction-after-recovery}).
    As the implementation of $\fVerifyCommitCertificate(\vTx, \vCommitCert)$ is essentially $\oGlobalStorage.\fKVSVerifyStoredCert('txs', \vTx, \vCommitCert)$, by the {\pKVSOutputValidity} property $\fVerifyCommitCertificate(\vTx, \vCommitCert) = \True$.
\end{proof}

\myparagraph{\pTransferSafety}

The {\pTransferSafety} is a type of property that is essential for any asset transfer system. It tells us that no account can overspend. We now show that {\Name} satisfies this property.

For the proof, let us consider an account $\vAcc$.
We say that a debit transaction $\vTx$ on $\vAcc$ is \emph{associated with an epoch $e$}, iff $e$ is the minimum epoch number such that:
\begin{itemize}
    \item Either there exists a certificate $\vCryptoCDCert$ such that \\ $\oCOD[\vAcc][e].\oVerifyCODCert(\vTx, \vCryptoCDCert) = \True$;
    \item Or there exists $\vRecoveryCert = \Tuple{*, *, *, e}$ such that \\ $\fVerifyRecoveryCert(\vTx, \vRecoveryCert) = \True$.
\end{itemize}

\begin{lemma}\label{lem:main:safety:init-debits}
For any epoch number $e$, $\oCOD[\vAcc][e].\vInitDebits$ includes all committed debit transactions associated with any epoch $e' < e$.
\end{lemma}
\begin{proof}
We prove this lemma by induction. 
The base case of the induction ($e = 1$) is trivially satisfied as there are no debit transactions associated with epoch numbers less than $1$ (note that genesis transactions are credits).

Now, assuming that the statement of this lemma holds up to the epoch number $e$, we prove that it also holds for the epoch number $e + 1$.
The initial state of the {\CODShort} object for epoch $e + 1$ (in particular, $\oCOD[\vAcc][e + 1].\vInitDebits$) is formed from the output value $\Tuple{\vNextCODState, *}$ of the $\oConsensus[\vAcc][e + 1].\oConsPropose$ operation.
Note that $\vNextCODState$ is also an input of $\oConsensus[\vAcc][e + 1]$ and it must satisfy $\pCODCloseSafety$ property to be accepted by a quorum of replicas (i.e., pass the check at line~\ref{line:verify-close-state-cert}).
Particularly, any transaction accepted by $\oCOD[\vAcc][e]$ is in $\vNextCODState.\vSelectedDebits$.
Furthermore, by the implementation of the $\fRecovery$ function, for any transaction $\vTx$, such that $\fVerifyRecoveryCert(\vTx, \vRecoveryCert) = \True$ and $\vRecoveryCert = \Tuple{*, *, *, e}$, the following holds: $\vTx \in \vNextCODState.\vSelectedDebits$.
This means that any transaction associated with epoch $e$ is in $\oCOD[\vAcc][e + 1].\vInitDebits$.
The fact that any transaction associated with epoch $e' < e$ is in $\oCOD[\vAcc][e + 1].\vInitDebits$ follows from the part of the {\pCODCloseSafety} property of the {\CODShort} abstraction saying that $\oCOD[\vAcc][e].\vInitDebits \subseteq \vNextCODState.\vSelectedDebits$.
By induction we know that any committed transaction associated with epoch $e' < e$ is in $\oCOD[\vAcc][e].\vInitDebits$.
Thus, if the statement of this lemma holds up to an epoch number $e$, then it also holds up to an epoch number $e + 1$, which concludes the induction.

\end{proof}

We say that account $\vAcc$ is in epoch $e$ at a given moment in time iff, at this moment, there exists a correct replica that initialized $\oCOD[\vAcc][e]$ object at line~\ref{line:main:initialize-next-cryptocd}, but no correct replica initialized $\oCOD[\vAcc][e + 1]$ yet.

\begin{theorem} \label{theorem:asset-transfer:safety}
    {\Name} satisfies the {\pTransferSafety} property of {\AssetTransfer}.
\end{theorem}
\begin{proof}
We prove this theorem by contradiction. 
Let us assume that {\Name} does not satisfy {\pTransferSafety} property, i.e., there exists an account $\vAcc$, such that\atadd{,} at some moment of time $t$\atadd{,} $\fBalance(C(t), \vAcc) < 0$.
Let us consider the first moment of time $t_0$ when it happens and an epoch $e$ account $\vAcc$ is in at time $t_0$.
We know that \yaadd{a} transaction can be committed if it obtains certificate via \yaadd{a} {\CODShort} object or via \yaadd{the} $\fRecovery$ procedure.
From the {\pCODProposeSafety} property of {\CODShort} object, we know that the for any time $t$ total value of $\oCOD[\vAcc][e].\vInitDebits$ and debits accepted by a {\CODShort} object by time $t$ for an account $\vAcc$ does not exceed total value of committed credits on account $\vAcc$ by time $t$.
By Lemma~\ref{lem:main:safety:init-debits}, we know that $\oCOD[\vAcc][e].\vInitDebits$ includes all committed debit transactions associated with any epoch $e' < e$.
Also, let us consider a set of committed transactions $\vSelectedDebits$ that are associated with an epoch $e' \le e$ and such that  $\forall \vTx \in \vDebits: \fVerifyRecoveryCert(\vTx, e, \vCert) = \True$. 
By the implementation, there exists a quorum of processes that signed a message  $\Message{\mCommitCODInitialStateResp, \Tuple{\vSelectedDebits, \vCredits\WithCert \vCancelledDebits}}$.
Then, $\Tuple{\vSelectedDebits, \vCredits\WithCert, \vCancelledDebits}$ should satisfy $\pCODCloseSafety$ property, in particular $\fTotalValue(\vSelectedDebits) \ge \fTotalValue(\fTransactions(\vCredits\WithCert))$. 
Also, note that $\vCredits\WithCert$ is a set of committed credits on account $\vAcc$ and $\fTransactions(\vCredits\WithCert) \subseteq \fCredits(C(t_0), \vAcc)$. 
In addition, from Lemma~\ref{lem:main:safety:init-debits} and the {\pCODCloseSafety} property of {\CODShort}, we know that $\vSelectedDebits$ includes all transactions that have been accepted by $\oCOD[\vAcc][e]$ and all committed transactions associated with an epoch $e' < e$.
This implies that $\fBalance(C(t_0), \vAcc) \ge 0$, which contradicts our assumption. Consequently, for all $t$ and for all $\vAcc$ $\fBalance(C(t), \vAcc) \ge 0$.

\end{proof}

\myparagraph{\pTransferConsistency}

Next, we prove that {\Name} satisfies the {\pTransferConsistency} property.

Let us briefly outline the proof structure.
Consider any execution $\Execution$.
We need to show that there exists a legal permutation of transactions in $\tT(\Execution,\vAcc)$ that is consistent with $\prec_{\Execution,\vAcc}$ for a correct account $\vAcc$.
First, we provide some formalism that we will use during the proof.
Then, we construct a permutation of transactions in $\tT(\Execution,\vAcc)$.
Using given definitions, by induction on the epoch number, we show that the constructed permutation is legal and consistent with $\prec_{\Execution,\vAcc}$.

Given a correct account $\vAcc$, we say that $n$ is the \emph{final epoch for $\vAcc$ in $\Execution$} iff this is the largest number such that at least one correct replica initialized $\oCOD[\vAcc][n]$ at line~\ref{line:main:initialize-next-cryptocd}.
Any epoch with a smaller number is said to be \emph{non-final}.

For the rest of this proof section, we consider a correct account $\vAcc$.

We say that transaction $\vTx$ \emph{belongs} to epoch $n$ iff it belongs to one of the following three groups:
\begin{itemize}
    \item Group $G_1(n)$:
        \begin{itemize}
            \item debit transactions accepted by $\oCOD[\vAcc][n]$, i.e., every $\vTx$, such that there exists $\oCOD[\vAcc][n].\oSubmit(\ldots)$ that returned $\tOK(\vAcceptedDebits\WithCert, *)$, such that $\Tuple{\vTx, *} \in \vAcceptedDebits\WithCert$, excluding the transactions from $\oCOD[\vAcc][n].\vInitDebits$;

            \item credit transactions returned from $\oCOD[\vAcc][n]$,  i.e., any $\vTx$ such that there exists $\oCOD[\vAcc][n].\oSubmit(\ldots)$ that returned $\tOK(*, \vCredits\WithCert)$, such that $\Tuple{\vTx, *} \in \vCredits\WithCert$, excluding the transactions from $\oCOD[\vAcc][n].\vInitCredits\WithCert$;
        \end{itemize}
    \item Group $G_2(n)$:
        \begin{itemize}
            \item debit and credit transactions selected by $\oConsensus[\vAcc][n + 1]$, i.e., any transaction $\vTx$, such that $\oConsensus[\vAcc][n + 1].\oConsPropose(\ldots)$ returns $\Tuple{\Tuple{\vAllCredits\WithCert, \vSelectedDebits, \vCancelledDebits}, *}$ and $\vTx \in \vSelectedDebits$ or $\Tuple{\vTx, *} \in \vAllCredits\WithCert$, excluding the transactions from $G_1(n)$, $\oCOD[\vAcc][n].\vInitDebits$ and $\oCOD[\vAcc][n].\vInitCredits\WithCert$;
        \end{itemize}
    \item Group $G_3(n)$:
        \begin{itemize}
            \item debit transactions canceled by $\oConsensus[\vAcc][n + 1]$, i.e., any debit transaction $\vTx$,  such that $\oConsensus[\vAcc][n + 1].\oConsPropose(\ldots)$ returns $\Tuple{\Tuple{ \vAllCredits\WithCert, \vSelectedDebits, \vCancelledDebits }, * }$ and  $\vTx 
            \in \vCancelledDebits$, excluding $\oCOD[\vAcc][n].\vRestrictedDebits$.
        \end{itemize}
\end{itemize}
Let $\Epoch(n)$ denote the set of transactions that belong to epoch $n$ (i.e., $\Epoch(n) = G_1(n) \cup G_2(n) \cup G_3(n)$).
Each transaction \textit{belongs} to at most one epoch, i.e., $\forall n_1 \neq n_2: \Epoch(n_1) \cap \Epoch(n_2) = \emptyset$.
Also, for any committed transaction $\vTx$ with $\vAcc \in \{ \vTx.\vSender, \vTx.\vRecipient\}$ $\exists n \ge 0$, for which $\vTx \in \Epoch(n)$.

For completeness, we also define $\Epoch(0)$ as a set that consists of only one special group $G_1(0)$, which contains only genesis transaction $\vTxInitForAcc$.

\begin{lemma} \label{lem:cod-init}
    Transaction $\vTx$ belongs to one of the sets $\oCOD[\vAcc][n].\vInitDebits$, $\fTransactions(\oCOD[\vAcc][n].\vInitCredits\WithCert)$, or $\oCOD[\vAcc][n].\vRestrictedDebits$
    iff $\vTx \in \Epoch(n')$ for some $n' < n$.
\end{lemma}
\begin{proof}
    The implication from left to right follows by induction from the definition of groups $G_2$ and $G_3$ and the implementation of the Recovery procedure.

    Similarly, the other direction follows by induction from the {\pCODCloseSafety} property and the implementation of the Recovery procedure, in a way analogous to \Cref{lem:main:safety:init-debits}.
\end{proof}

\begin{lemma}
    Each transaction $\vTx$ on account $\vAcc$ belongs to exactly one of the groups, i.e.,
    $\forall \vTx \in \tT(\Execution, \vAcc): \exists$ unique pair $(n, i)$ such that $\vTx \in G_i(n)$.
\end{lemma}
\begin{proof}
    Follows from \Cref{lem:cod-init} and the definition of groups.
\end{proof}

Let us now define a total order $<$ on the transactions in $\tT(\Execution, \vAcc)$ as follows:
\begin{enumerate}[label={(\roman*)}]
    \item First, we order the transactions by their epoch numbers, i.e.: $\forall k, m \text{ s.t. } k \neq m: \forall \vTx \in \Epoch(k), \vTx' \in \Epoch(m)$: $\vTx < \vTx'$ iff $k < m$;
    
    \item Within an epoch, by their group numbers: $\forall n,i,j \text{ s.t. } i \neq j: \forall \vTx \in G_i(n), \vTx' \in G_j(n)$: $\vTx < \vTx'$ iff $i < j$;

    \item Within each group, by $\fEnd_{\vAcc}$, giving the priority to the credit transactions: $\forall n,i: \forall \vTx, \vTx' \in G_i(n): \vTx < \vTx'$ iff $\fEnd_{\vAcc}(\vTx) < \fEnd_{\vAcc}(\vTx')$ or $\fEnd_{\vAcc}(\vTx) = \fEnd_{\vAcc}(\vTx')$, $\vTx$ is a credit transaction and $\vTx'$ is a debit transaction;

    \end{enumerate}
Let $H$ be the permutation of $\tT(\Execution, \vAcc)$ implied by the total order ``$<$''.
It is convenient to think of $H$ as a sequence of epoch sets, i.e., $H = (\Epoch(0), \Epoch(1), \ldots, \Epoch(n), \ldots)$ or a sequence of groups,  i.e., $H = (G_1(0), \ldots, G_1(n), G_2(n), G_3(n), \ldots)$.

Now, we need to prove that $H$ is both legal and consistent with real-time partial order $\prec_{\Execution, \vAcc}$. 
We are doing this by induction on the length of the permutation $H$.
We start with the base of the induction.

\begin{lemma}\label{lem:consistency:base}
$H$ up to $\Epoch(0)$ is consistent with $\prec_{\Execution, \vAcc}$ and legal.
\end{lemma}
\begin{proof}
This follows from the fact that $\Epoch(0)$ contains only genesis transaction $\vTxInitForAcc$ that deposits initial balance to the account, which by definition is non-negative.
\end{proof}

Now, assuming that $H$ is legal and consistent with $\prec_{\Execution, \vAcc}$ up to $\Epoch(k)$, let us prove that it is legal and consistent\atremove{ consistent} with $\prec_{\Execution, \vAcc}$ up to $\Epoch(k + 1)$.

First, we show that $H$ is consistent with $\prec_{\Execution, \vAcc}$.

\begin{lemma}\label{lem:consistency:consistent:group_all}
For any $i \in \{1,2,3\}$, for any $\vTx, \vTx' \in G_i(k+1)$ if $\vTx \prec_{\Execution,\vAcc} \vTx'$ then $\vTx < \vTx'$. 
\end{lemma}
\begin{proof}
Given $\fEnd_{\vAcc}(\vTx) < \fStart_{\vAcc}(\vTx')$, we need to show that $\fEnd_{\vAcc}(\vTx) < \fEnd_{\vAcc}(\vTx')$. This is obvious as $\fEnd_{\vAcc}(\vTx) < \fStart_{\vAcc}(\vTx') \le \fEnd_{\vAcc}(\vTx')$.
\end{proof}

\begin{lemma}\label{lem:consistency:consistent:group_1_2_3}
For any $\vTx \in G_1(k + 1)$ and  $\vTx' \in G_2(k + 1) \cup G_3(k + 1)$: $\vTx' \not \prec_{\Execution,\vAcc} \vTx$.
\end{lemma}

\begin{proof}
We prove this lemma by contradiction.
Let us assume that $\vTx'$ precedes $\vTx$. Consider two scenarios:
\begin{itemize}
    \item $\vTx$ is a debit transaction. Then, $\vTx$ could not be accepted by $\oCOD[\vAcc][k + 1]$ as it was closed before $\vTx'$ was submitted to the $\oConsensus[\vAcc][k + 2]$ according to the implementation. Contradiction.
    \item $\vTx$ is a credit transaction. According to {\pCODCloseSafety} and algorithm implementation, as $\vTx$ returned from $\oCOD[\vAcc][k + 1]$, then it is both consensus input and output. This implies that it was committed by the time $\vTx'$ was submitted to the $\oConsensus[\vAcc][k + 2]$. Contradiction.
\end{itemize}
We came to a contradiction in both cases, thus for any $\vTx \in G_1(k + 1)$ and  $\vTx' \in G_2(k + 1) \cup G_3(k + 1)$ $\vTx' \not \prec_{\Execution,\vAcc} \vTx$.
\end{proof}

\begin{lemma}\label{lem:consistency:consistent:group_2_3}
For any $\vTx \in G_2(k + 1)$ and  $\vTx' \in G_3(k + 1)$: $\vTx' \not \prec_{\Execution,\vAcc} \vTx$.
\end{lemma}
\begin{proof}
It follows from the fact that any pair of transactions $\vTx \in  G_2(k + 1)$ and $\vTx' \in G_3(k + 1)$ should have been submitted as a part of $\oConsensus[\vAcc][k + 2]$ input and both are part of its output, which implies that there should exist a time $t$ when both $\vTx$ and $\vTx'$ are active. 
\end{proof}

\begin{lemma}\label{lem:consistency:consistent:epochs}
For any transaction $\vTx \in \bigcup_{i=0}^k \Epoch(i)$, and $\vTx' \in \Epoch(k + 1)$: $\vTx' \not \prec_{\Execution,\vAcc} \vTx$.
\end{lemma}
\begin{proof}
Let $\fEpochStart(n)$ be the time when a process receives a value from $\oConsensus[\vAcc][n]$ for the first time. Note that $\fEpochStart(n) \le \fEpochStart(n + 1)$.

By {\pCODCloseSafety} property of {\CODShort}, and {\pConsValidity} of Consensus, for any $i \le k$, the output of $\oConsensus[\vAcc][i + 1]$ contains (in $\vAllCredits\WithCert$, $\vSelectedDebits$, or $\vCancelledDebits$) all transactions from $\Epoch(i)$. Hence, $\forall \vTx \in \Epoch(i): \fStart_{\vAcc}(\vTx) \le \fEpochStart(i + 1)$. Moreover, by definition of $\Epoch(k+1)$, for any $\vTx' \in \Epoch(k+1)$, $\fEnd_{\vAcc}(\vTx') \ge \fEpochStart(k+1)$.
Hence $\fStart_{\vAcc}(\vTx) \le \fEpochStart(i+1) \le \fEpochStart(k+1) \le \fEnd_{\vAcc}(\vTx')$.

\end{proof}

\begin{lemma}\label{lem:consistency:consistent:transition}
If $H$ is consistent with $\prec_{\Execution, \vAcc}$ up to $\Epoch(k)$, then it is  consistent with $\prec_{\Execution, \vAcc}$ up to $\Epoch(k + 1)$.
\end{lemma}
\begin{proof}
From Lemma~\ref{lem:consistency:consistent:group_all}, we know that transactions inside every group for epoch $k + 1$ are ordered such that if $\vTx \prec_{\Execution,\vAcc} \vTx'$, then $\vTx < \vTx'$.

Also, according to Lemma~\ref{lem:consistency:consistent:group_1_2_3} and Lemma~\ref{lem:consistency:consistent:group_2_3}, ordering of groups is consistent as well: i.e., $\forall \vTx_1 \in G_1(k + 1),  \vTx_2 \in G_2(k + 1),  \vTx_3 \in G_3(k + 1)$: $\vTx_1 < \vTx_2 < \vTx_3$ and it cannot be that a transaction from a higher group precedes a transaction from a lower group in the real-time order $\prec_{\Execution, \vAcc}$.

Finally, according to Lemma~\ref{lem:consistency:consistent:epochs}, ordering of transactions between epochs is consistent with the real-time order.

Taking all these facts into consideration together with the fact that $H$ is consistent with $\prec_{\Execution, \vAcc}$ up to $\Epoch(k)$, we conclude that $H$ is consistent with $\prec_{\Execution, \vAcc}$ up to $\Epoch(k + 1)$.

\end{proof}

Now, let us show that $H$ is also legal.

\begin{lemma}\label{lem:consistency:legal:order}
Consider a debit transaction $\vTx$ such that $\vTx \in G_1(k + 1)$. Consider the first invocation of $\oCOD[\vAcc][k + 1].\oSubmit$ that returns $\Tuple{\vDebits\WithCert, \vCredits\WithCert}$ such that $\Tuple{\vTx, *} \in \vDebits\WithCert$. Then, for any credit transaction $\vTx_c \in \vCredits\WithCert$: $\vTx_c < \vTx$.
\end{lemma}
\begin{proof}

Note that by definition of $G_1(k + 1)$, either $\vTx_c \in G_1(k+1)$ or $\vTx_c \in \fTransactions(\oCOD[\vAcc][k+1].\vInitCredits\WithCert)$.
In the latter case, by \Cref{lem:cod-init}, $\vTx_c \in \Epoch(n')$ for some $n' < n$ and, hence, $\vTx_c < \vTx$.

Consider the former case ($\vTx_c \in G_1(k + 1)$).
We need to prove that $\fEnd_{\vAcc}(\vTx_c) < \fEnd_{\vAcc}(\vTx)$.
Let $t$ be the moment when the invocation returned.
It is easy to see that $t < \fEnd_{\vAcc}(\vTx)$.
Moreover, for any $\vTx_c \in \vCredits\WithCert$: $\fEnd_{\vAcc}(\vTx_c) = \fCommitTime(\vTx_c) < t$ since $\vCredits\WithCert$ includes a valid commit certificate for $\vTx_c$.
Hence, $\fEnd_{\vAcc}(\vTx_c) < t < \fEnd_{\vAcc}(\vTx)$.

\end{proof}

Recall that $\rho_{\Execution, \vAcc}(\vTx)$ maps the debit transaction $\vTx$ to the return value of the corresponding $\oTransfer$ operation in $\Execution$.

Also, recall that, for a debit transaction $\vTx$ on $\vAcc$ and a permutation $H$ of $\tT(\Execution, \vAcc)$, $S(H,\vTx)$ denotes the set of credit and successful debit transactions in the prefix of $H$ up to, but not including, $\vTx$, i.e., $S(H,\vTx) = \{ \vTx' \in \tT(\Execution, \vAcc) \mid \vTx' < \vTx \text{ and either } \vTx'.\vRecipient = \vAcc \text{ or } \vTx'.\vSender = \vAcc \text{ and } \rho_{\Execution, \vAcc}(\vTx') = \tOK\}$.

\begin{lemma}\label{lem:consistency:legal:fail}
For any debit $\vTx_f \in \Epoch(k + 1)$ such that $\rho_{\Execution, \vAcc}(\vTx_f) = \tFAIL$, $\fBalance(S(H, \vTx_f), \vAcc) < \vTx_f.\vAmount$.
\end{lemma}
\begin{proof}
Note that $\vTx_f \in G_3(k+ 1)$.
Also, $\forall \vTx \in S(H, \vTx_f)$: either $\vTx \in \bigcup_{i=0}^{k} \Epoch(i)$ or $\vTx \in G_1(k + 1) \cup G_2(k + 1)$.

Now, consider the first moment of time when $\vTx_f$ was returned as a part of a consensus output \atreplace{at}{on} line~\ref{line:recovery:consensus-call} to some correct process, i.e., 
an invocation of $\oConsensus[\vAcc][k+2]$ returned $\Tuple{\Tuple{\vCredits\WithCert, \vSelectedDebits, \vCancelledDebits}, *}$ such that $\vTx_f \in \vCancelledDebits$. 

Consider any $\vTx \in S(H, \vTx_f)$. We want to prove that $\vTx \in \fTransactions(\vCredits\WithCert) \cup \vSelectedDebits$. Indeed, consider two cases:
\begin{itemize}
    \item $\vTx \in \bigcup_{i=0}^{k} \Epoch(i)$: by \Cref{lem:cod-init}, any such transaction $\vTx$ should be either present in $\oCOD[\vAcc][k + 1].\vInitDebits$ or $\oCOD[\vAcc][k + 1].\vInitCredits\WithCert$, and by {\pCODCloseSafety} $\vInitCredits\WithCert \subseteq \vCredits\WithCert$ and $\vInitDebits \subseteq \vSelectedDebits$.

    \item $\vTx \in G_1(k + 1) \cup G_2(k + 1)$: $\fTransactions(\vCredits\WithCert) \cup \vSelectedDebits$ by the definition of groups $G_1(k + 1)$ and $G_2(k + 1)$ and {\pCODCloseSafety} property.
\end{itemize}

By \Cref{lem:cod-init} and the definition of $G_2(k+1)$, $\forall \vTx \in \vSelectedDebits \cup \fTransactions(\vCredits\WithCert)$: $\vTx \in G_2(k+1), G_1(k+1), \text{ or } \left(\bigcup_{i=0}^{k} \Epoch(i)\right)$. Hence, $\vTx \in S(H, \vTx_f)$.
Thus, $S(H, \vTx_f) = \vSelectedDebits \cup \fTransactions(\vCredits\WithCert)$.

Now, recall that, according to the {\pCODCloseSafety} property, the resulted set $\vSelectedDebits$ should be ``maximal by inclusion'', i.e., $\nexists \vTx \in \vCancelledDebits$, such that $\fTotalValue(\fTransactions(\vCredits\WithCert)) \ge \fTotalValue(\vSelectedDebits \cup \{\vTx\})$.

Let us summarize all of the above: 
\begin{enumerate}
    \item $\vTx_f \in G_3(k + 1) \Leftrightarrow \vTx_f \in \vCancelledDebits$;
    \item $\nexists \vTx \in \vCancelledDebits$, such that $\fTotalValue(\fTransactions(\vCredits\WithCert)) \ge \fTotalValue(\vSelectedDebits \cup \{\vTx\})$;
    \item $S(H, \vTx_f) = \vSelectedDebits \cup \fTransactions(\vCredits\WithCert)$.
\end{enumerate}
Hence,  $\fBalance(S(H, \vTx_f), \vAcc) < \vTx_f.\vAmount$.
\end{proof}

\begin{lemma}
\label{lem:consistency:legal:success}
For any debit $\vTx_s \in \Epoch(k + 1)$ such that $\rho_{\Execution, \vAcc}(\vTx_s) = \tOK$, $\fBalance(S(H, \vTx_s), \vAcc) \ge \vTx_s.\vAmount$.
\end{lemma}
\begin{proof}

Consider a debit transaction $\vTx_s \in \Epoch(k + 1)$ such that $\rho_{\Execution, \vAcc}(\vTx_s) = \tOK$.
Note that $\vTx_s$ cannot belong to $G_3(k + 1)$ as this group only contains failed transactions.
Hence, $\vTx_s$ belongs to either $G_1(k + 1)$ or $G_2(k = 1)$.

Suppose $\vTx_s \in G_1(k + 1)$.
Let $t = \fEnd_\vAcc(\vTx_s)$.
Let $\{o_i\}_{i=1}^k$ be the set of all $\oCOD[\vAcc][k + 1].\oSubmit$ operations such that $o_i$ returned $\tOK(\vDebits\WithCert_i, \vOutCredits\WithCert_i)$ by the moment $t$.
By {\pCODProposeSafety}, $\fTotalValue(\vInitDebits \cup \fTransactions(\bigcup_{i=1}^k \vDebits\WithCert_i)) \le \fTotalValue(\fTransactions(\vInitCredits\WithCert) \cup \fTransactions(\bigcup_{i=1}^k \vOutCredits\WithCert_i))$.
Moreover, by \Cref{lem:cod-init} and the definition of $G_1(k+1)$, one can see that $\fDebits(S(H, \vTx_s), \vAcc) \cup \{\vTx_s\} \subseteq \vInitDebits \cup \fTransactions(\bigcup_{i=1}^k \vDebits\WithCert_i))$ and $\fDebits(S(H, \vTx_s), \vAcc) \supseteq \fTransactions(\vInitCredits\WithCert) \cup \fTransactions(\bigcup_{i=1}^k \vOutCredits\WithCert_i)$.
Hence, $\fTotalValue(\fDebits(S(H, \vTx_s), \vAcc) \cup \{\vTx_s\}) \le \fTotalValue(\fDebits(S(H, \vTx_s), \vAcc))$, which is equivalent to saying that $\fBalance(S(H, \vTx_s), \vAcc) \ge \vTx_s.\vAmount$.

Now consider the case when $\vTx_s \in G_2(k + 1)$.
Let us have a look at $\vSelectedDebits$ and $\vAllCredits\WithCert$ that are part of the consensus output (i.e., returned from $\oConsensus[\vAcc][k + 2].\oConsPropose$).
From {\pCODProposeSafety} and the implementation of the algorithm, we know that $\fTotalValue(\vSelectedDebits) \le \fTotalValue(\fTransactions(\vAllCredits\WithCert))$.
Moreover, by \Cref{lem:cod-init} and the definition of $G_2(k + 1)$, $\vDebits(S(H, \vTx_s), \vAcc) \cup \{\vTx_s\} \subseteq \vSelectedDebits$ and $\vCredits(S(H, \vTx_s)) = \fTransactions(\vAllCredits\WithCert)$.
Hence, $\fTotalValue(\fDebits(S(H, \vTx_s), \vAcc) \cup \{\vTx_s\}) \le \fTotalValue(\fDebits(S(H, \vTx_s), \vAcc))$, which is equivalent to saying that $\fBalance(S(H, \vTx_s), \vAcc) \ge \vTx_s.\vAmount$.
\end{proof}

\begin{lemma}\label{lem:consistency:legal:transition}
If $H$ is legal up to $\Epoch(k)$, then it is legal up to $\Epoch(k + 1)$.
\end{lemma}
\begin{proof}
This follows directly from \Cref{lem:consistency:legal:fail,lem:consistency:legal:success}.

\end{proof}

\begin{theorem}\label{theorem:consistency}
    {\Name} satisfies the {\pTransferConsistency} property of {\AssetTransfer}.
\end{theorem}
\begin{proof}
This fact follows from Lemma~\ref{lem:consistency:base}, Lemma~\ref{lem:consistency:consistent:transition} and  Lemma~\ref{lem:consistency:legal:transition}.
\end{proof}

\myparagraph{\pAccountTransactions}

Let us now show that  the  implementation of the $\oGetAccountTransactions$ operation correct, i.e., {\Name} satisfies the {\pAccountTransactions} property.

\begin{theorem} \label{theorem:asset-transfer:account-transactions}
{\Name} satisfies the {\pAccountTransactions} property of {\AssetTransfer}.
\end{theorem}
\begin{proof}
    The proof of this theorem follows from the implementation of the algorithm and the {\pKVSOutputValidity} property of the {\KeyValStorage}.
    Let us assume that operation $\oGetAccountTransactions()$ is invoked by an owner of a correct account $\vAcc$ at time $t_0$ and returns set $\vTxs\WithCert$.
    
    Consider the first part of the property:  $(\fDebits(C(t_0), \vAcc) \cup \fCredits(C(t_0), \vAcc) \subseteq \fTransactions(\vTxs\WithCert)$. 
    It directly follows from the definition of committed transactions, implementation of $\fVerifyCommitCertificate$ function (via verifying function of {\GlobalStorage}) and {\pKVSConsistency} property of the {\KeyValStorage} (in particular, {\GlobalStorage}).
    
    Now, let us consider the second part of the property:  $\forall \Tuple{\vTx, \vCert} \in \vTxs\WithCert: \fVerifyCommitCertificate(\vTx, \vCert) = \True$.
    This is follows from the implementation of $\oGetAccountTransactions$ and $\fVerifyCommitCertificate$ (both are implemented via {\GlobalStorage}), and from the {\pKVSOutputValidity} property of the {\KeyValStorage}.
\end{proof}

\myparagraph{\pTransferConcurrency}

We conclude the proof of correctness of {\Name} by showing that it satisfies the {\pTransferConcurrency} property.
Basically, this property states that if there are no overspending attempts on a correct account $\vAcc$ after some time $t_0$, 
then after some time $t_1 \ge t_0$, the owners of $\vAcc$ do not use consensus.
Since we only use consensus to perform the Recovery procedure (line~\ref{line:recovery:consensus-call}), the proof boils down to an argument that the number of the epochs is finite if the condition in the {\pTransferConcurrency} property holds.

Recall that $C(t)$ denotes the set of all transactions committed by time $t$ and $O(t, \vAcc)$ denotes the set of all active debit transaction of $\vAcc$ at time $t$, i.e., $\{\vTx \mid \vTx.\vSender = \vAcc \text{ and } \fStart_{\vAcc}(\vTx) \le t \le \fEnd_{\vAcc}(\vTx)\}$. 
In the proofs, we consider a correct account $\vAcc$ and a time $t_0$ such that for all $t \ge t_0$: $\fBalance(C(t), \vAcc) \ge \fTotalValue(O(t, \vAcc) \setminus C(t))$.
 
We say that epoch $e$ \emph{starts} at the moment when the first correct replica initializes $\oCOD[\vAcc][e]$ (line~\ref{line:main:initialize-next-cryptocd}) and \emph{ends} when the next epoch ($e + 1$) starts.
\newcommand{\tst}{\mathit{st}}
\newcommand{\tend}{\mathit{end}}
\newcommand{\vEpN}{\vEp_{\mathit{n}}}
\newcommand{\op}{op}
\newcommand{\opmax}{\op_{\mathit{max}}}
We call a debit transaction $\vTx$ \textit{interfering} iff $\oTransfer(\vTx)$ was invoked before $t_0$.
Also, we say that debit transaction $\vTx$ is \textit{active in epoch $e$} iff it was active at some time between the start and the end of epoch $e$.

\begin{lemma}\label{lemma:main:concurrecny:toxic-transactions}
There exists an epoch $e$, such that there are no interfering transactions active in epoch $e$, or the number of the epochs is finite. 
\end{lemma}
\begin{proof}
This lemma follows from the {\pTransferLiveness} property, which says that every $\oTransfer$ operation eventually terminates, and the fact that there can only be a finite number of interfering transactions.
\end{proof}

\begin{lemma}\label{lemma:main:concurrecny:finite-epoch}
The number of epochs is finite.
\end{lemma}
\begin{proof}
We prove this lemma by contradiction.
Let us assume that the number of the epochs is infinite.
Then, by Lemma~\ref{lemma:main:concurrecny:toxic-transactions}, there should exist an epoch $e^*$, such that there are no interfering transactions active in epoch $e^*$.
Note that $e^*$ must have started after $t_0$ (by definition of an interfering transaction).
As number of the epochs is infinite, there should exist an epoch $e^* + 1$.
We know from the implementation that correct account can progress into an epoch $e^* + 1$ only in case one of the clients returned $\tFAIL$ from $\oCOD[\vAcc][e^*].\oSubmit(\ldots)$.
According to the {\pCODProposeSuccess} of {\CODShort} object, it can only happen if there exists a set $S = \{\op_1, \ldots, \op_n\}$, where $op_i$ =  $\oCOD[\vAcc][e^*].\oSubmit(\vDebits_i, \vCredits\WithCert_i)$), such that $\fTotalValue(\vDebits) > \fTotalValue(\vCredits)$, where $\vDebits = \bigcup_{i \in S} \vDebits_i \cup \vInitDebits$ and $\vCredits = \fTransactions(\bigcup_{i \in S} \vCredits\WithCert_i \cup \vInitCredits\WithCert)$. 
Let us consider a minimal (by inclusion) such set $S$.
Each $\oSubmit$ operation can be naturally associated with a $\oTransfer$ operation by which it was invoked.
Let us sort the $\oSubmit$ operations in $S$ by the beginning time of the associated $\oTransfer$ operations in ascending order.
We consider the ``largest'' (w.r.t. the above sorting) operation $\opmax =  \oCOD[\vAcc][e^*].\oSubmit(\vDebits_{max}, \vCredits\WithCert_{max})$.
Note that $\vCredits\WithCert_{max}$ are committed credits and were read from the $\oGlobalStorage$.
From the condition imposed by $\pTransferConcurrency$, we also know that for all $t \geq t_0$, $\fBalance(C(t), \vAcc)\geq \fTotalValue(O(t,\vAcc)\setminus C(t))$. 
Combining these facts together, we can easily see that 
$\fTotalValue(\fTransactions(\vInitCredits\WithCert \cup \vCredits\WithCert_{max})) > \fTotalValue(\bigcup_{i \in S} \vDebits_i \cup \vInitDebits)$.
Then such set $S$ does not exist. A contradiction.

\end{proof}

\begin{theorem} \label{theorem:asset-transfer:concurrency}
    {\Name} satisfies the {\pTransferConcurrency} property of {\AssetTransfer}.
\end{theorem}
\begin{proof}
In Lemma~\ref{lemma:main:concurrecny:finite-epoch}, for a correction account $\vAcc$, we showed that the number of the epochs is finite assuming that there exist $t_0$, such that for all $t \ge t_0$: $\fBalance(C(t), \vAcc) \ge \fTotalValue(O(t, \vAcc) \setminus C(t))$.
According to the implementation, if number of the epochs is finite, then from some moment of time $t_1$, no consensus objects are invoked after time $t_1$ on $\vAcc$. Thus, {\Name} satisfies the {\pTransferConcurrency} property of {\AssetTransfer}.

\end{proof}

\newcommand{\sset}{\widetilde{S}}

\subsection{Proof of Correctness: {\CODLong}}\label{subsection:proof:cryptocd}

In this subsection, we demonstrate the correctness of our implementation of the {\CODLong} abstraction, the most important building block of {\Name}, as specified in \Crefrange{alg:crypto-la-state}{alg:crypto-la-replica}.

\myparagraph{\pCODProposeValidity}
The next two lemmas will help us to prove the {\pCODProposeValidity} property of {\CODShort}.

\begin{lemma}
\label{lemma:ccd:propose-validity-1}
If a correct client returns $\tOK(\vDebits\WithCert, \vOutCredits\WithCert)$ from $\oSubmit(\vDebits, \vCredits\WithCert)$, then $\vDebits = \fTransactions(\vDebits\WithCert)$.
\end{lemma}
\begin{proof}
This lemma directly follows from the implementation. 
When returning from the $\fAccept$ function, the client forms $\vDebits\WithCert$ from $\vDebits$ transactions passed to the function (line~\ref{line:ccd:confirm:form-result}). 
Note that a correct client passes $\vDebits$ transactions from the $\oSubmit(\vDebits, \vCredits\WithCert)$ to the $\fAccept$ function.
\end{proof}

\begin{lemma}
\label{lemma:ccd:propose-validity-2}
If a correct client returns $\tOK(\vDebits\WithCert, \vOutCredits\WithCert)$ from $\oSubmit(\ldots)$, then $\forall \Tuple{\vTx, \vAcceptCert} \in \vDebits\WithCert: \oVerifyCODCert(\vTx, \vAcceptCert) = \True$.
\end{lemma}
\begin{proof}
This lemma directly follows from the implementation. 
A correct client forms a certificate for each transactions that will be accepted by the verifying function $\oVerifyCODCert$.
\end{proof}

\begin{lemma}
\label{lemma:ccd:propose-validity-3}
If a correct client returns $\tOK(\vDebits\WithCert, \vOutCredits\WithCert)$ from $\oSubmit(\ldots)$, then $\forall \Tuple{\vTx, \vAcceptCert} \in \vOutCredits\WithCert: \oVerifyCODCert(\vTx, \vAcceptCert) = \True$.
\end{lemma}
\begin{proof}
This lemma directly follows from the implementation. 
A correct client uses only committed credit transactions.
\end{proof}

\begin{theorem}
    \label{theorem:ccd:validity}
Our implementation of {\CODLong} satisfies the {\pCODProposeValidity} property.
\end{theorem}
\begin{proof}
    This theorem follows directly from Lemma~\ref{lemma:ccd:propose-validity-1}, Lemma~\ref{lemma:ccd:propose-validity-2} and Lemma~\ref{lemma:ccd:propose-validity-3}.
\end{proof}

\myparagraph{\pCODProposeSafety}

The next property of {\CODShort} that we will address is the {\pCODProposeSafety}.

A certificate $\vPrepareCert$ is called a \textit{prepare certificate} for a set $\vDebits$ iff it is a threshold signature of a message $\Message{\mPrepareResp, \vDebits}$ (formed at line~\ref{line:ccd:prepare:threshold-sign}).
If a set of debit transactions $\vDebits$ has a prepare certificate, then it is called \textit{prepared}.

Let us consider all prepared sets of debits transactions for a $\oCOD$ object.

\begin{lemma}
\label{lemma:ccd:comparability}
Any two prepared debit sets $\vDebits_i$ and $\vDebits_j$ are comparable (w.r.t. $\subseteq$): either $\vDebits_i \subseteq \vDebits_j$ or $\vDebits_j \subset \vDebits_i$.
\end{lemma}
\begin{proof}
Let us consider a prepared set of debits $\vDebits_i$.
As it is prepared, there exists a prepare certificate $\vPrepareCert_i$.
As a prepare certificate is, essentially, a threshold signature, then there exists a quorum $Q$ that signed message $\Message{\mPrepareResp, \vDebits_i}$.

Similarly, there exists a quorum $Q'$ that signed a message $\Message{\mPrepareResp, \vDebits_j}$ for $\vDebits_j$.

Due to the quorum intersection property there exists a correct replica $r$, such that $r \in Q\cap Q'$.
This implies that $r$ signed both  $\Message{\mPrepareResp, \vDebits_i}$ and $\Message{\mPrepareResp, \vDebits_j}$.
Note that correct replicas sign only comparable sets of transactions (w.r.t.~$\subseteq$).
Thus, either $\vDebits_i \subseteq \vDebits_j$ or $\vDebits_j \subset \vDebits_i$. 
\end{proof}

\begin{lemma}
\label{lemma:ccd:propose-safety-2}
The total amount spent by $\vInitDebits$ and transactions accepted by a $\oCOD$ instance never exceeds the sum of the committed credits for the account $\oCOD.\vAcc$.
\end{lemma}
\begin{proof}
Let us note that $\vInitDebits$ is covered by $\vInitCredits\WithCert$ by definition.
Also, from the implementation it follows that any prepared set of debits contains $\vInitDebits$.

According to the implementation it holds that if $\vTx$ is accepted, then there exists a prepared set of debit transactions $\vDebits$, such that $\vTx \in \vDebits$.
By Lemma~\ref{lemma:ccd:comparability}, all prepared set of debits are related by containment. 
Then, for any finite set $S$ of accepted transactions, there exists a prepared set $\vDebits_{all}$, such that it contains all transactions from $S$.
A corresponding prepare certificate for $\vDebits_{all}$  is a threshold signature formed from signatures of  of $\Message{\mPrepareResp, \vDebits_{all}}$ of some quorum $Q$. 
In every quorum there are at least $f$ signatures made by correct replicas.
Note that correct replicas only sign $\Message{\mPrepareResp, \vDebits_{all}}$ if they saw enough committed credits (line~\ref{line:ccd:prepare:check-replica}).
Consequently, the total amount spent by $\vInitDebits$ and transactions accepted by a $\oCOD$ instance never exceeds the sum of committed credits for the account $\oCOD.\vAcc$.
\end{proof}

Now, we want to prove that if $\vAcc$ is a correct account, then the total amount of initial debits and debits accepted by the {\CODShort} by time $t$, does not exceed the total amount of all credits returned by the $\oSubmit$ operation by time $t$. 
We prove this by showing that our implementation satisfies even stronger property in the following lemma.

\begin{lemma}\label{lemma:ccd:propose-safety-1}
For any subset $S = \{o_1, \dots, o_k\}$ of $\oSubmit$ operations invoked by correct clients such that $\forall i \in \{1,\dots,k\}: o_i$ returns $\tOK(\vDebits\WithCert_i, \vOutCredits\WithCert_i)$, it holds that $\fTotalValue(\vInitDebits \cup \bigcup_{i=1}^k \vDebits\WithCert_i) \le \fTotalValue(\vInitCredits\WithCert \cup \bigcup_{i=1}^k \vCredits\WithCert_i)$.
\end{lemma}
\begin{proof}
Let us consider a set of operations $S$ from the lemma condition.
From the {\pCODProposeValidity} property, we know that if there exists $i \in \{1,\dots,k\}$, such that  $\vTx \in \vDebits\WithCert_i$, then $\vTx$ is accepted.
Let us match any operation $o_i \in S$ with prepared set of debit transactions $\vPreparedDebits_i$ that was collected during execution of a given operation.
By Lemma~\ref{lemma:ccd:comparability}, all prepared sets of debit transaction are comparable (w.r.t., $\subseteq$). Thus, in the set $\{\vPreparedDebits_i\}_{i=1}^{i=k}$, there should exist maximum set $\vPreparedDebits_m$.

A corresponding prepare certificate for $\vPreparedDebits_m$  is a threshold signature formed from signatures of  of $\Message{\mPrepareResp, \vPreparedDebits_m}$ of some quorum $Q$.
In every quorum there are at least $f$ signatures made by correct replicas.
Note that correct replicas only sign $\Message{\mPrepareResp, \vPreparedDebits_m}$ if they accounted enough committed credits (line~\ref{line:ccd:prepare:check-replica}) to cover all transactions from there.
As $\vOutCredits\WithCert_m$ contains all credits correct replicas saw before signing the message, $\fTotalValue(\fTransactions(\vOutCredits\WithCert_i)) \ge \fTotalValue(\vPreparedDebits_m)$. 
From the implementation, $\fTransactions(\vDebits\WithCert_m) \subseteq \vPreparedDebits_m$.
Even more, $\forall i \in \{1, \ldots, k\}:  \fTransactions(\vDebits\WithCert_i) \subseteq \vPreparedDebits_i$.
Also, recall that $\forall i \in \{1,\ldots,k\}: \vPreparedDebits_i \subseteq \vPreparedDebits_m$.
Hence, $\fTotalValue(\vInitDebits \cup \bigcup_{i=1}^k \vDebits\WithCert_i) \le \fTotalValue(\vInitCredits\WithCert \cup \bigcup_{i=1}^k \vCredits\WithCert_i)$.
\end{proof}

\begin{lemma}
\label{lemma:ccd:propose-safety-3}
If $\vTx \in \vRestrictedDebits$, then $\vTx$ is never accepted by a  {\CODShort} object.
\end{lemma}
\begin{proof}
This lemma follows from the implementation. 
Correct replicas do not respond to messages containing transactions from $\vRestrictedDebits$ during Prepare phase (line~\ref{line:ccd:prepare:no-restricted}).
\end{proof}

\begin{theorem}
    \label{theorem:ccd:safety}
Our implementation of {\CODLong} satisfies the {\pCODProposeSafety} property.
\end{theorem}
\begin{proof}
    This theorem follows from Lemma~\ref{lemma:ccd:propose-safety-2}, Lemma~\ref{lemma:ccd:propose-safety-1} and Lemma~\ref{lemma:ccd:propose-safety-3}.
\end{proof}

\myparagraph{\pCODProposeSuccess}

The next step is to show that our {\CODShort} protocol satisfies the {\pCODProposeSuccess} property.

\begin{theorem}
    \label{theorem:ccd:propose-success}
Our implementation of {\CODLong} satisfies the {\pCODProposeSuccess} property.
\end{theorem}

\begin{proof}
We prove this theorem by contradiction. Let us assume that  $\vAcc$ is a correct account, no client invokes the $\oClose$ operation, and for every possible subset $S$ of invoked operations $\oSubmit(\vDebits_i, \vCredits\WithCert_i)$: $\fTotalValue(\vDebits) \le \fTotalValue(\vCredits)$ (where $\vDebits = \bigcup_{i \in S} \vDebits_i \cup \vInitDebits$ and $\vCredits = \fTransactions(\bigcup_{i \in S} \vCredits\WithCert_i \cup \vInitCredits\WithCert)$); however there exists an operation $\oSubmit(\vDebits_p, \vCredits\WithCert_p)$  invoked by a client $p$ that returns $\tFAIL$.
Before we continue with the proof, let us note that any valid reply $\Message{\mPrepareResp, \vCredits\WithCert, \vDebits\WithCert, \vSignature}$ from a replica should satisfy the following condition: $\forall \Tuple{\vTx, \Tuple{\vDeps_{\vTx}, \vSignature_{\vTx}}} \in \vDebits\WithCert$, $\forall \vId \in \vDeps_{\vTx}: \exists \Tuple{\vTx', \vCert_{\vTx'}} \in \vCredits\WithCert$ such that $\vId = \vTx'.\vId$.
Also, $\forall i: \vInitDebits \subseteq \fTransactions(\vDebits\WithCert_i)$ and $\vInitCredits\WithCert \subseteq \vCredits\WithCert_i$.

Now, consider operation $op = \oSubmit(\vDebits_p, \vCredits\WithCert_p)$ executed by a correct client $p$.
Note that $op$ can only return $\tFAIL$ from the $\fPrepare$ function, as $\fAccept$ returns $\tFAIL$ only after receiving $\mClosed$ reply, and this is impossible, since no client invokes the $\oClose$ operation.
Thus, client $p$ should have collected a set of messages $M = \{ \Message{\mPrepareResp, \vDebits\WithCert_k, \vCredits\WithCert_k, \vSignature_k} \}_{k \in \{1,\ldots,|M|\}}$ from a quorum $Q$.
Otherwise, either $\oClose$ operation was invoked by some client, which is impossible by
Thus, $\tFAIL$ is returned by the $\fPrepare$ function. 
To be more precise it is returned at line~\ref{line:ccd:prepare:balance-fail}.
It means that, naturally, there should exist a set $S'$ of $\oSubmit(\vDebits_i, \vCredits\WithCert_i)$ operations and a set of credits $\vCredits\WithCert_{extra}$, such that $\bigcup_{i\in S'} \vDebits_i \cup \vInitDebits =  \fTransactions(\bigcup_{k\in M} \vDebits\WithCert_k)$ and $\bigcup_{i\in S'} \vCredits_i  \cup \vInitCredits\WithCert = \fTransactions(\bigcup_{k\in M} \vCredits\WithCert_k \cup \vCredits\WithCert_{extra})$.
As we assume that $p$ returns $\tFAIL$, then $ \fTotalValue(\fTransactions(\bigcup_{k\in M} \vDebits\WithCert_k)) > \fTotalValue(\fTransactions(\bigcup_{k\in M} \vCredits\WithCert_k \cup \vCredits\WithCert_{extra}))$.
Then, it also means that $\fTotalValue(\bigcup_{i\in S'} \vDebits_i \cup \vInitDebits) > \fTotalValue(\bigcup_{i\in S'} \vCredits_i  \cup \vInitCredits\WithCert)$.
However, this contradicts with the assumption that 
for any subset $S'$ of invoked $\oSubmit(\vDebits_i, \vCredits\WithCert_i)$, such that $\fTotalValue(\vDebits) > \fTotalValue(\vCredits)$ (where $\vDebits = \bigcup_{i \in S'} \vDebits_i \cup \vInitDebits$ and $\vCredits = \fTransactions(\bigcup_{i \in S'} \vCredits\WithCert_i \cup \vInitCredits\WithCert)$).
Consequently, our assumption that there exists a $\oSubmit(\vDebits_p, \vCredits\WithCert_p)$ operation, which returns $\tFAIL$, is wrong. 
This means that our implementation of {\CODShort} satisfies {\pCODProposeSuccess} property.  
\end{proof}

\myparagraph{\pCODCloseValidity}
Now we continue with the proof of the {\pCODCloseValidity} property of the {\CODShort} object, which is relatively simple.

\begin{theorem}
    \label{theorem:ccd:close-validity}
Our implementation of {\CODLong} satisfies the {\pCODCloseValidity} property.
\end{theorem}
\begin{proof}
Trivially follows from the algorithm implementation.
\end{proof}

\myparagraph{\pCODCloseSafety}

\begin{lemma}
\label{lemma:ccd:close-safety-1}
If a correct client returns $\Tuple{\vCODState, \vStateCert}$ from $\oClose(\vPendingDebits)$, then:
    \begin{itemize}
        \item $\forall \vTx \in \vPendingDebits: \vTx \in \vCODState.\vSelectedDebits$ or  $\vTx \in \vCODState.\vCancelledDebits$;
        \item $\vInitDebits \subseteq \vCODState.\vSelectedDebits$ and $\vRestrictedDebits \subseteq \vCODState.\vCancelledDebits$;
        \item $\vCODState.\vSelectedDebits \cap \vCODState.\vCancelledDebits = \emptyset$;
        \item $\fTotalValue(\fTransactions(\vCODState.\vCredits\WithCert)) \ge \fTotalValue(\vCODState.\vSelectedDebits)$
    \end{itemize}
\end{lemma}
\begin{proof}
This lemma trivially follows from the implementation of the $\fExtractSafeTransactions$ function (\crefrange{line:cod:split-debits:start}{line:cod:split-debits:end}) and the way it is used in the implementation of the $\oClose$ operation (\crefrange{line:cod:close:start}{line:cod:close:end}).
\end{proof}

\begin{lemma}
\label{lemma:ccd:close-safety-2}
If a correct client $p$ returns $\Tuple{\vCODState, \vStateCert}$ from $\oClose(\vPendingDebits)$, then for every transaction $\vTx$ accepted by this $\oCOD$: $\vTx \in \vCODState.\vSelectedDebits$.  
\end{lemma}
\begin{proof}
Let us consider an accepted transaction $\vTx$.
If $\vTx$ is accepted, then there exists a certificate $\vAcceptCert$, such that $\oVerifyCODCert(\vTx, \vAcceptCert) = \True$.
This means that $\vAcceptCert$ contains a threshold signature formed by a client $q$ from valid signatures $\{\vSignature_r\}_{r \in Q}$ made by a quorum of replicas $Q$.

As $p$ is a correct client, then during $\oClose$ operation it should have collected valid $\mCloseResp$ responses from a quorum $Q'$.
By quorum intersection property there exists a correct replica $r$, such that  $r \in Q \cap Q'$. 
Then, one of the following should have happened before the other: (i) $r$ produced a signature $\vSignature_r$ and sent $\Message{\mAcceptAck, \vSignature_r}$ to $q$; (ii) $r$ sent to $p$ a valid $\mCloseResp$ reply.
Note that as $r$ is correct, (ii) could not happen before (i). In this case, $r$ would send $\mClosed$ message to $q$.
Thus, (i) happened before (ii).
As $r$ is correct, it signed $\vPreparedDebits$ (such that $\vTx \in \vPreparedDebits$) and attached it to $\mCloseResp$ that was sent to $p$.
As $p$ is correct and due to the way $\fExtractSafeTransactions$ is implemented: $\vPreparedDebits \subseteq \vCODState.\vSelectedDebits$.

The above reasoning is valid for any accepted transaction $\vTx$ and any $\oClose$ operation performed by a correct client.
\end{proof}

\begin{lemma}
\label{lemma:ccd:close-safety-3}
If a correct client obtains $\Tuple{\vCODState, \vStateCert}$  from $\oClose(\vPendingDebits)$, then for any $\oSubmit$ operation that returns  $\tOK(\vDebits\WithCert, \vOutCredits\WithCert)$ to a correct client: $\vOutCredits\WithCert \subseteq \vCODState.\vCredits\WithCert$ and $\forall \Tuple{\vTx, \vCert} \in \vCODState.\vCredits\WithCert$: $\fVerifyCommitCertificate(\vTx, \vCert) = \True$. Moreover, $\vInitCredits\WithCert \subseteq \vCODState.\vCredits\WithCert$.
\end{lemma}
\begin{proof}
The first part of the lemma follows from the implementation of $\fAccept$ phase. 
Upon receiving $\mAcceptRequest$ message, correct replicas add $\vOutCredits\WithCert$ to their local set of credit transactions (line~\ref{line:ccd:accept:add-credits}). 
For any such set there exist at least $f + 1$ correct replicas that store these credit transactions.
Then, by quorum intersection property, during any execution of $\oClose$ operation invoked by client $c$, $\forall \vTx \in \vOutCredits\WithCert$ will be read by $c$.

The second part of the lemma follows from the fact that replicas ignore messages if credits do not come together with commit certificates (line~\ref{line:ccd:accept:ignore-fake-credits}).

\end{proof}

\begin{theorem}
    \label{theorem:ccd:close-safety}
Our implementation of {\CODLong} satisfies the {\pCODCloseSafety} property.
\end{theorem}
\begin{proof}
    This theorem follows from Lemma~\ref{lemma:ccd:close-safety-1}, Lemma~\ref{lemma:ccd:close-safety-2} and Lemma~\ref{lemma:ccd:close-safety-3}.
\end{proof}

\myparagraph{\pCODLiveness}
Finally, we want to show that any operation invoked by a correct process eventually returns and, hence, to prove that our implementation satisfies the  {\pCODLiveness} property. 

Let us recall that once a correct client returns $\tFAIL$ from an invocation of $\oSubmit$, it never invokes this operation on a given {\CODShort} object again.

\begin{lemma}
\label{lemma:ccd:liveness-1}
Every $\oClose$ operation invoked by a correct client eventually returns.
\end{lemma}
\begin{proof}
The lemma follows from the fact that the operation makes a constant number of steps, all {\WaitFor} conditions will be satisfied as the client waits for a quorum of replies and there exists a quorum that consists of only correct replicas that eventually responds.
\end{proof}

\begin{lemma}
    \label{lemma:ccd:propose-close}
    If a correct client $p$ does not return from $\oSubmit$ operation, then $\oClose$ is never invoked.
\end{lemma}
\begin{proof}
 Indeed, otherwise, if $\oClose$ is invoked, then $p$ will eventually receive $\mClosed$ reply and return $\tFAIL$
\end{proof}

\begin{lemma}
    \label{lemma:ccd:wait-freedom}
If some correct client invokes $\oSubmit$, then some (not necessarily the same) correct client eventually returns from $\fPrepare$ function.
\end{lemma}
\begin{proof}
We prove this lemma by contradiction. 
Let us assume that all processes that invoked $\oSubmit$ operation of a given ${\CODShort}$ object never return from $\fPrepare$ function.
Note, that in this case $\oClose$ is never invoked (Lemma~\ref{lemma:ccd:propose-close}).
As an account is shared by a finite number of clients, there is a finite number of $\oSubmit(\vDebits_i, \vCredits\WithCert_i)$ invocations.
Due to the fact that the number of invocations is finite, eventually every client will receive a set of equal $\Message{\mPrepareResp, \vDebits\WithCert, \vCredits\WithCert}$ replies from some quorum $Q$. 
No client can return $\tFAIL$ (by the assumption), however as a client obtains a set of equal replies, the condition at line~\ref{line:ccd:prepare:equal-sets} will be satisfied and it will return $\tOK(\ldots)$.
This contradicts with our initial assumption.
Thus, if some correct client invokes $\oSubmit$, then some correct client eventually returns from $\oSubmit$ operation.
\end{proof}

\begin{lemma}\label{lemma:ccd:accept-liveness}
If a correct client invokes the $\fAccept$ function while executing the $\oSubmit$ operation, it eventually returns from it.
\end{lemma}
\begin{proof}
The proof of this lemma directly follows from the implementation: i.e., it should eventually get a quorum of valid $\mAcceptAck$ messages or at least one valid $\mCloseResp$ message.
\end{proof}

\begin{lemma}\label{lemma:ccd:liveness-2}
If a correct client  $p$ invokes $\oSubmit(\vDebits, \vCredits\WithCert)$, then it eventually returns from it.
\end{lemma}
\begin{proof}
We prove this lemma by contradiction. 
Let us assume that $p$ never returns from the $\oSubmit$ function. The only place it can stuck is while executing $\fPrepare$ function: indeed, correct processes always return from $\fAccept$ by Lemma~\ref{lemma:ccd:liveness-2}. 

Now, let us consider two scenarios: (i) there exists a time $t_0$, starting from which no client returns from the $\oSubmit$ operation, (ii) no such time $t_0$ exists, i.e., for any $t$, there always exists some client $q$, which returns from $\oSubmit$ function, after time $t$.

We start with the first scenario.
As an account is shared by a finite number of clients, there is a finite number of $\oSubmit(\vDebits_i, \vCredits\WithCert_i)$ invocations.
Let us consider all operations $\oSubmit(\vDebits_i, \vCredits\WithCert_i)$ that are active after time $t_0$.
There should exist time $t_1 \ge t_0$, such that all operations that do not return from $\fPrepare$ are active at all times $t > t_1$.
There exists at least one operation that is active at that time -- one that was invoked by client $p$.
Due to the fact that the number of invocations is finite, eventually every client will receive a set of equal $\Message{\mPrepareResp, \vDebits\WithCert_i, \vCredits\WithCert_i, \vSignature_i}$ replies from some quorum $Q$.
No client can return $\tFAIL$ (by the assumption that clients do not exit $\fPrepare$ function), however as a client obtains a set of equal replies, the condition at line~\ref{line:ccd:prepare:equal-sets} will be satisfied and it will return $\tOK(\ldots)$ from the $\fPrepare$ function.
This way, client $p$  returns from the $\oSubmit$ operation. This contradicts our assumption, and, consequently, (i) is impossible.

Let us consider now the second scenario: for any $t$, there always exists some client $q$, which returns from $\oSubmit$ operation, after time $t$.
Note that the first $\Message{\mPrepare, \vDebits\WithCert, \vCredits\WithCert, \vSubmitDebits\WithCert}$ message client $p$ sent during the execution of $\oSubmit$ operation will eventually reach all correct replicas and they will add all transactions from $\vSubmitDebits\WithCert$ to their local set of debit transactions (line~\ref{line:ccd:prepare:add-debits-replica}).
Let denote the time it happens as $t_1$.
From the scenario we consider, there should be an infinite number of $\oSubmit$ invocations on a given {\oCOD} object: indeed, otherwise there should exists a time $t_n$, such that no operation returns after $t_n$.
As a consequence, there should exist a $\oSubmit$ operation, which was invoked after $t_1$ and returned $\tOK(\ldots)$ (recall that correct client do not invoke $\oSubmit$ once it returns $\tFAIL$). 
As it started after $t_1$, client should have collected a set $S$ of  $\Message{\mPrepareResp, \vDebits\WithCert_i, \vCredits\WithCert_i, \vSignature_i}$, such that (i) $\forall i$: $\vSubmitDebits\WithCert \subseteq \vDebits\WithCert_i$ (ii) every message in $S$ has equal set of $\vDebits\WithCert_i$.
Then, there exists a prepare set of debits that includes all of the transactions from $\vSubmitDebits\WithCert$. Hence, eventually client $p$ should return from the $\fPrepare$ function at line~\ref{line:prepare:fast-return}, and then complete $\oSubmit$ operation after returning from the $\fAccept$ function. However, this contradicts with our assumption, this means that (ii) is impossible.

We considered two potential scenarios and came to a contradiction in both of them. 
Note that the set of scenarios is exhaustive. Hence, 
\end{proof}

\begin{theorem}
\label{theorem:ccd:liveness}
Our implementation of {\CODLong} satisfies the {\pCODLiveness} property.
\end{theorem}
\begin{proof}
This theorem follows from Lemma~\ref{lemma:ccd:liveness-1} and Lemma~\ref{lemma:ccd:liveness-2}.
\end{proof}

\begin{theorem}
\label{theorem:ccd:correctness}
Algorithm~\ref{alg:crypto-la-client} and Algorithm~\ref{alg:crypto-la-replica} correctly implement {\CODLong}.
\end{theorem}
\begin{proof}
All properties of {\CODLong} hold:
\begin{itemize}
    \item {\pCODProposeValidity} follows from Theorem~\ref{theorem:ccd:validity};

    \item {\pCODProposeSafety} follows from Theorem~\ref{theorem:ccd:safety};
    
    \item {\pCODProposeSuccess} follows from Theorem~\ref{theorem:ccd:propose-success};
    
    \item {\pCODCloseValidity} follows from Theorem~\ref{theorem:ccd:close-validity};
    
    \item {\pCODCloseSafety} follows from Theorem~\ref{theorem:ccd:close-safety};
    
    \item {\pCODLiveness} follows from Theorem~\ref{theorem:ccd:liveness}. 
\end{itemize}
\end{proof}

\subsection{Proof of Correctness: {\KeyValStorage}}\label{subsection:proof:keyval}
In this subsection we show that our implementation of the {\KeyValStorage} is correct, i.e., it satisfies {\pKVSConsistency}, {\pKVSInputValidity}, {\pKVSOutputValidity} and {\pKVSLiveness}.

\begin{lemma}
\label{lemma:kvs:consistency}
For any value $v$ and a key $k$, if there exists $\vKVSCert$ such that $\fKVSVerifyStoredCert(k, v, \vKVSCert)= \True$ at the moment when $\oReadKey(k)$ was invoked by a correct client, then the output of this operation $\vVs\WithCert_{out}$ will contain $v$ (paired with a certificate), i.e., $\Tuple{v, *} \in \vVs\WithCert_{out}$.
\end{lemma}
\begin{proof}
If there exists $\vKVSCert$ such that $\fKVSVerifyStoredCert(k, v, \vKVSCert)= \True$, then there should exist a quorum of processes $Q$ and $\vVs\WithCert$, such that every process $r \in Q$ signed a message $\Message{\mAppendKeyResp, k, \fMerkleTree(\{v \mid \Tuple{v, *} \in \vVs\WithCert\}).\vMTRoot}$.
Note that any correct process $r \in Q$ should have also added $\vVs\WithCert$ in their $\vLog[k]$.
Let us consider a correct client $p$ that returns $\vVs\WithCert_{out}$ from $\oReadKey(k)$ invoked when there existed $\vKVSCert$ such that  $\fKVSVerifyStoredCert(k, v, \vKVSCert)= \True$.
Then, there should exist a quorum of processes $Q'$ that responded with $\Message{\mReadKeyResp, \vVs\WithCert_i}$ messages.
According to the quorum intersection property, there should exist a correct replica $r$, such that $r \in Q \cap Q'$.
As $r$ added $v$ (with its certificate) to its $\vLog[k]$, it should have also included it into $\Message{\mReadKeyResp, \vVs\WithCert_r}$ that it replied with to $p$.
By the implementation, $\vVs\WithCert_{out} = \bigcup_i \vVs\WithCert_{i}$. Thus $p$ includes $v$ (with its certificate) in $\vVs\WithCert_{out}$.
\end{proof}

\begin{theorem}
\label{thm:kvs:correctness}
Algorithm~\ref{alg:keyval-storage} is a correct implementation of the {\KeyValStorage}.
\end{theorem}
\begin{proof}
All properties of the {\KeyValStorage} hold:
\begin{itemize}
    \item {\pKVSInputValidity} follows from the implementation;
    \item {\pKVSOutputValidity} follows from the implementation;
    \item {\pKVSConsistency} follows from Lemma~\ref{lemma:kvs:consistency};
    \item {\pKVSLiveness} follows from the fact that each operation makes a constant number of steps and that any $\WaitFor$ condition will be satisfied, as there exist $n-f$ correct processes that form a quorum and must eventually respond.
\end{itemize}

\end{proof}

\begin{ppreview}
\section{Latency Proofs} \label{app:proof-latency}

\begin{lemma}
\label{lemma:latency:aos-write}
Latency of the $\oAppendKey$ operation of {\AppendOnlyStorage} invoked by a correct client is $1$ round-trip.
\end{lemma}
\begin{proof}
Follows directly from the implementation (\Cref{alg:keyval-storage}, \crefrange{line:aos:append-key-start}{line:aos:write-values-end}).
The client simply sends the request to the replicas and waits for signed acknowledgments from a quorum.
\end{proof}

\begin{lemma}
\label{lemma:latency:aos-read}
    Latency of the $\oReadKey$ operation of {\AppendOnlyStorage} invoked by a correct client is $2$ round-trips.
\end{lemma}
\begin{proof}
Similarly, this fact follows directly from the implementation (\Cref{alg:keyval-storage}). In the algorithm, the client first waits for a quorum of $\mReadKeyResp$ replies ($1$ round-trip) and then performs a write-back. The latency of the latter is $1$ RTT as shown in \Cref{lemma:latency:aos-write}.
\end{proof}

\begin{theorem} \label{thm:latency}
    {\Name} exhibits $k$-overspending-free latency (as defined in \Cref{sec:problem-statement}) of $k+4$ round-trips.
\end{theorem}
\begin{proof}
    In the definition of $k$-overspending-free latency, we only consider operations that start after the system stabilizes from all overspending attempts, i.e., when no epoch changes happen during the operation execution.
    In this case, the client will do the following sequence of actions:
    \begin{itemize}
        \item Execute $\oReadKey$ on \cref{line:main:fetch-cryptocd-state,line:main:read-global-storage,line:main:read-debits} and $\oAppendKey$ on \cref{line:main:put-tx-to-account-storage} concurrently. 
        By \Cref{lemma:latency:aos-write,lemma:latency:aos-read}, this step will take at most 2 round-trips;

        \item Send the $\mInitCOD$ message to replicas (\cref{line:main:init-cryptocd}). This step does not affect the latency, as the client does not wait for replicas' replies and moves on directly to the next step;

        \item Execute $\oCOD[\vAcc][\vEp].\oSubmit$ on \cref{line:main:crypto-la-propose}. As we discuss in the next paragraph, this step takes at most $k+1$ round-trips;

        \item Finally, execute $\oGlobalStorage[\vAcc].\oAppendKey$ on \cref{line:main:write-transaction-after-crypto-la}. By \Cref{lemma:latency:aos-write}, this step will take 1 round-trip.
    \end{itemize}

    All we have left to show is that\atremove{, in this context,} each client will return from the $\oSubmit$ operation of\atremove{ the} $\oCOD[\vAcc][\vEp]$ after at most $k + 1$ RTTs.
    We consider the Prepare phase first and show that the upper bound \atreplace{for}{on} the number of loop iterations (line~\ref{line:prepare:while-loop}) is $k$. 
    The client $c$ can return from the loop when it either (i) detects an overspending attempt (line~\ref{line:prepare:balance-check}) or (ii) converges on the debit sets received from a quorum of replicas (line~\ref{line:ccd:prepare:equal-sets}).
    Note that (i) is impossible given the conditions of this theorem, so we can safely consider only (ii). 

    \atrev{For all debit transactions that terminated before the client started its $\oTransfer$ operation, the client will observe them in the {\AccountStorage} and, thus, will include them in its input to {\CODShort}.
    Hence, each time, after receiving a quorum of valid replies from the replicas, either the client learns about at least one new concurrent debit transaction, or terminates.
    Thus, after at most $k$ round-trips, there will be no new transactions to learn and the client will exit the Prepare phase on \cref{line:prepare:return-ok-quorum}.}

    Finally, the latency of the Accept phase is always $1$ round-trip, which results in a latency of $k + 1$ RTTs for $\oCOD[\vAcc][\vEp].\oSubmit$ operation.
\end{proof}
\end{ppreview}

\begin{atreview}
\begin{proof}[Proof of \Cref{thm:main} (\Cref{sec:problem-statement})]
    Follows from the theorems claiming that {\Name} satisfies all the properties of an asset transfer system with {\pTransferConcurrency} (\Cref{theorem:asset-transfer:safety,theorem:asset-transfer:validity,theorem:consistency,theorem:asset-transfer:account-transactions,theorem:asset-transfer:liveness,theorem:asset-transfer:concurrency}) as well as \Cref{thm:latency} asserting the $k$-overspending-free latency of {\Name}.
\end{proof}
\end{atreview}

\end{document}